\documentclass[a4paper,12pt]{article}
\usepackage{natbib}
\usepackage[english]{babel}
\usepackage{amsmath,amssymb,enumerate,multirow}
\usepackage{bm,latexsym,mathrsfs}
\usepackage{graphicx}
\usepackage[T1]{fontenc}
\usepackage{subfig}
\usepackage{morefloats}
\usepackage{setspace}
\usepackage[dvipsnames]{xcolor}
\usepackage{threeparttable}

\usepackage{authblk}

\usepackage{tikz}
\usetikzlibrary{shapes, backgrounds}

\usepackage{rotating}
\usepackage{changepage}

\usepackage{url}

\usepackage{array}
\newcolumntype{C}[1]{>{\centering\let\newline\\\arraybackslash\hspace{0pt}}m{#1}}

\newcommand{\iid}{\stackrel{\mathrm{i.i.d.}}{\sim}}
\newcommand{\ind}{\stackrel{\mathrm{ind}}{\sim}}

\begin{document}

\title{\bf A Bayesian approach to study synergistic interaction effects in \\ in-vitro drug combination experiments}
\date{}

\author[1,2,5]{Andrea Cremaschi}
\author[1,3]{Arnoldo Frigessi}
\author[2,4]{Kjetil Task\'{e}n}
\author[1]{Manuela Zucknick}

\affil[1]{\footnotesize Oslo Centre for Biostatistics and Epidemiology (\textbf{OCBE}), University of Oslo, Oslo, Norway}
\affil[2]{\footnotesize Department of Cancer Immunology, Institute of Cancer Research, Oslo University Hospital, Oslo, Norway}
\affil[3]{\footnotesize Oslo Centre for Biostatistics and Epidemiology (\textbf{OCBE}), Oslo University Hospital, Oslo, Norway}
\affil[4]{\footnotesize Institute of Clinical Medicine, University of Oslo, Oslo, Norway}
\affil[5]{\footnotesize Yale-NUS College, Singapore.}

\maketitle

\small{
\textbf{Abstract}

	\noindent
	In cancer translational research, increasing effort is devoted to the study of the combined effect of two drugs when they are administered simultaneously. In this paper, we introduce a new approach to estimate the part of the effect of the two drugs due to the interaction of the compounds, i.e. which is due to synergistic or antagonistic effects of the two drugs, compared to a reference value representing the condition when the combined compounds do not interact, called \emph{zero-interaction}. We describe an in-vitro cell viability experiment as a random experiment, by interpreting cell viability as the probability of a cell in the experiment to be viable after treatment, and including information related to different exposure conditions. We propose a flexible Bayesian spline regression framework for modelling the viability surface of two drugs combined as a function of the concentrations. Since the proposed approach is based on a statistical model, it allows to include replicates of the experiments, to evaluate the uncertainty of the estimates, and to perform prediction. We test the model fit and prediction performance on a simulation study, and on an ovarian cancer cell dataset. Posterior estimates of the zero-interaction level and of the synergy term, obtained via adaptive MCMC algorithms, are used to compute interpretable measures of efficacy of the combined experiment, including relative volume under the surface (rVUS) measures to summarise the zero-interaction and synergy terms and a bi-variate alternative to the well-known EC$_{50}$ measure.

}

\clearpage
\section{Introduction}

Drugs are usually administered to cancer patients following a protocol designed for the specific cancer. When a drug reveals to be no longer effective, due to the development of resistance, its concentration might be increased, or the drug itself might be changed. The effectiveness of a drug against a particular malignancy can, in some cases, be estimated in pre-clinical in-vitro concentration-response experiments, such as viability assays or other cell count assays, in a set-up called cancer drug sensitivity screening (CDSS). Primary cancer cells, derived from patient material, or cell-lines derived from patients, are treated with drugs in-vitro, thus providing an estimate of the number of viable cells after exposure to treatment. These studies are often performed for one drug at a time (\emph{monotherapy}) in high-throughput screenings. The data produced by these assays present different sources of variability, related to biological factors such as cell growth or the drug's mechanism of action, in addition to measurement errors. The mitigation of such variability components is difficult to tackle, and often viability data are analysed by simply fitting a parametric concentration-response curve to the data, after removal of background noise elements. The most commonly used function for this purpose is the Hill equation \citep{Hill_1910}, or median-effect law \citep{Chou_Talalay_1984}, also called four-parameter log-logistic (4LL) curve:
\begin{equation}\label{eq:2LL}
f(x | m, \lambda) = a + (b - a)\left(1 + 10^{\lambda(\log_{10}(x) - m))} \right)^{-1}, \quad x\in \mathbb{R}^+,
\end{equation}
where $a$ and $b$ are the lower and upper asymptotes of the curve, respectively. The parameter $\lambda$ represents the steepness of the fitted curve: positive values of $\lambda$ are associated with cell survival (viability). The parameter $m$ is the popular $\log_{10}$-transformed \emph{Half-Maximal Effective Concentration} (or $\text{EC}_{50}$), providing the amount of compound needed to observe a response equal to 50\%. 

Many studies have explored the possibility of testing multiple drugs simultaneously in-vitro, with the aim of understanding the interactions between them and the consequent effect on cancer progression. An interaction can strengthen the effect (\emph{synergistic}) of each drug, or weaken it (\emph{antagonistic}) \citep[for recent studies, see e.g.][]{ONeil_etal_2016, Kashif_etal_2017}. The synergistic or antagonistic effect of a combination of drugs is defined in relation to a non-interaction baseline, representing the condition in which the two drugs do not interact with each other. This condition is not testable experimentally, so that hypotheses need to be made, based on appropriate mathematical definitions. Cell viability observations from monotherapy experiments are used to define the \emph{zero-interaction} level. Many different models have been proposed in the literature, in particular \cite{Loewe_Muischnek_1926, Loewe_1953, Bliss_1939, Webb_1963, Chou_Talalay_1984, Berenbaum_1989, Tallarida_Porreca_Cowan_1989, Greco_etal_1995}; see also the review \cite{Fouquier_Guedj_2015}. More recently, the interest has moved towards trying to understand, by means of statistical methods, how reliable the interaction assessments are. Some studies are based on viability experiments, as in \cite{Tallarida_1992, Yadav_etal_2015}. Others have proposed model-based analysis \citep{Boik_etal_2008, Lee_Kong_2009, Whitehead_etal_2013, Hautaniemi_etal_2018}, and some such as \cite{Johnstone_etal_2016, Hennessey_etal_2010, Li_etal_2007} are set in a Bayesian framework.

This paper presents a novel model-based approach to the study of drug-drug interaction, modelling the drug combination surface using a flexible Bayesian spline regression approach. The viability surface, depending on the concentrations of the two drugs, is described by a simple stochastic model, which allows to discriminate between the non-interaction and the interaction parts of the viability. Our modelling approach allows for posterior inference on both the monotherapy concentration-response curves (included in the non-interaction part), and the interaction part. Importantly, we are able to quantify the uncertainty of our Bayesian estimates. To the best of our knowledge, this is the first time that a similar probabilistic description of the viability experiment is proposed in the literature.

The main model is introduced in Section \ref{sec:Prob_Model}, by providing a probabilistic interpretation of the viability experiment, and by defining both the zero-interaction and the interaction terms. Section \ref{sec:Results} proposes a performance evaluation of the new methodology with an extensive simulation, as well as in an in-house produced ovarian cancer dataset. Section \ref{sec:Concl} concludes.

\section{Modelling Responses from Combined Experiments}\label{sec:general_approach}

In a viability combination experiment, two compounds at concentrations $x_{1i}$, for $i = 0, \dots, n_1$, and $x_{2j}$, for $j = 0, \dots, n_2$, are dispensed. Every experiment is repeated $n_{rep}$ times at the selected concentrations. We will sometimes write $\bm x_1 = (x_{10},\dots,x_{1n_1})$ and $\bm x_2 = (x_{20},\dots,x_{2n_2})$. Specifically, $i = 0$ and $j = 0$ correspond to the absence of the compounds, i.e. $x_{10} = 0$ and $x_{20} = 0$. The response is measured as the fluorescence (or luminescence) level $F^r_{ij}$, assumed proportional to the number of viable cells present at time of observation, denoted by $Z^r_{ij} \in \mathbb{N}$, for $(i,j)$ as above and $r = 1, \dots, n_{rep}$. The proportionality constant (i.e., the fluorescence of a single viable cell) is assumed to be the same in each experimental condition, so that $Z^r_{ij}$ represents the measured viable cell count. Given the pair of indices $(i,j)$, consider the well characterized by the concentrations $(x_{1i}, x_{2j})$. Let $N^r_{0,ij}$ be the total number of cells - viable or not - present in this well at the time of measuring, and indicate with $p_{ij} \in (0,1)$ the probability that a cell is viable in the well, independently of the experimental replicate. This probability can be estimated using the cell counts via the ratio $\frac{Z^r_{ij}}{N^r_{0,ij}}$, for $i = 0, \dots, n_1$ and $j = 0, \dots, n_2$. However, in practice the value of $N^r_{0,ij}$ is not known, and it is estimated using specific control wells, obtaining an estimate $\tilde{N}^r_0$ by averaging the control counts. Due to the inherited variability of the estimate $\tilde{N}^r_0$, the so-obtained random variables $Y^r_{ij} = \frac{Z^r_{ij}}{\tilde{N}^r_0}$ can be interpreted as a noisy version of the underlying probabilities of success $p_{ij}$, subject to an error term dependent on both biological and experimental factors. For this reason, we model the observed quantities $y^r_{ij}$ as follows:
\begin{equation}\label{eq:Gaussian_yij}
y^1_{ij}, \dots, y^{n_{rep}}_{ij} \sim N(p_{ij}, \sigma^2_{\epsilon}), \ \  i = 0,\dots, n_1, \mbox{ and }  j = 0,\dots, n_2,
\end{equation}
where $\sigma^2_{\epsilon}$ in a homoschedastic variance term. Notice how this model accommodates observations which lie outside the range $(0,1)$, which is the range of admissible values for the mean term $p_{ij}$. This is a desired feature of our model, since the observations $y^r_{ij}$ often lie outside the range $(0,1)$. In fact, this can be observed as an effect of normalization with respect to controls.

In the current literature on drug-drug interaction quantification, the responses are assumed to be influenced by two main elements: a \emph{zero-interaction} term describing the impact of the individual drugs on the outcome as if they were not interacting, and a residual term describing the amount of interaction present in the experiment. The first term is usually obtained by describing  mathematical models suitable for the experiment under study, while the second - the \emph{interaction} term - is quantified as the residual of the observations and the expected values under the chosen zero-interaction model. Therefore, it is clear how the definition of the zero-interaction model plays a crucial role in the interpretation and estimation of the interaction between two drugs. In the next two sections, we provide details on these two terms and show how they can be included jointly in a Bayesian model.

\subsection{Models for the Zero-Interaction Term}\label{sec:Prob_Model}

We now move on to modelling the mean viability probability $p_{ij}$. We start by providing insights on the mechanism of action of the combination of compounds. The drugs used in the experiment can act on the same sites of the targeted molecule, or affect the same signaling pathway, and therefore can be considered \emph{mutually non-exclusive}. If they act on different sites or pathways, are instead called \emph{mutually exclusive}. In the latter scenario, the Bliss independence model \citep{Bliss_1939} is an appropriate model for zero-interaction \citep{Fitzgerald_2006}. This model is introduced more rigorously, by defining the following event:
\begin{itemize}
\item[$\text{A}_{ij}$]  := ``a cell survives both drugs at concentrations $(x_{1i}, x_{2j})$ in the combined experiment''.
\end{itemize}
Referring to equation \eqref{eq:Gaussian_yij}, our interest lies in modelling $p_{ij} = \mathbb{P}(\text{A}_{ij})$. It is useful to define the single-drug events:
\begin{itemize}
\item[$\text{D}_{1i}$] := ``a cell survives the first drug at concentration $x_{1i}$ in the combined experiment'';

\item[$\text{D}_{2j}$] := ```a cell survives the second drug at concentration $x_{2j}$ in the combined experiment''.
\end{itemize}
We have that $p_{ij} = \mathbb{P}(\text{A}_{ij}) = \mathbb{P}(\text{D}_{1i} \cap \text{D}_{2j})$. In order to describe $\mathbb{P}(\text{A}_{ij})$, we introduce the zero-interaction component $\mathbb{P}^0(\text{A}_{ij})$ in the analysis. Then, the difference between $\mathbb{P}(\text{A}_{ij})$ and $\mathbb{P}^0(\text{A}_{ij})$ is the interaction term, i.e. the quantity representing the amount of information in the combined experiment that is not explained by the zero-interaction model: $\Delta_{ij} := p_{ij} - \mathbb{P}^0(\text{A}_{ij})$. 

Following \cite{Bliss_1939}, we interpret the zero-interaction of the two compounds as probabilistic independence of the single-drug events, so that $\mathbb{P}^0(\text{A}_{ij}) = \mathbb{P}(\text{D}_{1i}) \mathbb{P}(\text{D}_{2j})$. We point out that the events $\text{D}_{1i}$ and $\text{D}_{2j}$ are defined in the combined experiment, and their probabilities can be estimated from monotherapy experiments. We take this approach in the present paper, and model $\mathbb{P}(\text{D}_{1i})$ and $\mathbb{P}(\text{D}_{2j})$ by assuming \eqref{eq:2LL} after setting the boundaries $(a, b) = (0, 1)$. The resulting parametric function is the 2-parameter log-logistic curve (2LL). This approach is analogous to the zero-interaction potency (ZIP) model introduced by \cite{Yadav_etal_2015}. We write $p^0_{ij} :=\mathbb{P}^0(\text{A}_{ij}) = f(x_{1i}|\theta_1)f(x_{2j}|\theta_2)$, where $\theta_1 = (m_1,\lambda_1)$ and $\theta_2 = (m_2,\lambda_2)$. Notice that $\theta_1$ and $\theta_2$ are assumed to be constant across replicates and different concentrations.

Alternative specifications for the zero-interaction term exist in the literature. Very popular are the highest single agent (HSA) model by \cite{Berenbaum_1989} and the Loewe additivity model by \cite{Loewe_1953}. The first one is based on the idea that the zero-interaction effect is equal to the most effective compound taken alone:
\begin{equation*}
y^{\text{HSA}}_{ij} = \max \{ y_{1i} , y_{2j} \}.
\end{equation*}
It can be easily shown that $y^{\text{HSA}}_{ij} \leq p^0_{ij}$, where $p^0_{ij}$ is the Bliss zero-interaction term just introduced \cite[see also][]{Tang_etal_2015}.

When the two compounds have a similar mechanism of action and are \emph{mutually non-exclusive}, 
the Loewe additive model is often used \citep{Fitzgerald_2006}. This situation can be interpreted as the drugs acting on the same site of the targeted molecule, or along the same pathway. Therefore, it would be sensible to assume that the two compounds should compensate each other when varying their concentrations. Under this assumption, and further assuming that a functional form is available to describe the monotherapy experiments $f_1(x_{1i})$, $f_2(x_{2j})$, then the zero-interaction response $y^{\text{Loewe}}_{ij}$ is defined as the solution of the following equation:
\begin{equation}\label{eq:Loewe_2compounds}
\frac{x_{1i}}{f^{-1}_1(y^{\text{Loewe}}_{ij})} + \frac{x_{2j}}{f^{-1}_2(y^{\text{Loewe}}_{ij})} = 1,
\end{equation}
for each pair of doses $(x_{1i}, x_{2j})$, with $i = 1, \dots, n_1$ and $j = 1, \dots, n_2$. The functions $f_1$ and $f_2$ are the monotherapy dose-response curves for the compounds tested, and are often assumed to be of form \eqref{eq:2LL}. Synergism or antagonism are detected when the value of the left term in \eqref{eq:Loewe_2compounds} is smaller or greater than 1, respectively.

\subsection{The Interaction Term}\label{sec:Inter}
Next we model the interaction term $\Delta_{ij}$. Because $p_{ij} = p^0_{ij} + \Delta_{ij} \in (0,1)$, and the co-domain of the 2LL curve is $(0,1)$, so that $p^0_{ij} \in (0,1)$ for any value of $(\theta_1, \theta_2)$, the range of admissible values for the interaction term $\Delta_{ij}$ is the interval $I_{\Delta_{ij}} := (-p^0_{ij}, 1 - p^0_{ij})$. While monotonicity in concentration is assumed for the monotherapy response curves and for the zero-interaction surface, this is too restrictive for the interaction term $\Delta_{ij}$, as we want to be able to capture both synergistic and antagonistic behaviours. To allow higher flexibility, we use natural cubic splines to model the interaction between the compounds \citep[see][for a review]{De_Boor_2001}. The use of splines in the analysis of drug combination surfaces has been proposed in \cite{Wheeler_2017}, where an approach based on Gaussian processes is pursued, but without differentiating between the zero-interaction and the interaction terms. In this paper, we specify a tensor spline obtained as the cross-product of two univariate cubic B-splines, each defined over a set of $K_1$ and $K_2$ equally spaced knots to cover the $\log_{10}$ concentration ranges of $\bm x_1$ and $\bm x_2$, respectively. We use truncated power polynomials to produce the B-spline basis, following \cite{Eilers_2010}. The spline coefficients are parameterized by a matrix $\bm C$, for which a suitable prior distribution is chosen. We include the tensor-spline values for each combination in a regression setting, with the use of additional linear coefficients $\gamma_0$, $\gamma_1$, and $\gamma_2$. These coefficients do not have a direct interpretation in the model description, but allow for more flexibility in the posterior inference for the interaction term $\Delta_{ij}$:
\begin{align*}
B_{ij} &= \gamma_0 + \gamma_1 x_{1i} + \gamma_2 x_{2j} + \mathcal{B}(x_{1i} , x_{2j}), \\
\mathcal{B}(x_{1i} , x_{2j}) &= \sum\limits_{l,m} \bm C_{lm} B_l(x_{1i})B_m(x_{2j}),
\end{align*}
with $B_l(x)$ a univariate cubic spline at knot $l$ and evaluated at $x \in \mathbb{R}$. We cannot merely assume $\Delta_{ij} = B_{ij}$, because $B_{ij} \in \mathbb{R}$, while $\Delta_{ij} \in I_{\Delta_{ij}}$, for each $(i,j)$. Therefore, we introduce a suitable link function $g:\mathbb{R} \rightarrow I_{\Delta_{ij}}$. Several different choices for $g$ can be considered, for instance a truncation term or a linear transformation. Here, we use the following:
\begin{equation*}
g(B_{ij}) = - p^0_{ij} (1 + e^{b_1 B_{ij}})^{-1} + (1 - p^0_{ij}) (1 + e^{-b_2 B_{ij}})^{-1}.
\end{equation*}
The link function $g$ is applied to each spline term and pair of concentrations tested, yielding $\Delta_{ij} = g(B_{ij})\mathbb{I}_0(i,j)$, where $\mathbb{I}_0(i,j) = 0$ if $i = 0$ or $j = 0$, and $\mathbb{I}_0(i,j) = 1$ otherwise. This indicator prevents any interaction in the absence of either of the compounds (or at such a low level that we do not expect any compound activation), i.e. for $x_{10}$ or $x_{20}$. Two extra parameters $(b_1,b_2)$ are introduced with $g$, regulating the behaviour of the transformation. In particular, by imposing that $b_1, b_2 > 0$, we can ensure that $g$ is monotonically non-decreasing and surjective. Observe how imposing $b_1 = b_2 = b > 0$ yields $p_{ij} = (1 + e^{-bB_{ij}})^{-1}$, corresponding to a tensor-product spline logistic regression model, losing the interpretability of the terms $p_{ij}^0$ and $\Delta_{ij}$. Therefore, we will assume that $b_1 \neq b_2$, a condition that is verified by assuming, for instance, that $b_1$ and $b_2$ are continuous random variables a priori.

\subsection{Full Bayesian Model for Drug Interaction}\label{sec:Bayesian_model}
Summarising, the proposed model has the form:
\begin{align}\label{eq:Final_Model}
&Y^r_{ij} = p^0_{ij} + \Delta_{ij} + \epsilon^r_{ij},\ \ \{\epsilon^r_{ij}\} \iid N(0,\sigma^2_{ij}), \nonumber\\
&p^0_{ij} = f(x_{1i}|m_1, \lambda_1)f(x_{2j}|m_2, \lambda) = (1 + 10^{\lambda_1 (\log_{10}(x_{1i}) - m_1)})^{-1} (1 + 10^{\lambda_2 (\log_{10}(x_{2j}) - m_2)})^{-1}, \nonumber\\
&\Delta_{ij} = g(B_{ij})\mathbb{I}_0(i,j), \quad g(B_{ij}) = - p^0_{ij} (1 + e^{b_1 B_{ij}})^{-1} + (1 - p^0_{ij}) (1 + e^{-b_2 B_{ij}})^{-1}, \nonumber\\
&B_{ij} = \gamma_0 + \gamma_1 x_{1i} + \gamma_2 x_{2j} + \mathcal{B}(x_{1i} , x_{2j}), \nonumber\\
&\mathcal{B}(x_{1i} , x_{2j}) = \sum_{l,m} \bm C_{lm} B_l(x_{1i})B_m(x_{2j}), \\
&\bm C \sim \text{Matrix-N}_{_{K_1, K_2}}(\bm 0, \Psi_{K_1}, \Psi_{K_2}), \nonumber\\
&\psi|\alpha_{\psi}, \beta_{\psi} \sim \Gamma(\alpha_{\psi}, \beta_{\psi}), \quad \psi \in \{\lambda_1, \lambda_2, b_1, b_2\} \nonumber\\
&\phi | \sigma^2_{\phi} \sim \text{N}(0, \sigma^2_{\phi}), \quad \phi \in \{ m_1, m_2, \gamma_0, \gamma_1, \gamma_2 \}, \nonumber \\
&\sigma^2_{m_1}, \sigma^2_{m_2}, \sigma^2_{\gamma_0}, \sigma^2_{\gamma_1}, \sigma^2_{\gamma_2}, \sigma^2_{ij} \sim \pi_{\sigma^2}, \nonumber
\end{align}
for $i = 0,\dots,n_1$, $j = 0,\dots,n_2$, and $r = 1,\dots,n_{rep}$. The last four lines of \eqref{eq:Final_Model} describe our prior distributions. The matrix of spline coefficients $\bm C$ is a priori distributed as a matrix-variate normal of dimensions $(K_1, K_2)$ centred on the zero matrix $\bm 0$. Second order difference matrices $\Psi_{K_1}$ and $\Psi_{K_2}$ are used to penalise the jumps at the knot values of the tensor product spline. We assume vague Gamma prior distributions for the parameters $\lambda_1$, $\lambda_2$, $b_1$, and $b_2$ (mean = 1, variance = 100), and normal prior distributions with zero mean and variance parameter $\sigma^2_{m_1}$ and $\sigma^2_{m_2}$ for $m_1$ and $m_2$. We explored a range of possibilities for the hyper-priors on the variability parameters $\sigma^2_{\phi}$, denoted in the last line of \eqref{eq:Final_Model} as $\pi_{\sigma^2}$: half-Cauchy distribution $\text{HC}(h)$, $h \in \{0.1, 0.5, 1, 5, 10\}$, or inverse-gamma distribution $\text{IG}(\alpha,\beta)$, $(\alpha,\beta) \in \{(3, 2), (0.1, 0.1), (0.05, 0.05), (0.025, 0.025), (0.01, 0.01)\}$. We point out that the choice of assigning the same hyper-prior to all the variance terms $\sigma^2_{\phi}$, for $\phi$ as in \eqref{eq:Final_Model}, is motivated by easing the computational burden, and that other options can be easily explored (e.g., including prior information about the variability of the parameters $\phi$). Finally, the variability of the error terms $\{\epsilon^r_{ij}\}$ is assigned a prior distribution $\pi_{\sigma^2_{ij}}$, in accordance with the choices specified for $\pi_{\sigma^2}$, restricting our analysis to the homoscedastic case, for which $\sigma^2_{ij} = \sigma^2_{\epsilon}$ for each $(i,j)$. This framework is easily extendible to recover the heteroscedastic model depicted in \eqref{eq:Final_Model}, without significantly increasing the computational burden. However, the heteroscedastic model would require a more detailed prior elicitation analysis, representing a possibility for further extensions of the proposed framework.

\section{Applications}\label{sec:Results}
\subsection{Computational details}

The posterior estimates of the parameters of the model are obtained by MCMC sampling. Due to the presence of several non-conjugate parameters in the model, we resort to an adaptive version of the Metropolis-within-Gibbs algorithm \cite[see][]{Griffin_Stephens_2013}. The class of adaptive MCMC methods involves using the samples obtained during the sweeps of the Gibbs sampler in order to produce a candidate for the current Metropolis-Hastings step. In particular, the proposal density of a parameter of interest, say $q(\theta^{curr},\theta')$ - where $\theta'$ represents a new value of the parameter $\theta$, and $\theta^{curr}$ is the value at the current iteration - can be defined to include additional information deriving from the values of $\theta$ visited so far. A description of the algorithm is provided in Section 2 of the Supplementary Materials.

\subsection{Simulation study}\label{sec:Simul_Study}

In order to provide a performance evaluation of the proposed model, we simulated datasets with different features. In particular, the mean surface of the true model is split into two components, as to represent the zero-interaction term $p^0$ and the interaction term $\Delta$, and it is characterised by the same error term $\sigma^2_{\epsilon}$, for each $(i,j)$, as follows:
\begin{align*}
Y^r_{ij} | p_{ij}, \sigma^2_{\epsilon}, \nu &\ind \text{F}(\cdot | p_{ij}, \sigma^2_{\epsilon}, \nu), \quad r = 1, \dots, n_{rep}, \\
p_{ij} &= p^0_{ij} + \Delta_{ij}, \\
p^0_{ij} &= \bm \Phi \left( 2\bm \mu - (x_{1 i},x_{2 j}), \bm \mu = (0,5), \Sigma = \begin{bmatrix}
5 & 0\\
0 & 5
\end{bmatrix} \right), \\
\sigma^2_{\epsilon} &= 0.05^2,
\end{align*}
where the concentrations are $\log_{10}(\bm x_1) = (-\infty, -4, \dots, 4, 5)$ and $\log_{10}(\bm x_2) = (-\infty, -3.5, \dots,$ $3.5, 5.5)$. For computational reasons, the smallest concentration will be replaced by a value equal to -2 times in $\log_{10}$-scale from the next minimum concentration when needed in the algorithm. The term $\text{F}(\cdot | p_{ij}, \sigma^2_{\epsilon}, \nu)$ represents the sampling model chosen for the simulations, which depends on the mean term $p_{ij}$ at each concentration pair $(x_{1i}, x_{2j})$, the error $\sigma^2_{\epsilon}$, and a set of additional parameters $\nu$. In particular, we adopt a normal distribution, s.t. $\text{F}(\cdot | p_{ij}, \sigma^2_{\epsilon}, \nu) = N(\cdot | p_{ij}, \sigma^2_{\epsilon})$, for which no additional parameters are needed, or a location-scale $t$-Student distribution with $\nu = 5$ degrees of freedom, s.t. $\text{F}(\cdot | p_{ij}, \sigma^2_{\epsilon}, \nu) = t_5(\cdot | p_{ij}, \sigma^2_{\epsilon})$. The function $\bm \Phi((x,y), \bm \mu , \Sigma)$ represents the c.d.f. of the bi-variate normal distribution with mean $\bm \mu$ and covariance matrix $\Sigma$, evaluated at $(x,y)$. The zero-interaction term reflects specific monotherapy behaviours at the boundaries, for which one drug is more effective than the other. We assess the efficacy of the individual drugs via the \emph{drug sensitivity scores} (DSS) \citep{Yadav_etal_2014}. This quantity is a normalized area under the concentration-response curve, taking into account the range of the data, and interpretable as the percentage of efficacy of the drug tested. In order to compute the DSS scores for the individual compounds, we first estimate the parameters of the concentration-response curves by fitting a 2LL regression to each monotherapy set of points via the \texttt{drm} function of the R package \texttt{drc}, yielding $(\hat{m}_1, \hat{\lambda}_1) = (0, 0.33)$ and $(\hat{m}_2, \hat{\lambda}_2) = (4.96, 0.31)$, respectively. These estimates were used to obtain DSS scores of 64.02 and 13.25, respectively, indicating higher efficacy of the first drug. The interaction term $\Delta$ was specified in three different ways throughout the simulation study:
\begin{align*}
\mbox{\textbullet} \ \Delta^{(1)}_{ij} &= 0, \forall (i,j), \\
\mbox{\textbullet} \ \Delta^{(2)}_{ij} &= \bm \Phi \left( (x_{1i},x_{2j}), \bm\mu = (5,5), \Sigma = \begin{bmatrix}
10 & 0\\
0 & 10
\end{bmatrix} \right),\\
\mbox{\textbullet} \ \Delta^{(3)}_{ij} &= 0.5\left[ \bm \Phi \left( (x_{1i},x_{2j}), \bm \mu_1, \Sigma_1 = \begin{bmatrix}
1 & 0\\
0 & 1
\end{bmatrix} \right) \right. \\
& \hspace{0.9cm} \left. - \bm \Phi \left(2 \bm \mu_2 - (x_{1i},x_{2j}), \bm \mu_2, \Sigma_2 = 
\begin{bmatrix}
1 & 0\\
0 & 1
\end{bmatrix} \right) \right],
\end{align*}
where $\bm \mu_1 = (1,1)$ and $\bm \mu_2 = -\bm \mu_1$. We considered datasets with $n_{rep} \in \{1,3,5,10\}$. We fit model \eqref{eq:Final_Model} to each simulated dataset for every choice of prior distribution $\pi_{\sigma^2}$, as mentioned in Section \ref{sec:Bayesian_model}. The adaptive MCMC algorithm is run for 100.000 iterations, of which the first half is discarded as burn-in period, and the second half is thinned every 10-th iteration, yielding a final sample of size 5.000. We point out that the non-conjugate parameters of the model are updated adaptively only after the first 1.000 iterations, serving as initial burn-in for the computation of the adaptive quantities.

\begin{figure}[!ht]
\vspace{-2.5cm}
\centering
\subfloat[$\Delta^{(1)}$]{\includegraphics[height=.39\textwidth]{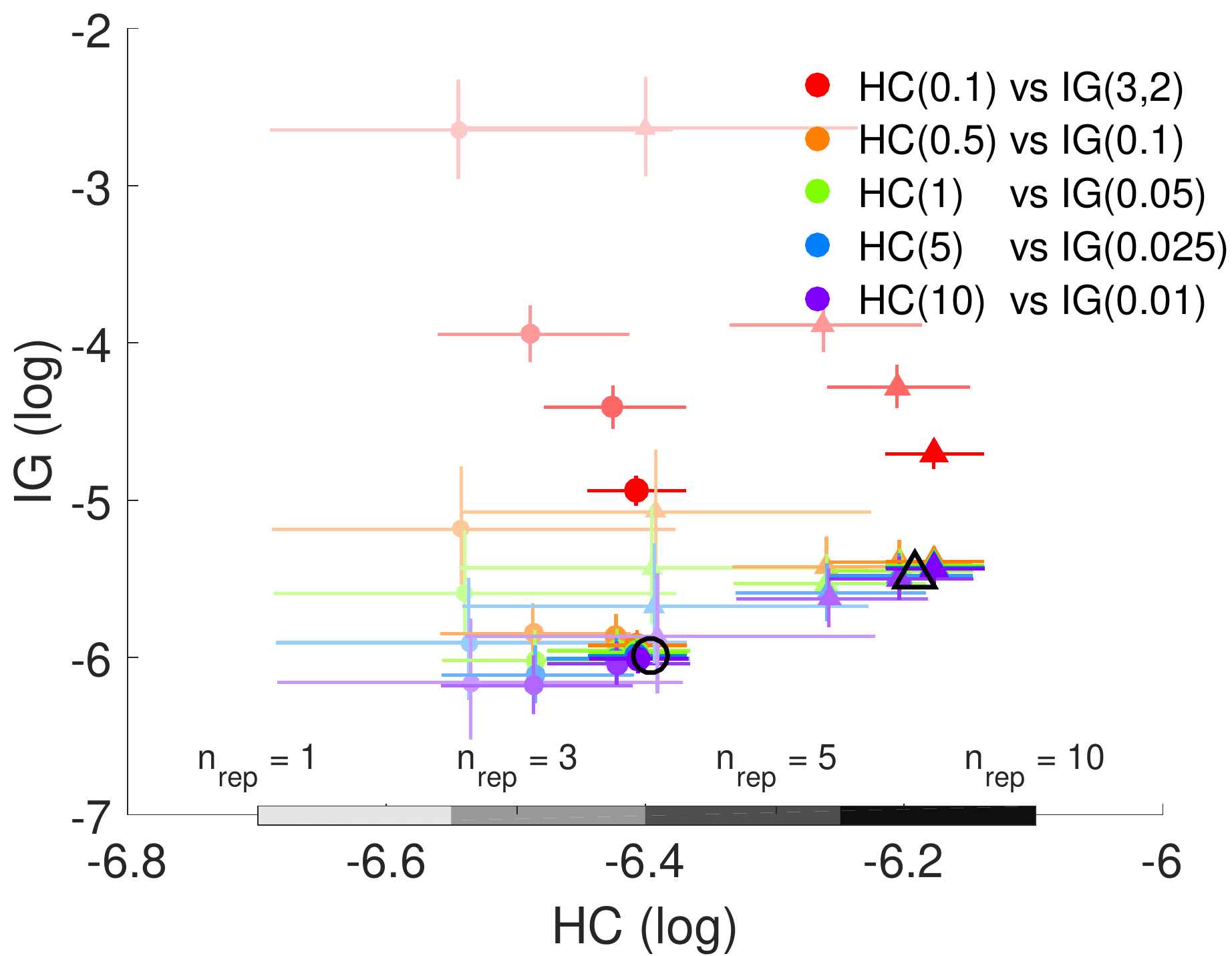}}
\subfloat[$\Delta^{(1)}$]{\includegraphics[height=.39\textwidth]{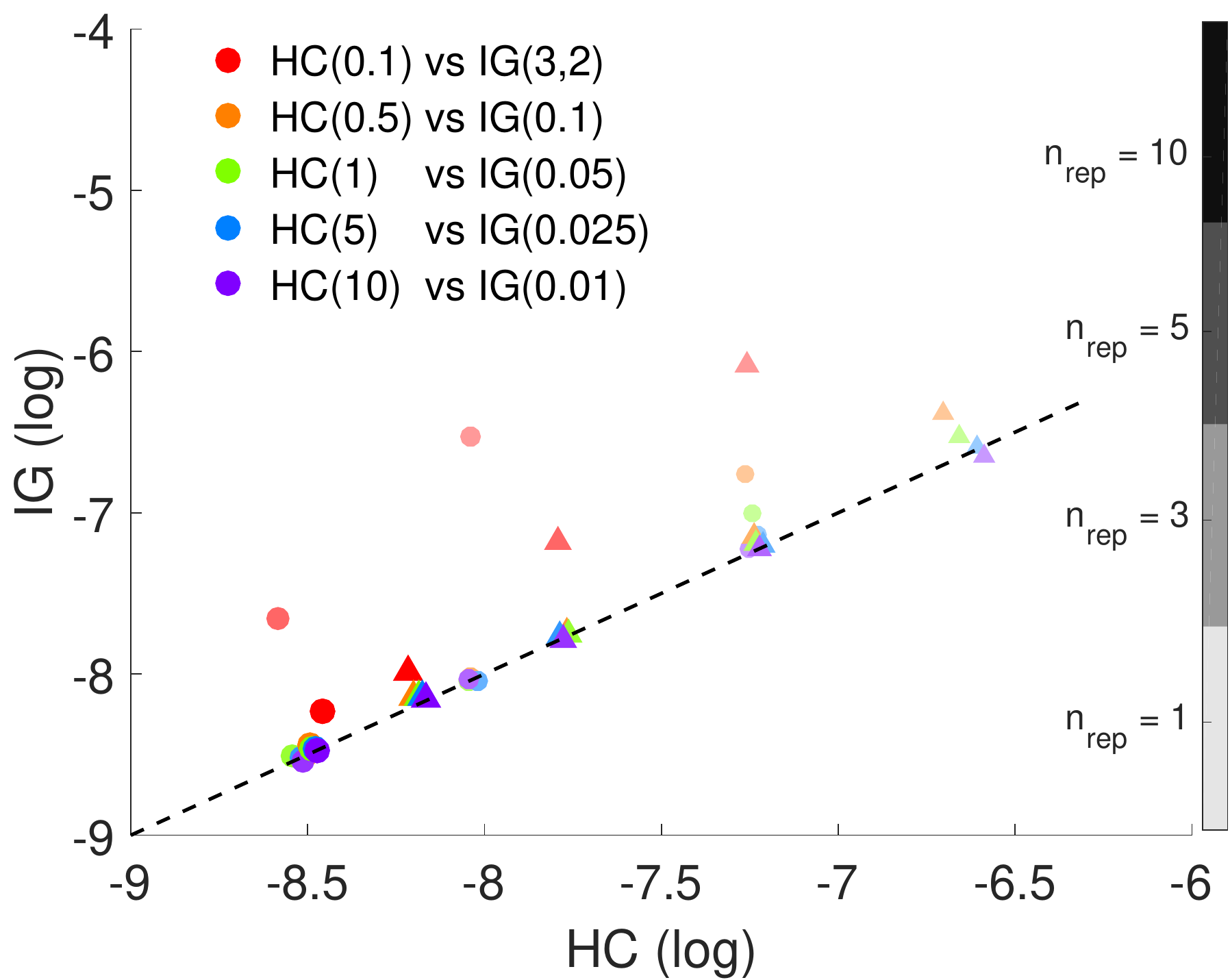}}

\centering
\subfloat[$\Delta^{(2)}$]{\includegraphics[height=.39\textwidth]{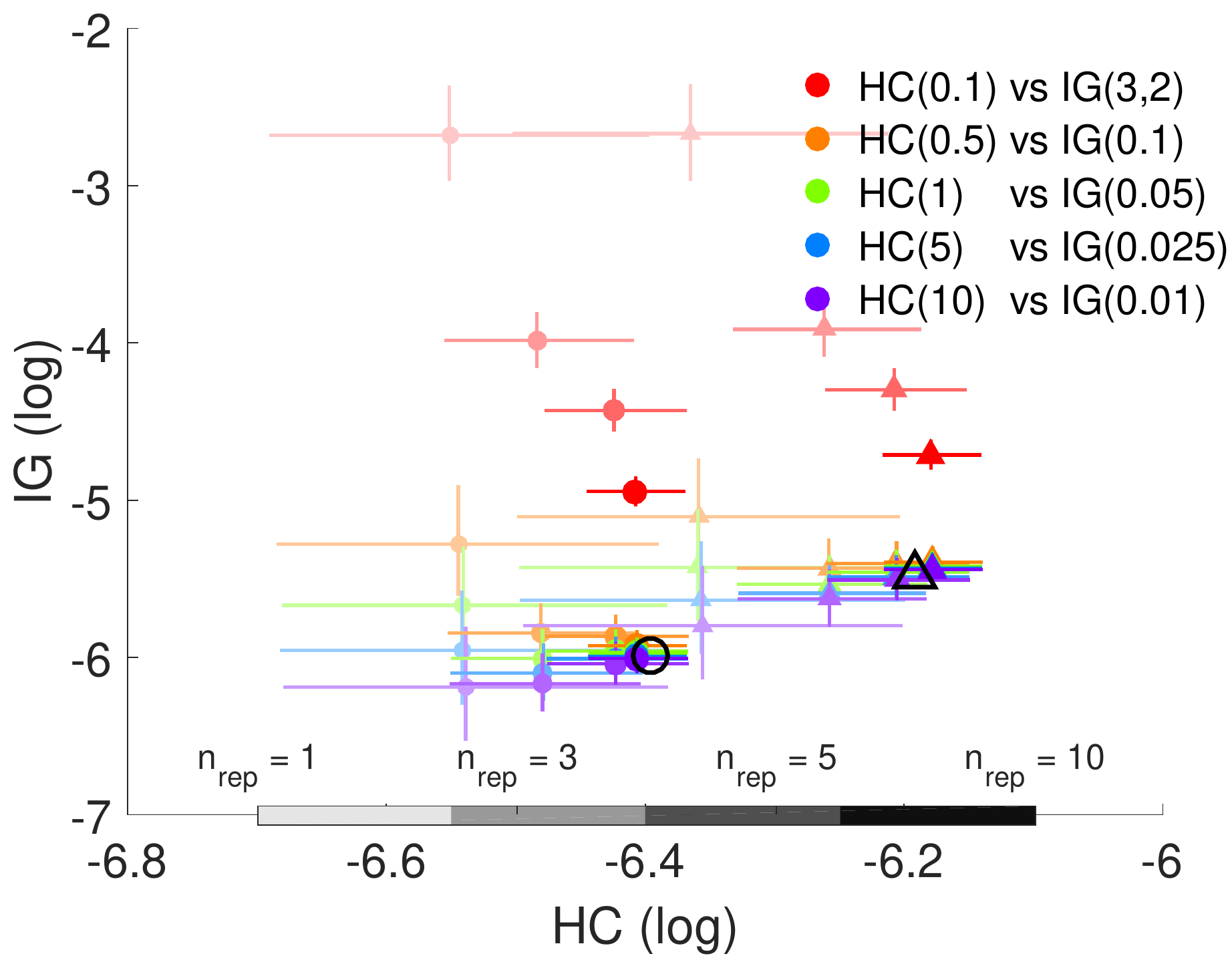}}
\subfloat[$\Delta^{(2)}$]{\includegraphics[height=.39\textwidth]{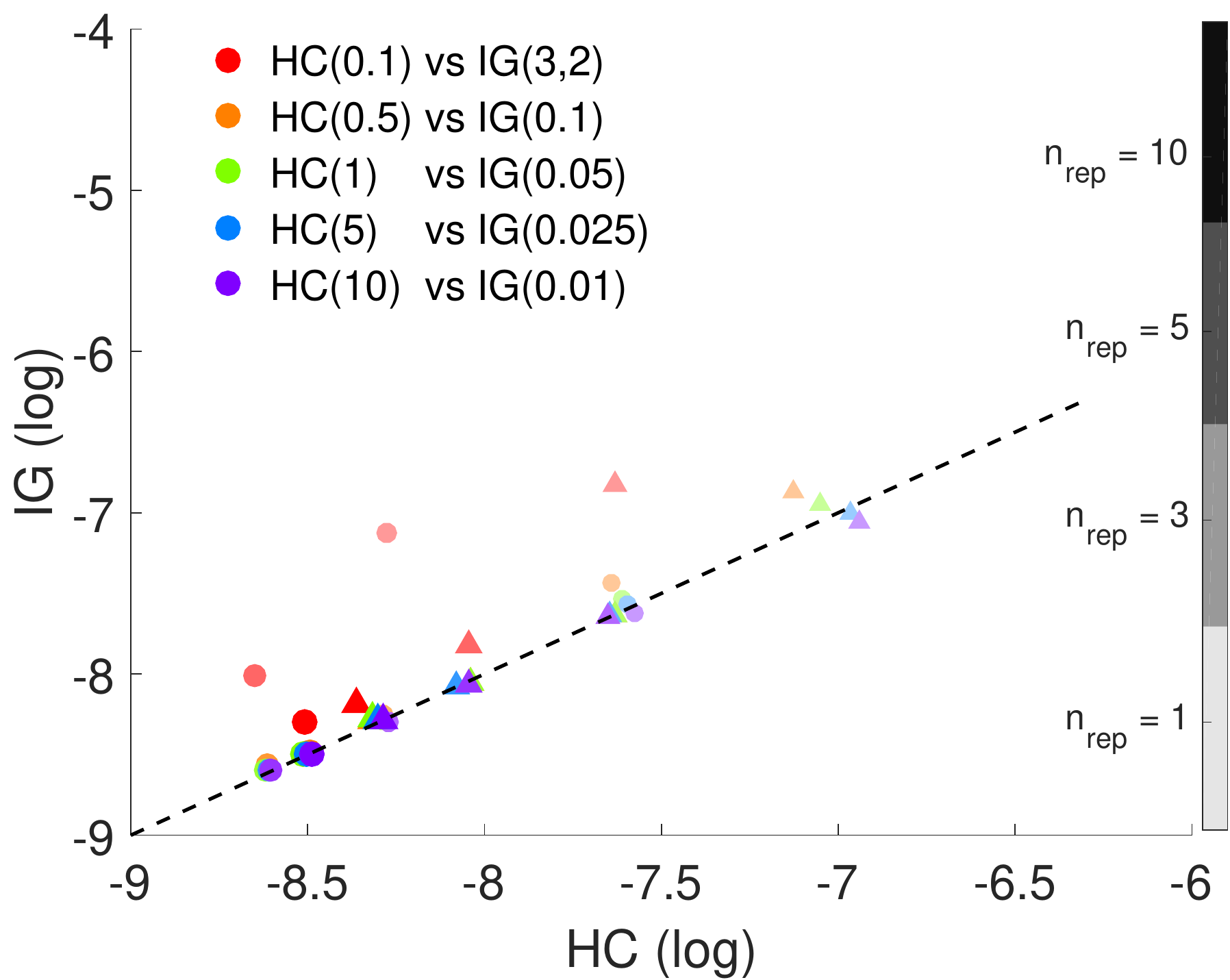}}

\centering
\subfloat[$\Delta^{(3)}$]{\includegraphics[height=.39\textwidth]{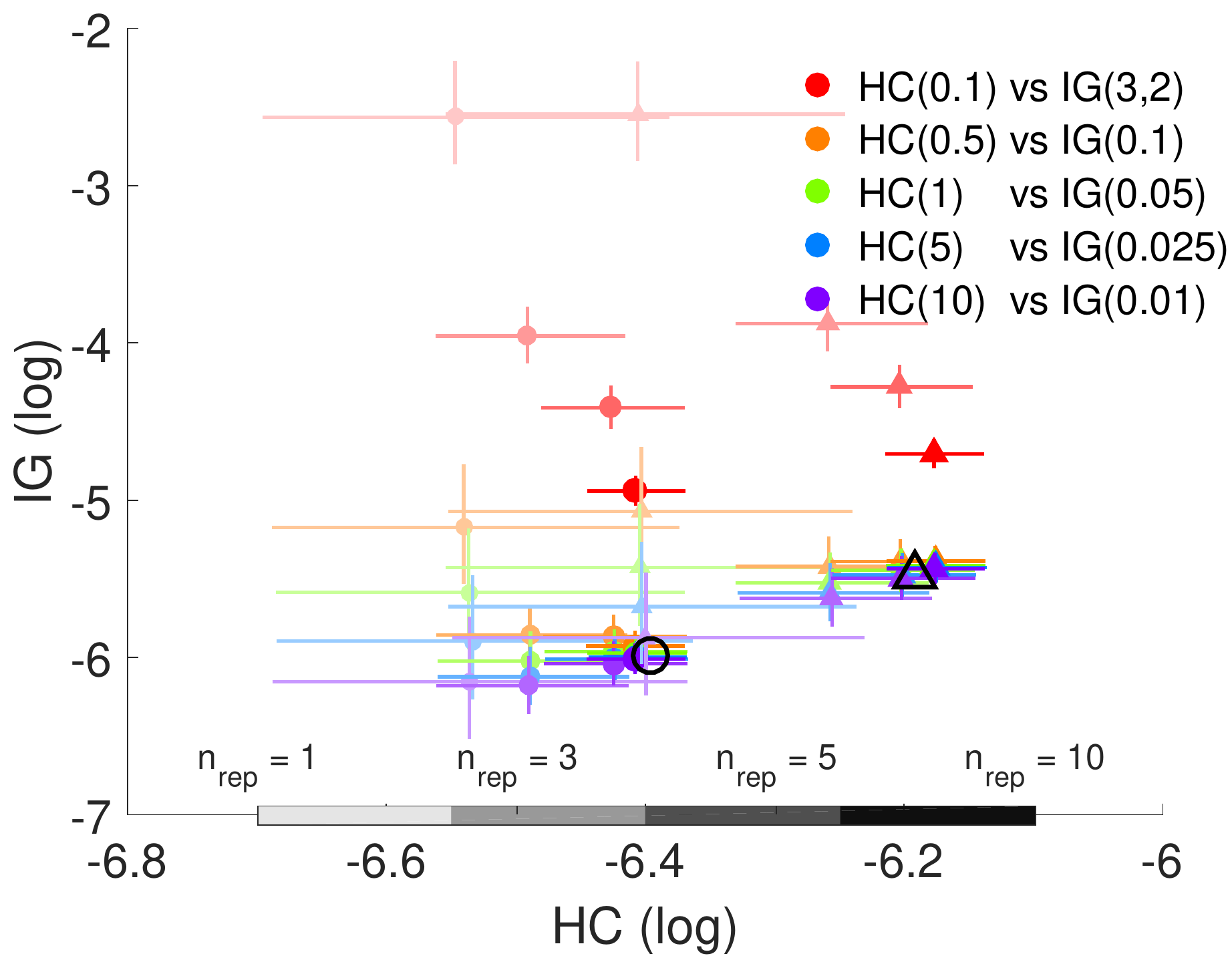}}
\subfloat[$\Delta^{(3)}$]{\includegraphics[height=.39\textwidth]{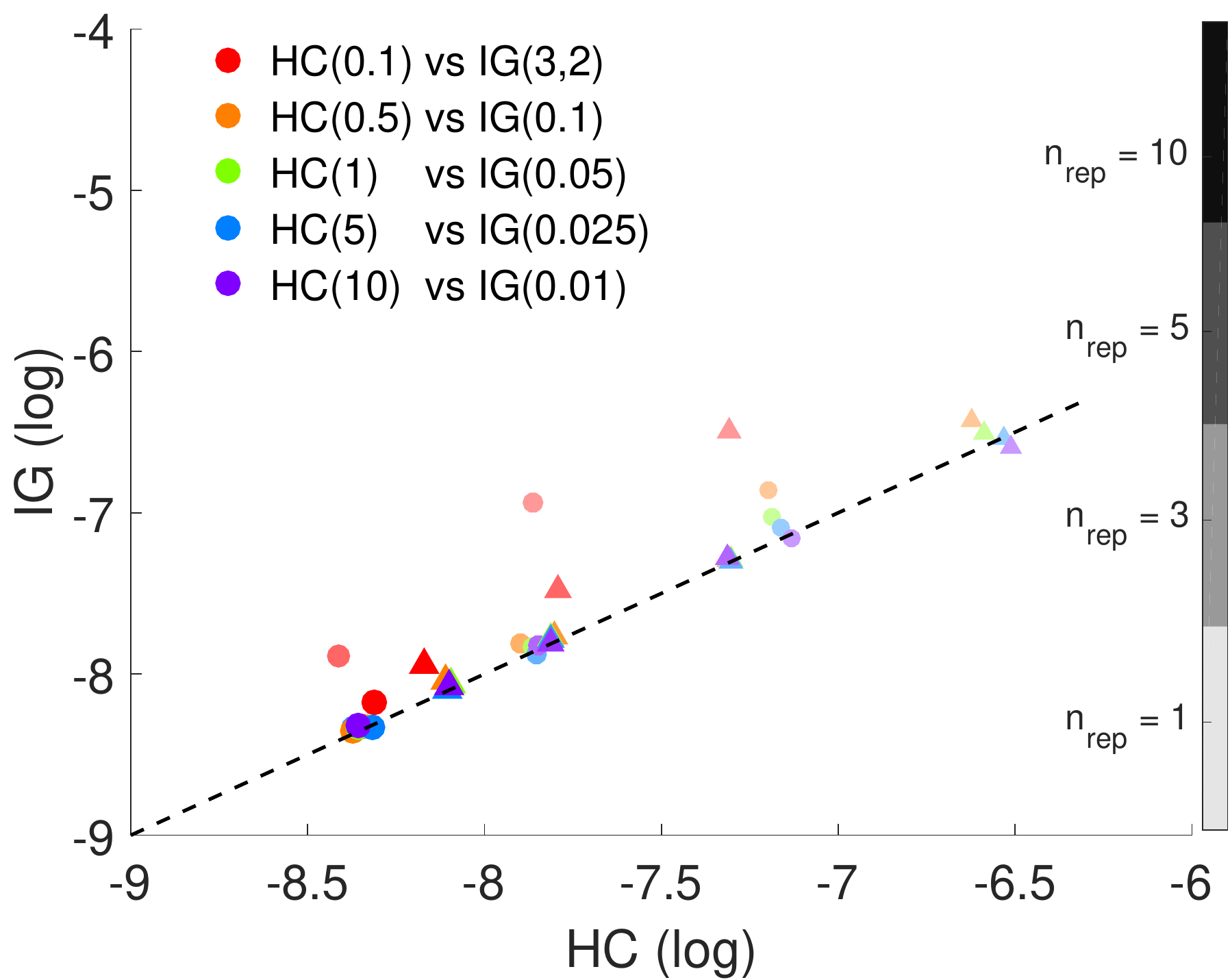}}
\caption{(a,c,e): Posterior medians and 95\% posterior credibility intervals (crossing lines) of $\sigma^2_{\epsilon}$. (b,d,f): $\text{MSE}_{\Delta}$ (log) (dashed lines indicate the identity). Each row refers to a different interaction term. The colour estimates are organised by pairing the prior distributions used, half-Cauchy with hyperparameter $h$ (HC$(h)$), or inverse-gamma with identical shape and rate parameters $\alpha$ (IG$(\alpha)$). The intensity of the colour and the size of the dots increase with $n_{rep}$. Circles indicate scenarios with $\epsilon^r_{ij} \sim N(p_{ij}, \sigma^2_{\epsilon})$, triangles with $\epsilon^r_{ij} \sim t_5(p_{ij}, \sigma^2_{\epsilon})$. The black symbols (circles or triangles) in (a,c,e) correspond to the true values of $\sigma^2_{\epsilon}$.}
  \label{fig:Errors_MSE_INTER}
\end{figure}

We first assess the model performance by studying the posterior distribution of $\sigma^2_{\epsilon}$. The three panels in the first column of Figure \ref{fig:Errors_MSE_INTER}, reporting the posterior medians of $\sigma^2_{\epsilon}$ for the different simulation settings, do not show noticeable differences, indicating that the choice of the interaction term used in the simulations does not affect the estimation of $\sigma^2_{\epsilon}$. Not surprisingly, by increasing the number of replicates in the model, the posterior estimates of all the scenarios approach the true values used in the simulations. However, the choice of the prior distribution, as well as the hyperparameters, has a clear effect on the estimation. In particular, the inverse-gamma prior produces more biased results with less variability, while the half-Cauchy behaves in the opposite way. In terms of coverage, in general we observe larger 95\% posterior credibility intervals in the half-Cauchy setting that include the true value for most of the scenarios. Coverage in the inverse-gamma case is observed only for larger prior variances.

We assessed the ability of the model to identify the interaction term $\Delta$ by evaluating the mean square error of its estimate, indicated as $\textbf{\text{MSE}}_{\Delta}$ in Figure \ref{fig:Errors_MSE_INTER}(b,d,f). The $\text{IG}(3,2)$ case is associated with the highest $\textbf{\text{MSE}}_{\Delta}$ values, in agreement with the biased posterior estimates of $\sigma^2_{\epsilon}$ reported earlier. This result supports the choice of a weakly informative prior. A tabular summary of these values is reported in Supplementary Table 1. Of particular notice is the high robustness of the half-Cauchy prior to the choice of the hyperparameter $h$. We also provide in Supplementary Table 2 the values of $\textbf{\text{MSE}}_{\Delta}$ obtained by estimating the interaction surface with some of the most popular methods in the literature, described in Section \ref{sec:Prob_Model}. The latter are computed by using the R package \texttt{synergyfinder} \citep{Yadav_etal_2015}. The R package does not handle the presence of replicates, hence we averaged the resulting surfaces when $n_{rep} > 1$. Overall, the standard methods are outperformed by the proposed model. Also, as expected, the estimated errors decrease with the number of replicates. Finally, the $t$-Student case yields poorer results. We provide additional goodness-of-fit measures in the Supplementary Materials, namely the \emph{Log Pseudo-Marginal Likelihood} (LPML) of \citet{Geisser_Eddy_1979} in Supplementary Table 3, and the  for the mean surface in Supplementary Table 4. The LPML values do not vary largely between the different scenarios, maintaining the same magnitude when different prior settings are used. A slight departure from this consistent behaviour is observed for the IG$(3,2)$ case, as previously observed.

\begin{figure}[!ht]
\centering{
\subfloat[$p^0$ and $\Delta$]{\includegraphics[height=.39\textwidth]{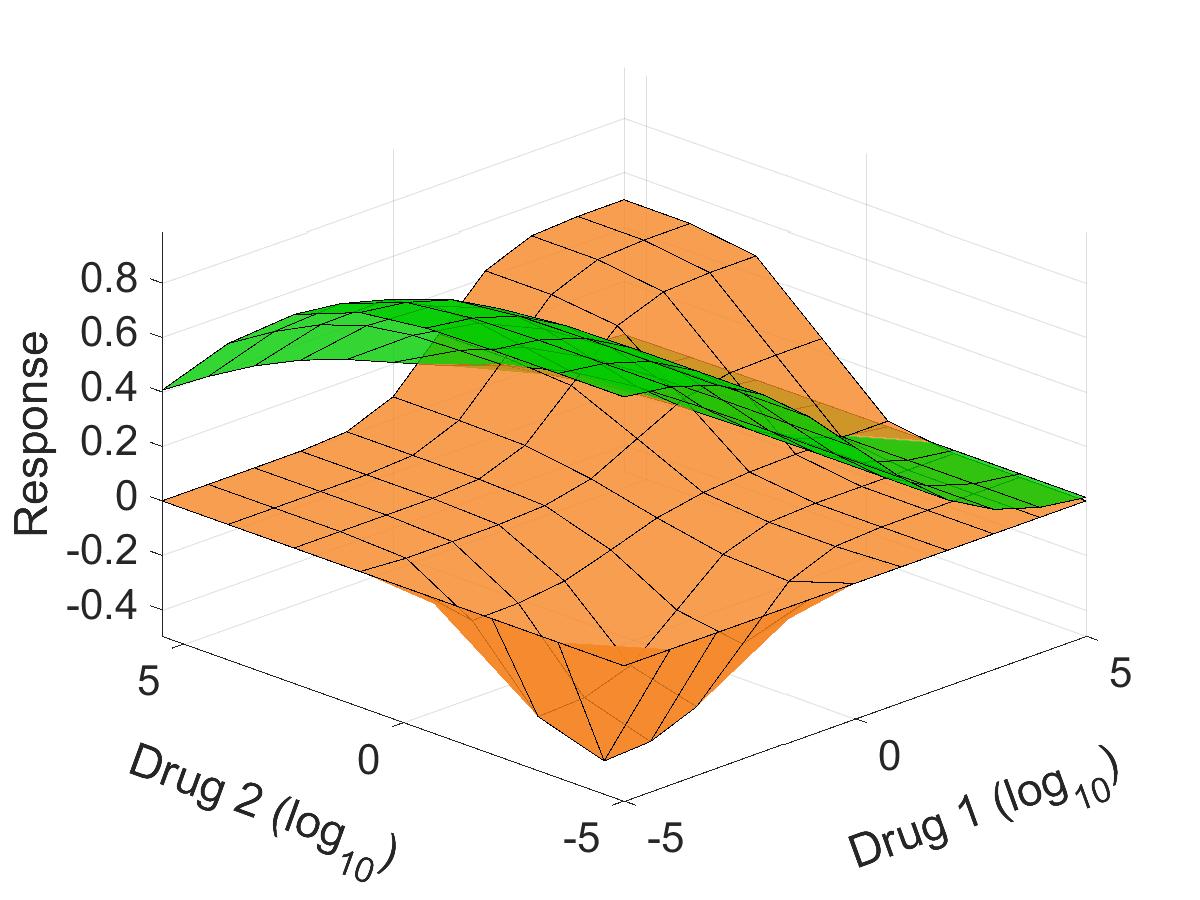}}
\subfloat[$p$]{\includegraphics[height=.39\textwidth]{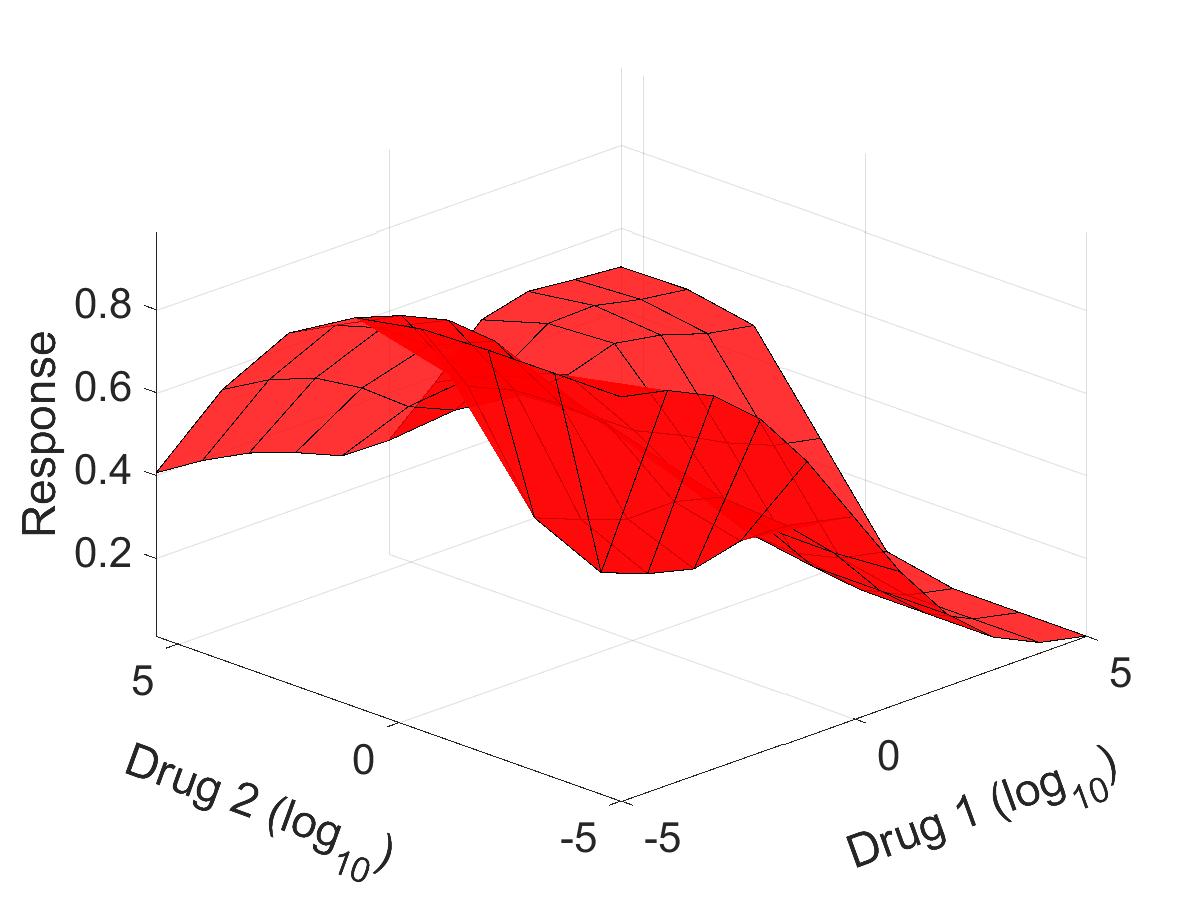}}

\subfloat[$\hat{p}^0$ and $\hat{\Delta}$]{\includegraphics[height=.39\textwidth]{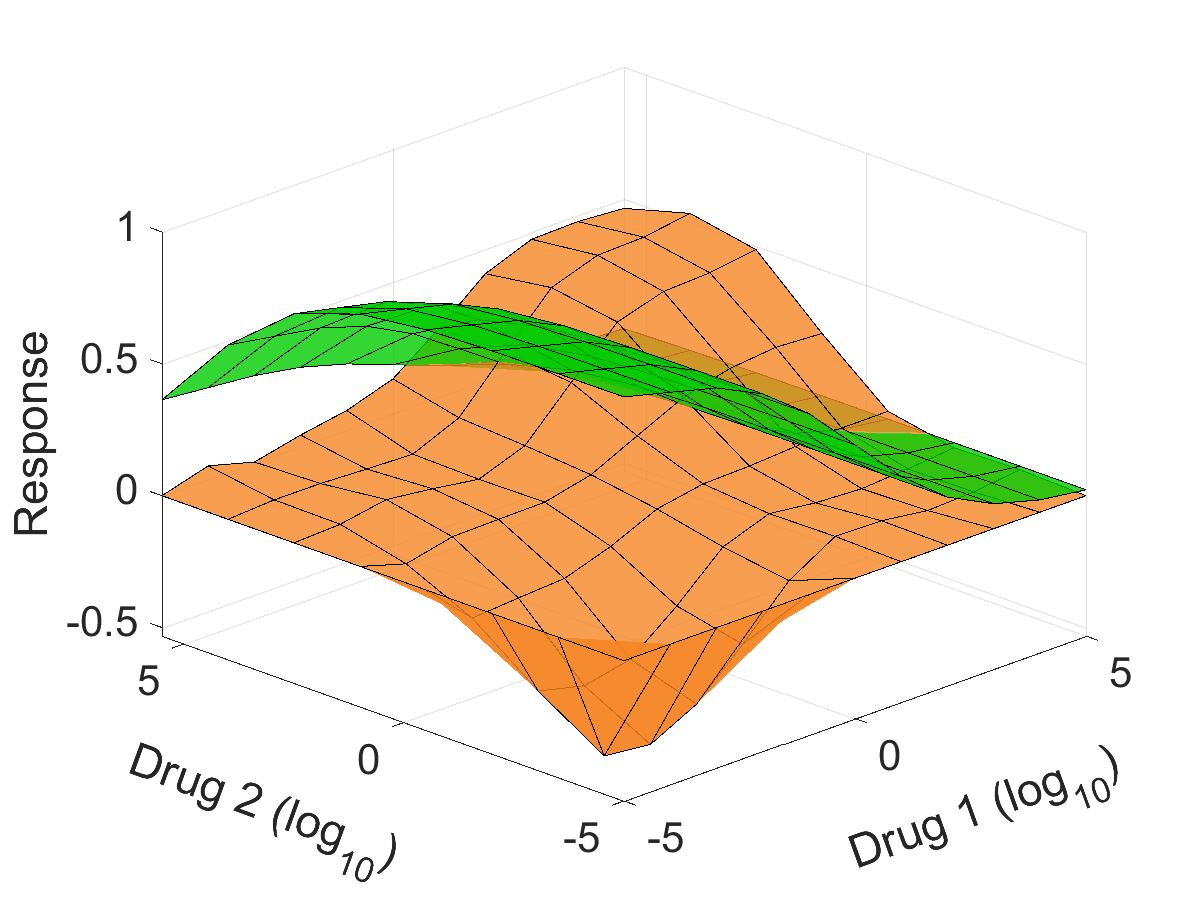}}
\subfloat[$\hat{p}$]{\includegraphics[height=.39\textwidth]{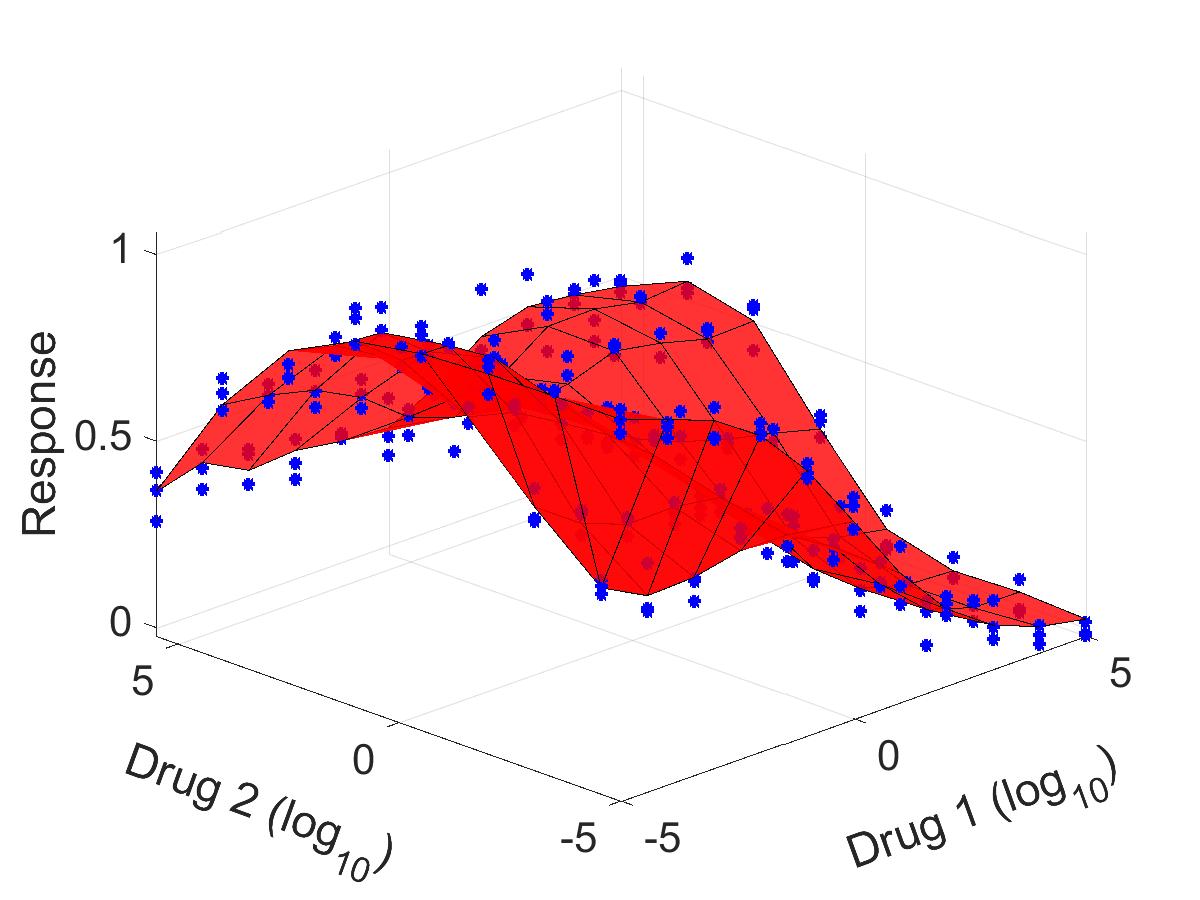}}
}
\caption{Simulation study: ($\Delta^{(3)}$, $\sigma^2_{\phi} \sim \text{HC}(1), \epsilon^r_{ij} \sim N(p_{ij}, \sigma^2_{\epsilon}), n_{rep} = 3)$. (a,b): True surfaces. (c,d): Estimated surfaces and simulated data. The zero-interaction surfaces are depicted in green, while the interaction surfaces in orange. The red surfaces represent the mean surface and its estimates.}
  \label{fig:selected_surfaces}
\end{figure}

\begin{figure}[!ht]
\centering{
\subfloat[]{\includegraphics[width=.5\textwidth]{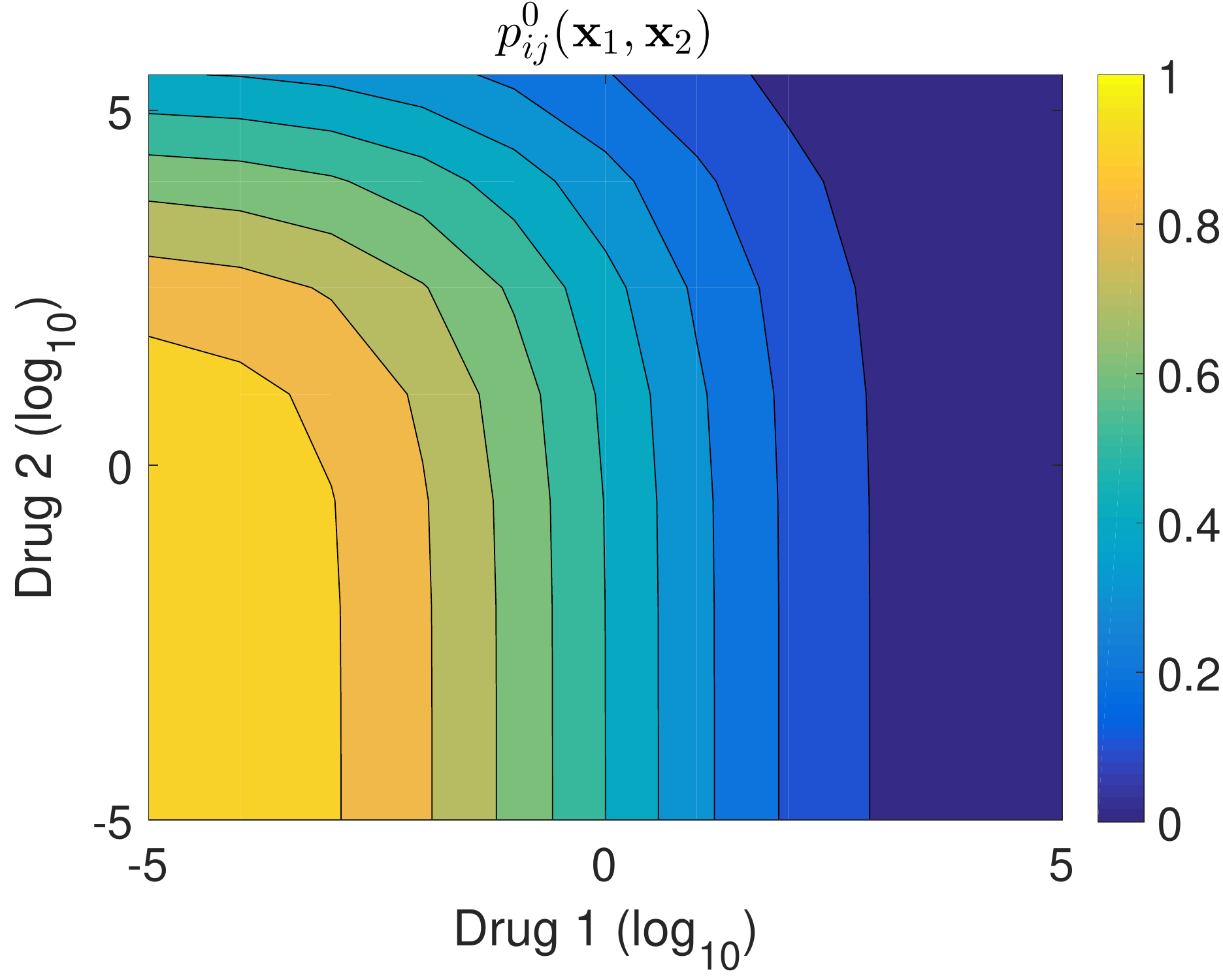}}
\subfloat[]{\includegraphics[width=.5\textwidth]{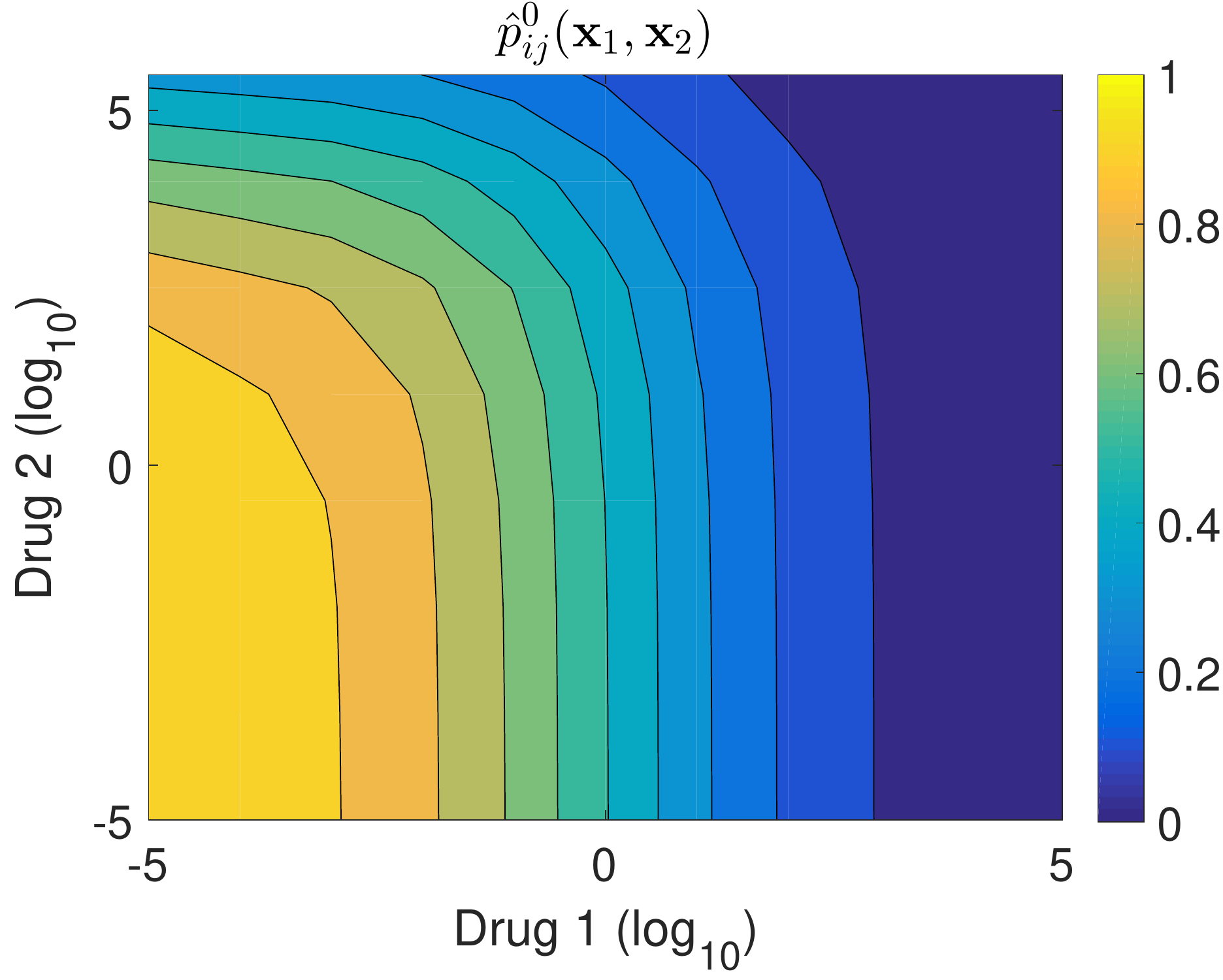}}

\subfloat[]{\includegraphics[width=.5\textwidth]{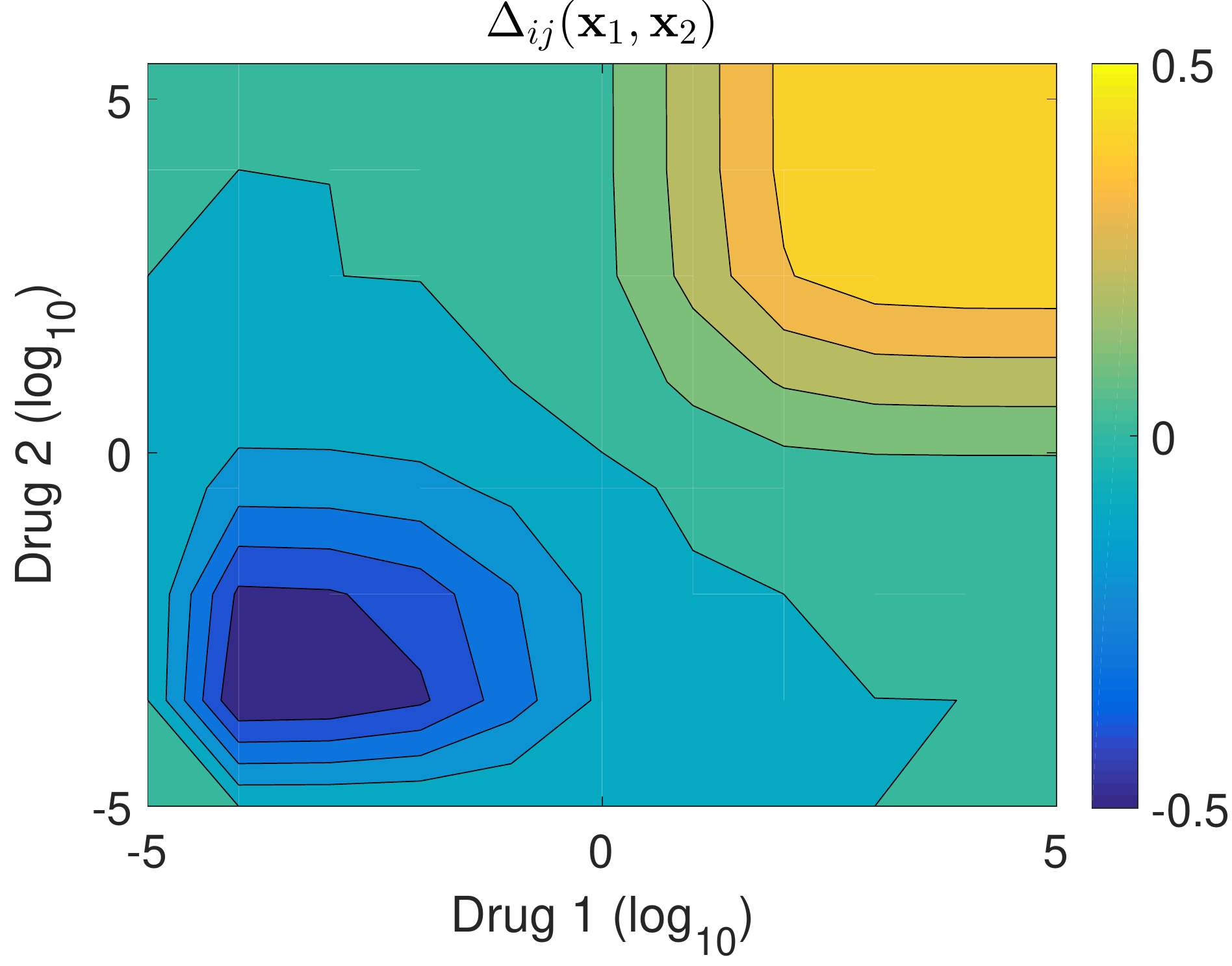}}
\subfloat[]{\includegraphics[width=.5\textwidth]{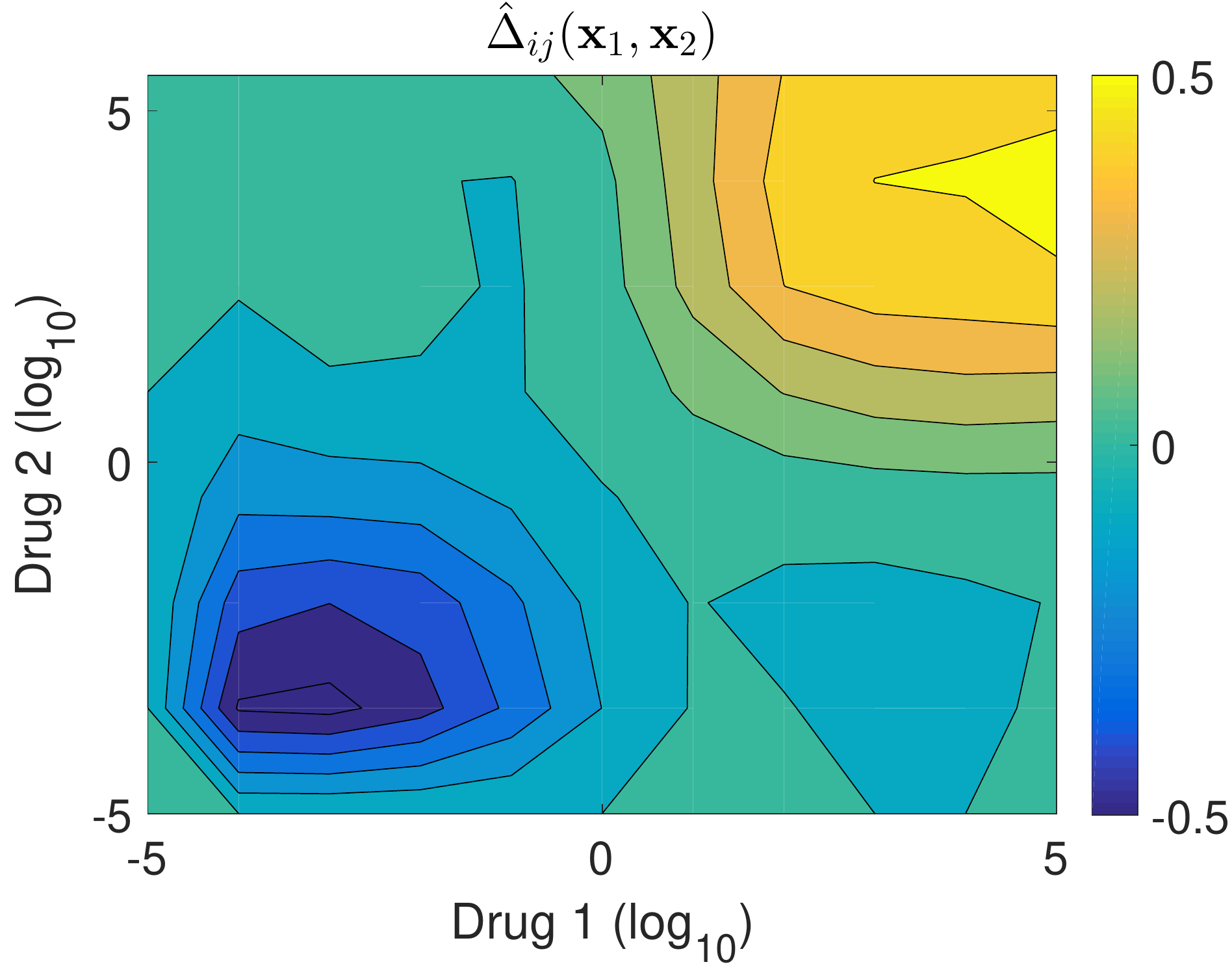}}

\subfloat[]{\includegraphics[width=.5\textwidth]{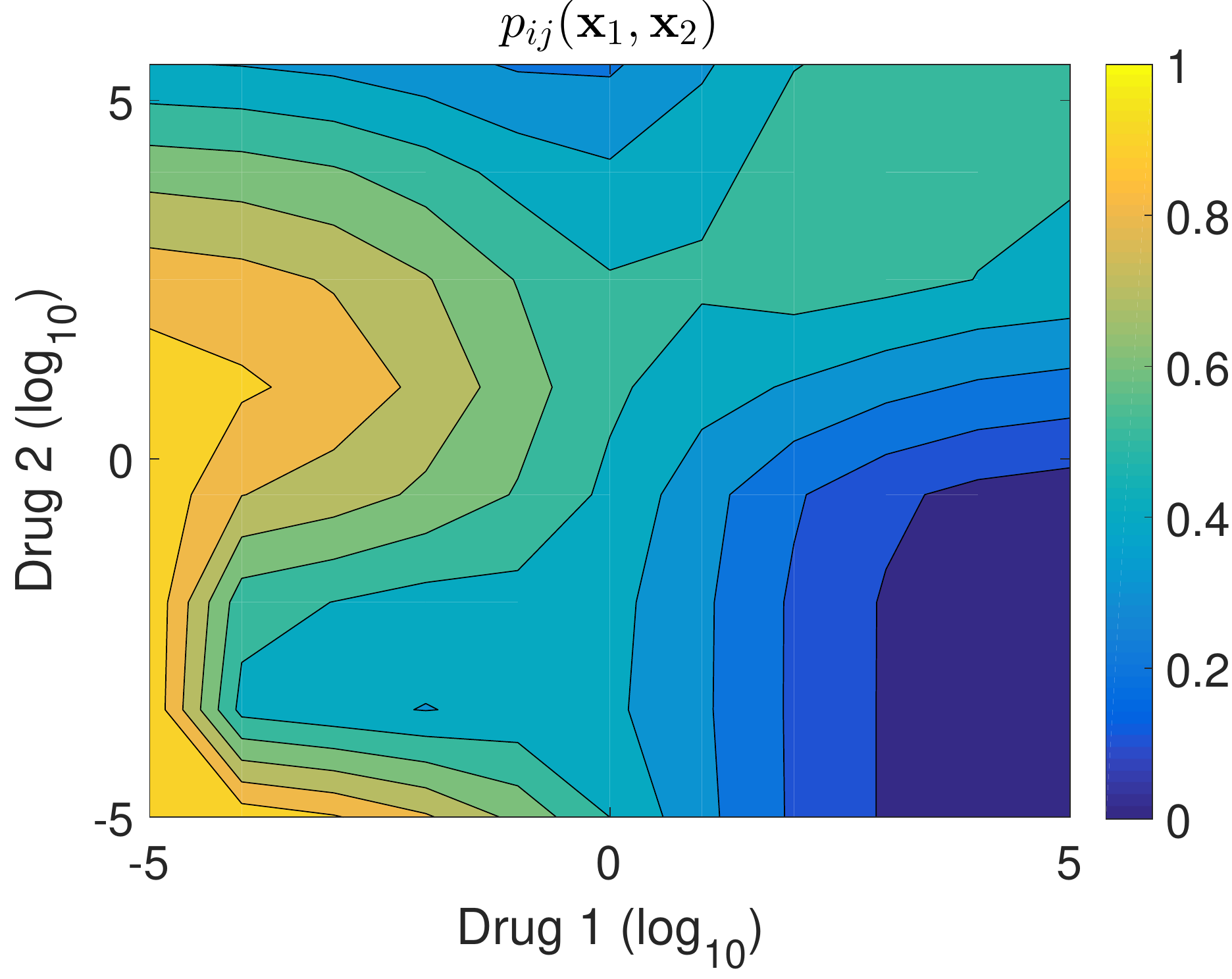}}
\subfloat[]{\includegraphics[width=.5\textwidth]{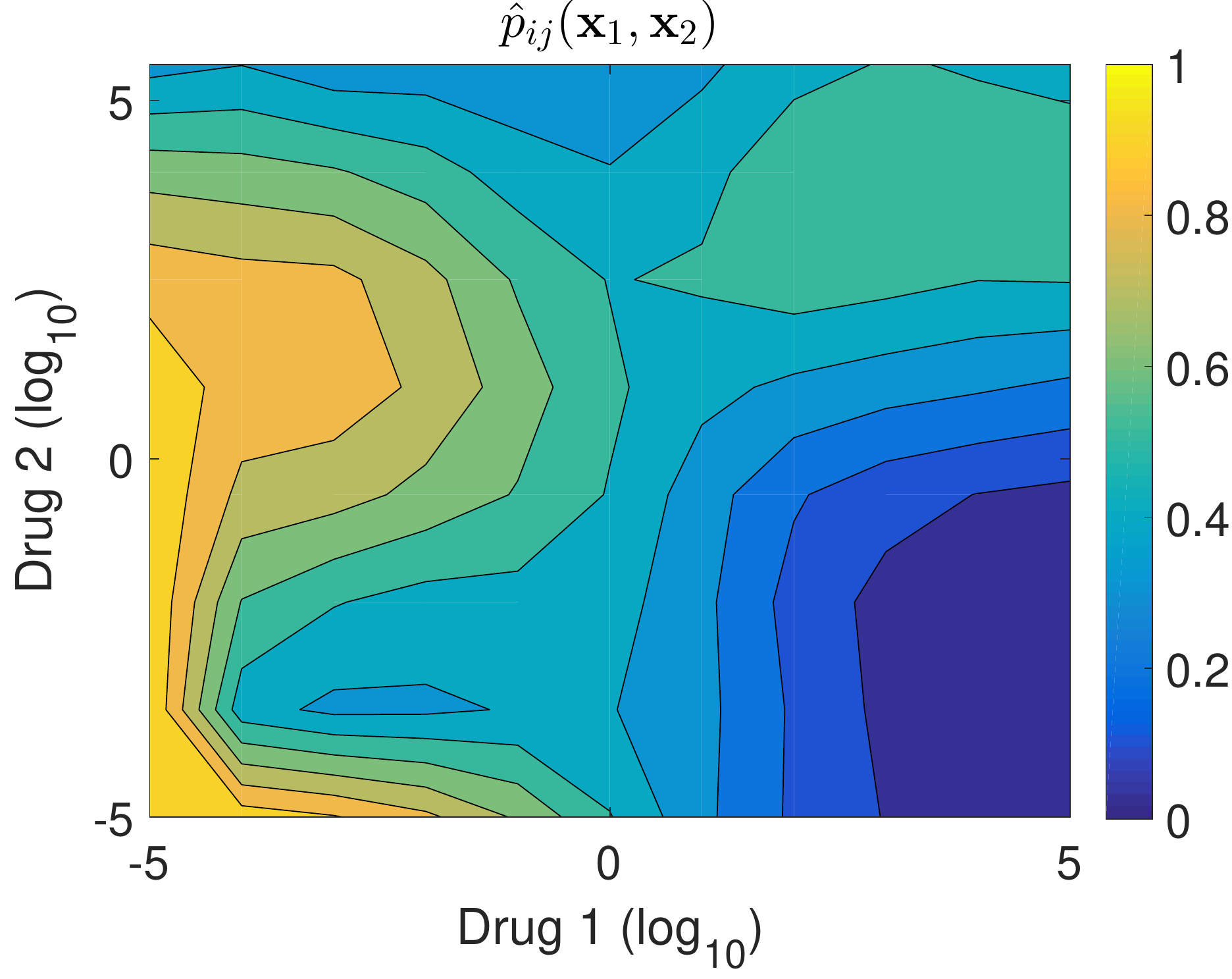}}
}
\caption{Simulation study: ($\Delta^{(3)}$, $\sigma^2_{\phi} \sim \text{HC}(1), \epsilon^r_{ij} \sim N(p_{ij}, \sigma^2_{\epsilon}), n_{rep} = 3)$. Contour plots of simulated (left column) versus estimated (right column) surfaces (zero-interaction $p^0_{ij}$, interaction $\Delta_{ij}$, and mean $p_{ij}$).}
  \label{fig:selected_contours}
\end{figure}

To further illustrate the adequacy of the results, we select the simulation highlighted in red in Supplementary Table 1, for which the interaction surface is simulated as the non-monotone $\Delta^{(3)}$, the variability parameters $\sigma^2_{\phi}$ are assigned a half-Cauchy prior HC$(1)$, the errors are simulated from a normal distribution, and $n_{rep} = 3$. Figure \ref{fig:selected_surfaces} shows the posterior estimates of the zero-interaction and interaction surfaces in comparison with the ground truth used to simulate the data. Furthermore, Figure \ref{fig:selected_contours} shows the contour plots of such surfaces. It is clear from this comparison, how the model is able to recover the main features of the surfaces of interest.

\subsection{Application to Ovarian Cancer Cell test data}\label{sec:OC}

We apply the proposed model to data obtained by in-vitro combination experiments involving two ovarian cancer cell-lines, namely OVCAR8 and SKOV3 described at \url{https://dtp.cancer.gov/discovery_development/nci-60/}. Human ovarian adenocarcinoma (commonly known as ovarian cancer) is the seventh most common cancer diagnosed in women worldwide, and it is characterized by a five years survival rate of only 30\% for advanced tumors. Late stage diagnosis is the main reason for the high mortality rate \cite{Reid_2017, Coleman_2013}. At later stages, the tumor presents an invasion of a tumor-triggered inflammatory fluid, called \emph{ascites}, into the abdominal cavity. This fluid is of heterogeneous composition, including tumorigenic factors, growth factors and bioactive lipids, that can favour the growth of the malignancy \cite{Kim_2016}. In-vitro studies on cell-lines showed a tendency to resistance of the tumor to platinum-based standard-of-care drugs \cite{OC_inpreparation}. This motivates the interest in studying the effect of drug combinations on ovarian cancer cell cultures, in particular comparing the results when ascites material is added to the culture.

The cell-lines were cultured in the presence of medium alone, or of medium and ascites. The two drug pairs tested in these experiments were Niclosamide and WP1066, in combination with Nilotinib. Nilotinib targets the Bcr-Abl tyrosin kinase and the c-Kit pathway, as well as JAK-kinases. On the other hand, WP1066 and Niclosamide target the transcription factor STAT3 (signal transducer and activator of transcription 3). Despite Niclosamide being commonly used for the treatment of tapeworm, it has been used in drug re-purposing studies as STAT3 inhibitor. The different mechanisms of action of the combined drugs, acting on the same signaling pathway but at different levels, supports the use of our model, which is based on assumptions analogous to the Bliss independence \cite[see][]{Fitzgerald_2006}. For each cell-line and drug combination, an experiment with $n _{rep} = 3$ replicates has been performed, testing a 6x6 matrix of drug concentrations. The resulting viability observations are then fitted using the presented model. In particular, having observed high robustness in the simulation study, we select half-Cauchy prior distributions with $h = 1$ for the variance terms in the model. We produce posterior chains of 5.000 samples from initial chains of length 50.000, after discarding the first half as burn-in, and by applying a thinning of 5 iterations to the second half.

\subsection{Analysis of the monotherapy responses}\label{sec:OC_Mono}
\begin{figure}[!ht]
\hspace{-0.5cm}
\subfloat[]{\includegraphics[height=.3\textwidth]{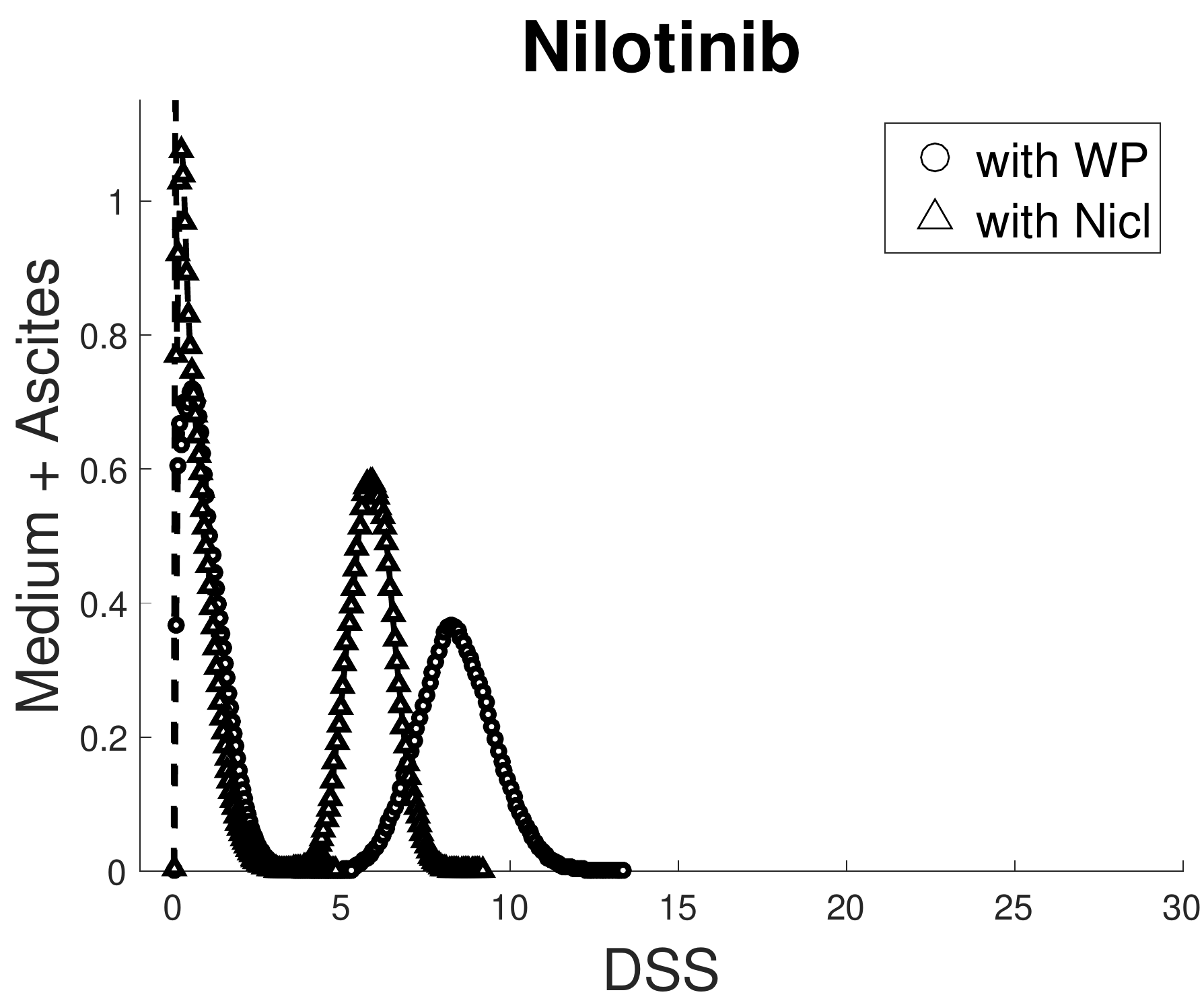}}
\subfloat[]{\includegraphics[height=.3\textwidth]{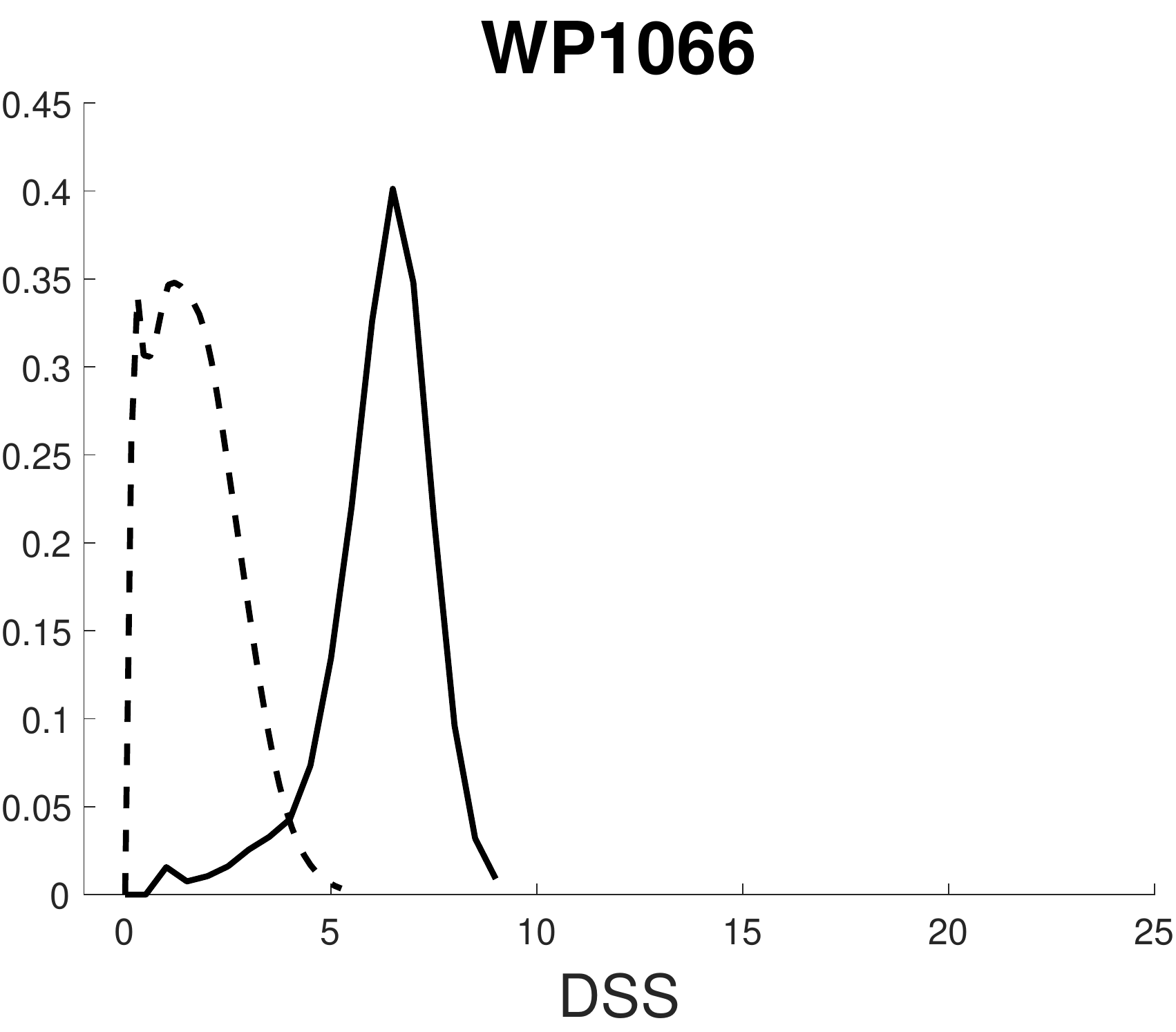}}
\subfloat[]{\includegraphics[height=.3\textwidth]{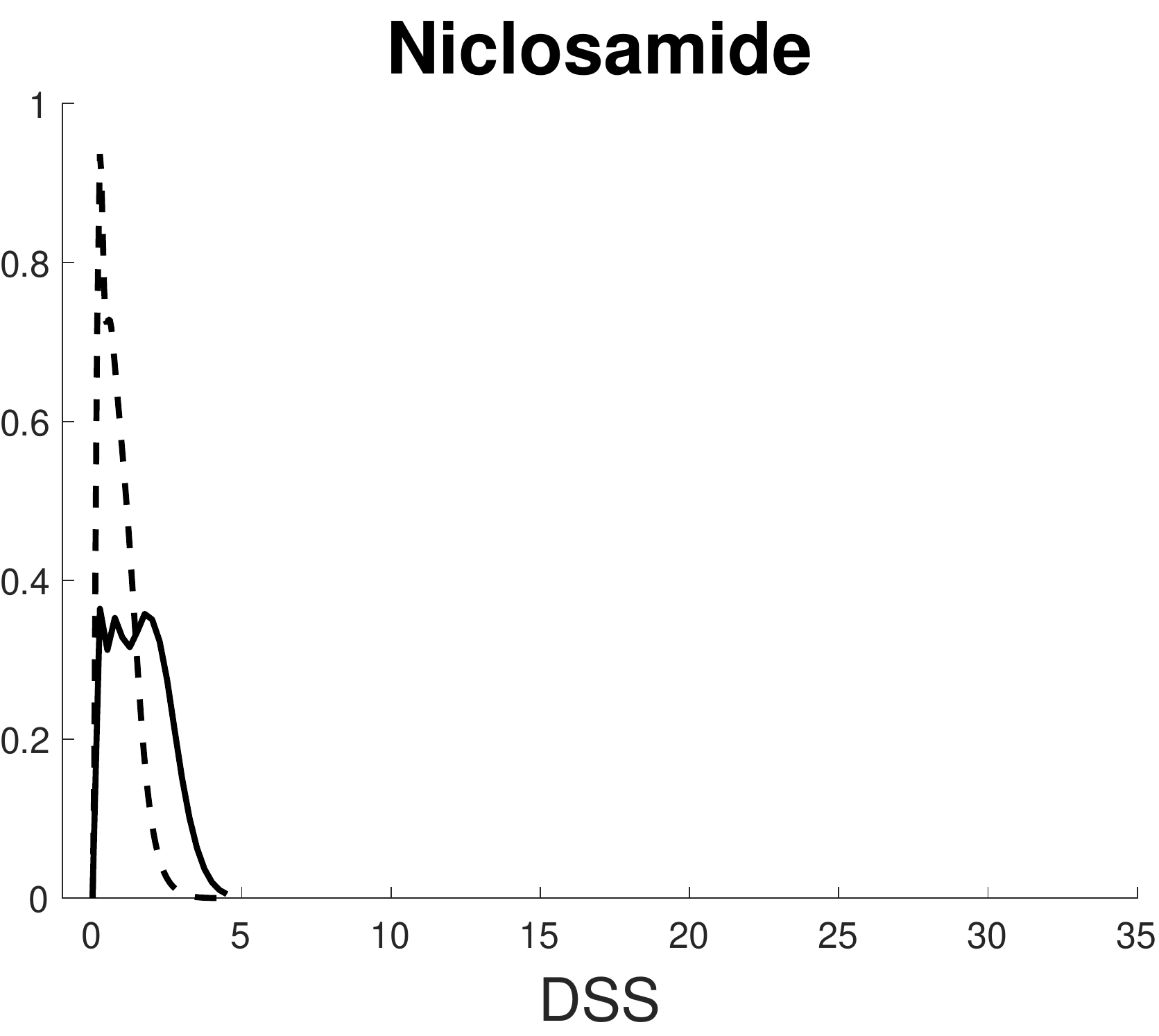}}

\hspace{-0.5cm}
\subfloat[]{\includegraphics[height=.3\textwidth]{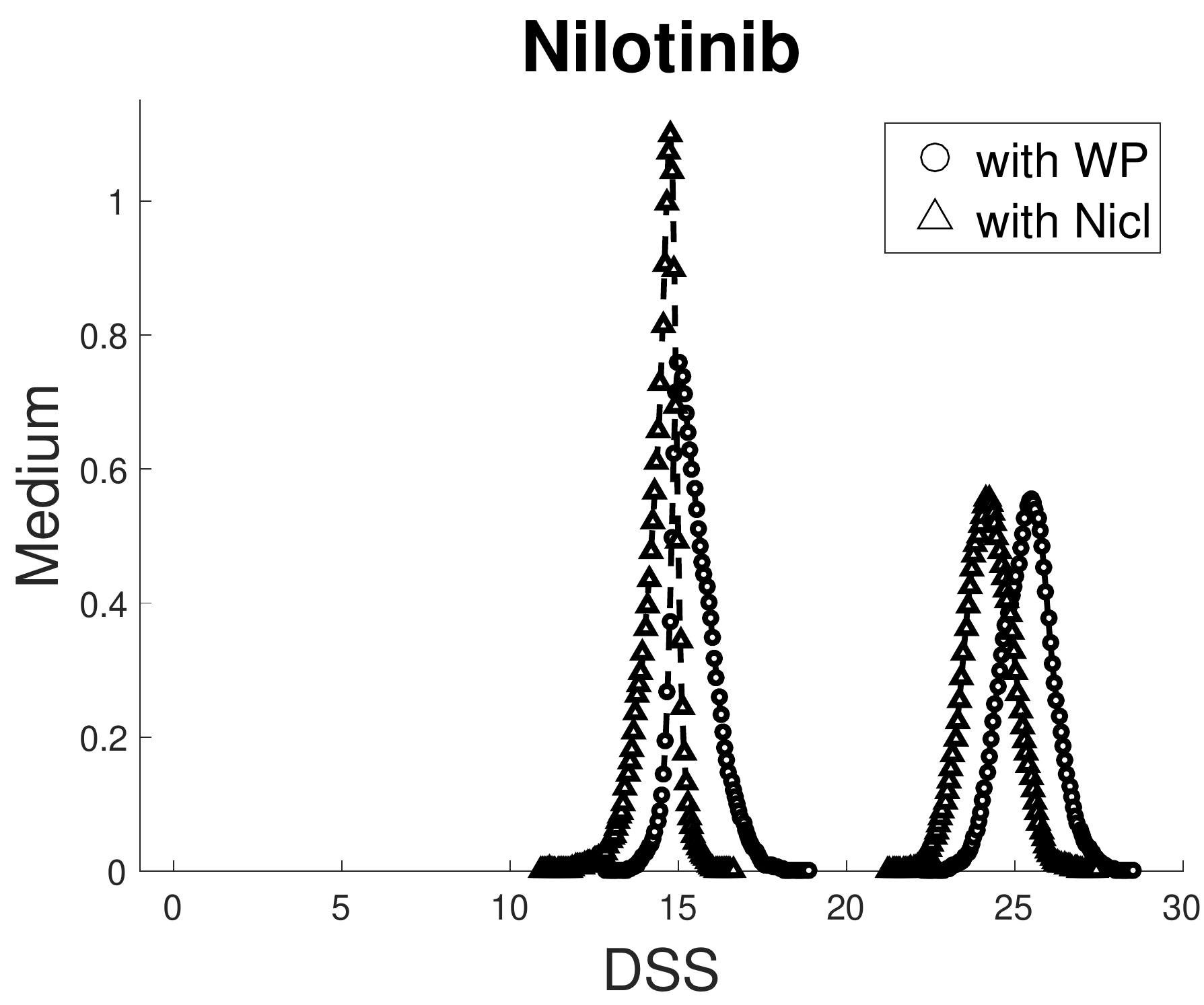}}
\subfloat[]{\includegraphics[height=.3\textwidth]{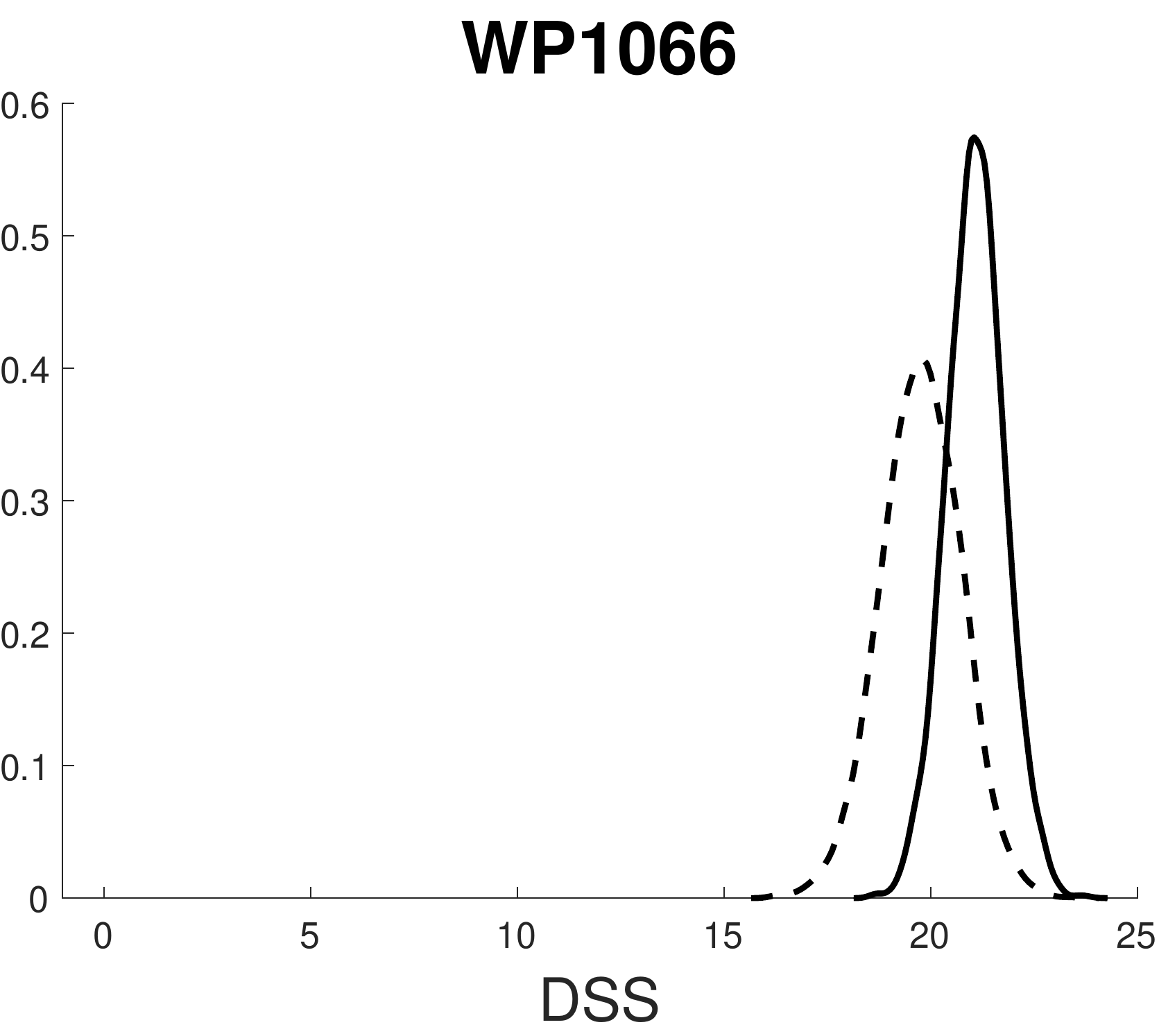}}
\subfloat[]{\includegraphics[height=.3\textwidth]{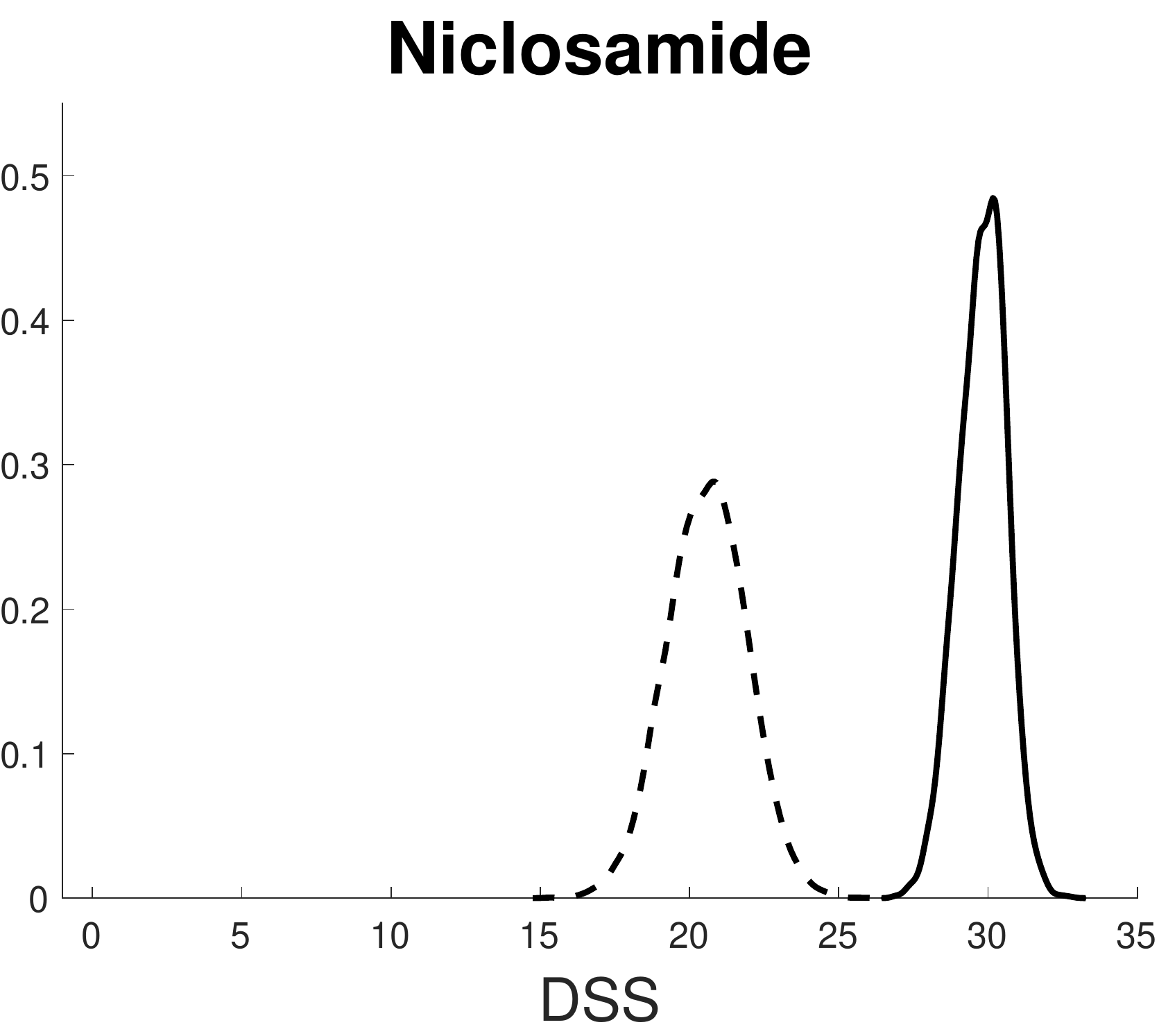}}
\caption{Application to Ovarian Cancer data -- Posterior distribution of the drug sensitivity scores (DSS) for the single drugs, under different experimental settings. The curves are obtained obtained via kernel-density estimation from the MCMC posterior samples of the DSS scores. Top row: medium and ascites; bottom row: medium alone. Continuous lines refer to the cell-line OVCAR8, while dashed lines to the cell-line SKOV3. The number of curves for Nilotinib is double, since this drug is present in every experiment.}
  \label{fig:OC_MT}
\end{figure}

One of the advantages of our model is the ability to perform joint inference on different parameters of interest. We begin with a description of some measures related to the monotherapy behaviour of the combined drugs (i.e., at the margins of the matrix of concentrations). Figure \ref{fig:OC_MT} shows the posterior distribution of the DSS, computed for each of the drugs employed in the combination experiments. The figure is divided into two rows, one for each experimental condition (medium and ascites or medium alone), while within each row, we show a sub-figure for each drug tested. Regardless of the combination experiment considered, it is striking to notice how the compounds are more effective in the absence of ascites, supporting the fact that the presence of ascites in the tumor micro-environment reduces the effectiveness of the drugs \cite{OC_inpreparation}. Furthermore, by looking at the centrality of the posterior distributions of the DSS scores, we can see that all compounds are more effective when used on the cell-line OVCAR8 (continuous lines) than with SKOV3 (dashed lines), which may relate to their different resistance profiles. As expected, the posterior distributions relative to the DSS scores for Nilotinib, which is the compound common to all experiments, do not show considerable differences for the same experimental setting (i.e., for the same culture and cell-line characteristics, see Figure \ref{fig:OC_MT}(a,d)).

\subsection{Analysis of the combination responses}\label{sec:OC_Combos}
\begin{figure}[!ht]
\hspace{-0.5cm}
\subfloat[$p^0$]{\includegraphics[scale=0.21]{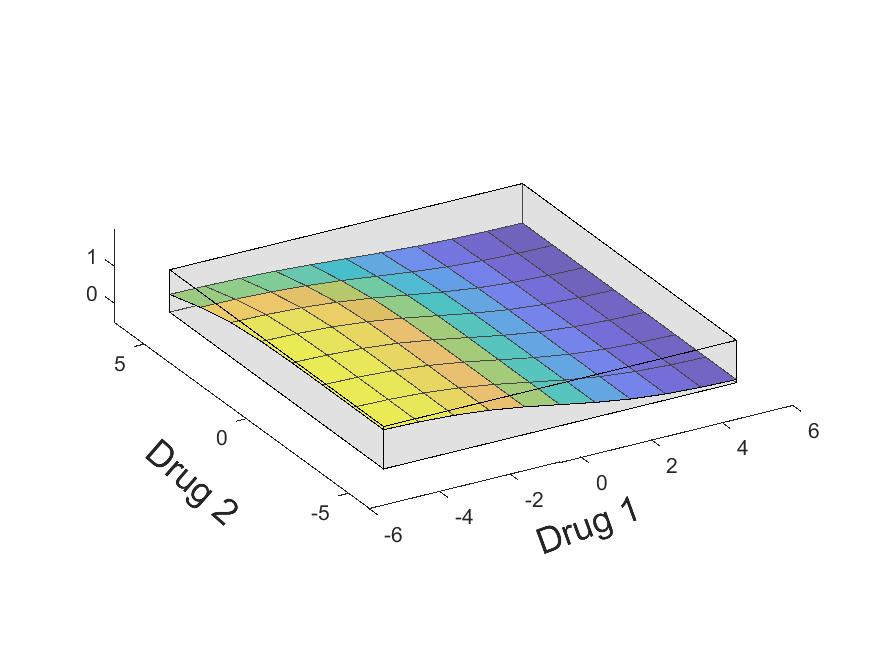}}
\subfloat[$|\Delta|$]{\includegraphics[scale=0.24]{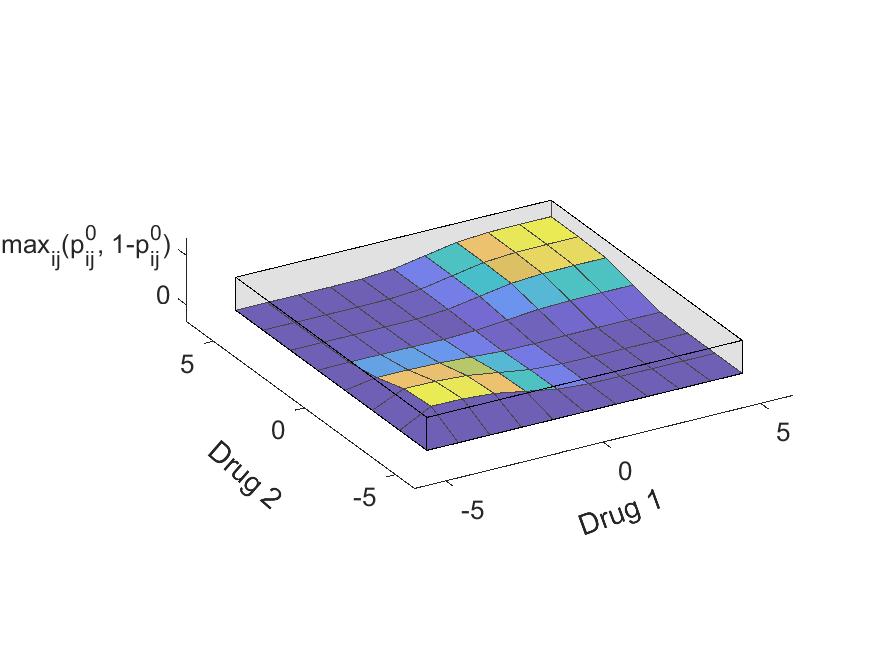}}
\caption{Illustration of the computation of the relative volume under the surface (rVUS) -- This quantity is computed as the ratio between the volume under a given surface S and that of the cube limited by the support of S (grey shapes in the figures above). The limits of the shape depend on the surface being analysed: the zero-interaction $p^0$ takes values in the range $(0,1)$ \textbf{(a)}, while the total interaction surface $|\Delta|$ takes values in $(0,\max_{ij}(p^0_{ij},1 - p^0_{ij}))$, due to the model constraints \textbf{(b)}. Throughout the paper, we also consider $\text{rVUS}(\Delta^+) = \text{rVUS}(|\min(0,\Delta)|)$ and $\text{rVUS}(\Delta^-) = \text{rVUS}(\max(0,\Delta))$ as the synergistic and antagonistic contributions to the total interaction volume, respectively. The computation of the volume under a surface S is obtained by applying the trapezoid method along each dimension in succession.}
  \label{fig:rVUS_ex}
\end{figure}

\begin{figure}[!ht]
\hspace{-0.5cm}
\subfloat{\includegraphics[height=.4\textwidth]{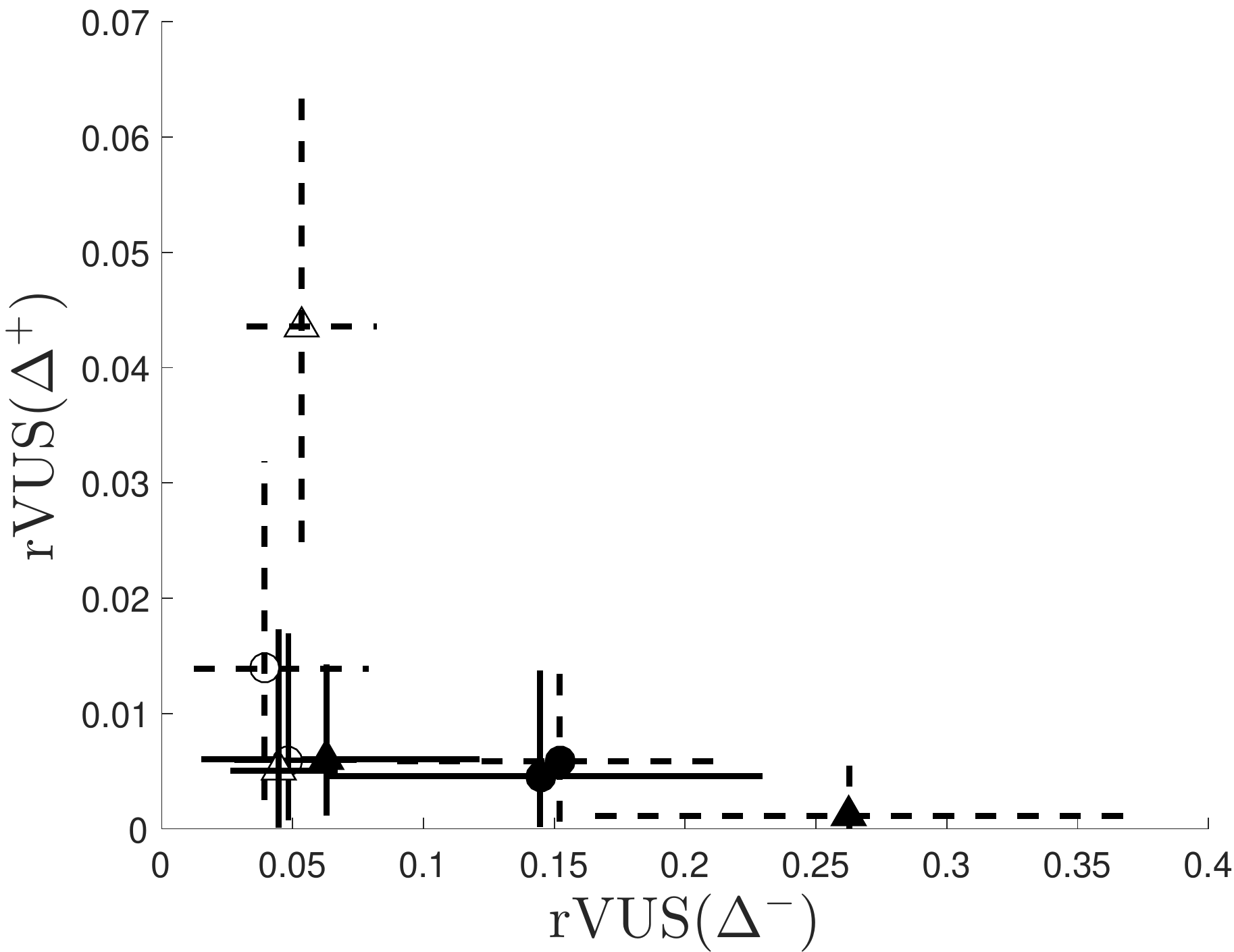}}
\subfloat{\includegraphics[height=.4\textwidth]{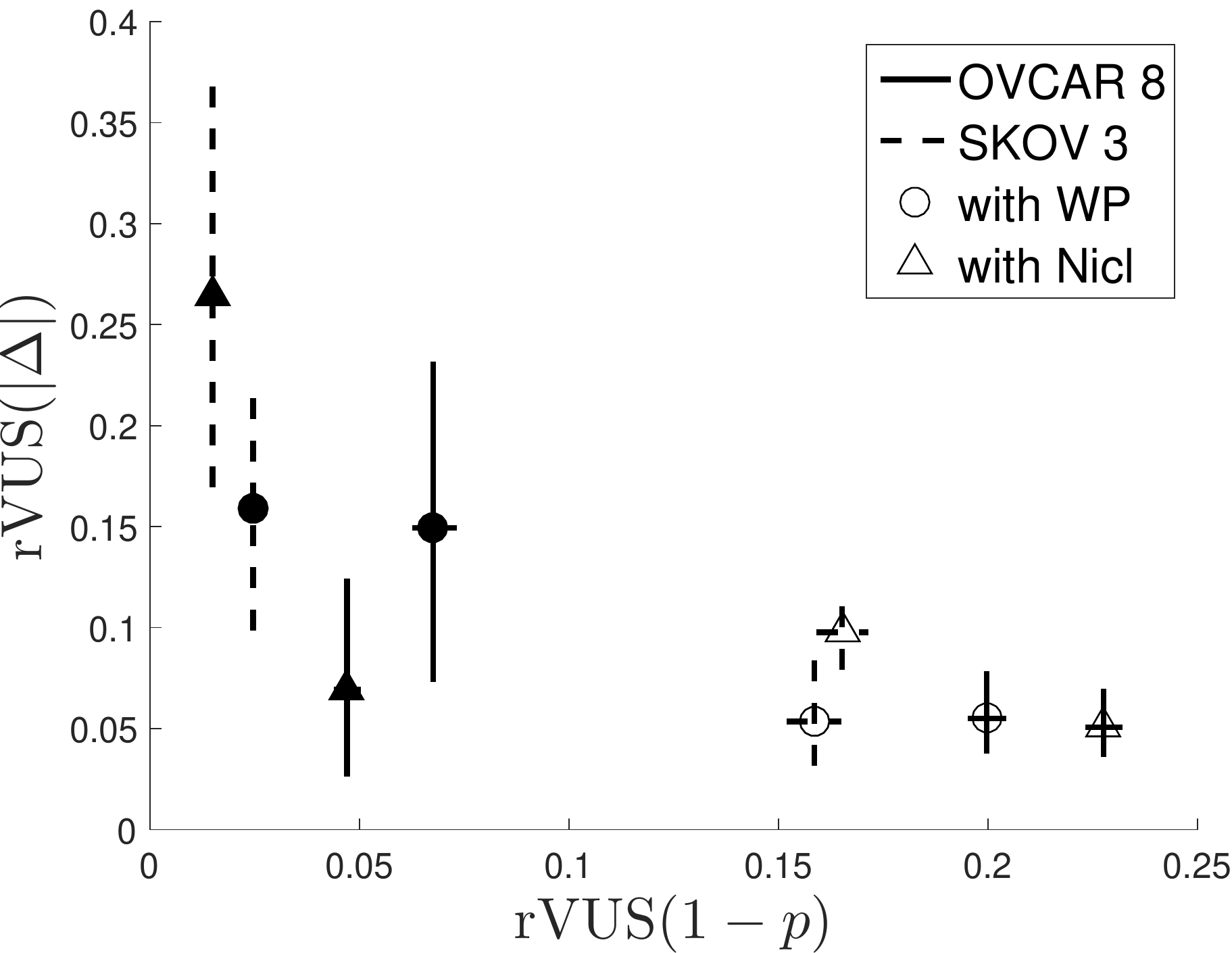}}
\caption{Application to Ovarian Cancer data -- \textbf{(a)}: Posterior estimates and 95\% credible intervals for the relative volume under the interaction surface, distinguishing between synergistic (rVUS($\Delta^+$)) and antagonistic (rVUS($\Delta^-$)) components. \textbf{(b)}: Posterior estimates and 95\% credible intervals for the total relative volume under the interaction surface (rVUS($|\Delta|$)) VS the relative volume under the complementary mean surface (rVUS($1 - p$)). In both panels, black markers indicate that ascites fluid was dispensed.}
  \label{fig:OC_rVUS}
\end{figure}

We present a summary of the posterior estimates of the zero-interaction and interaction terms by means of the relative volume under the surface (rVUS), devised to quantify the contribution of each term of the model the to the drug combination experiment. The computation of rVUS for a given surface is schematically described in Figure \ref{fig:rVUS_ex} and its corresponding caption. Figure \ref{fig:OC_rVUS} reports the posterior medians and 95\% credibility intervals of different rVUS, in the different experimental conditions. In particular, we present the rVUS for the interaction surface, specifying its synergistic and antagonistic components in panel (a), and the rVUS of the total interaction versus the rVUS of the complementary mean surface $(1 - p)$ in panel (b). The last quantity is of particular interest, as it represents an overall measure of efficacy of the combination experiments. Both combinations for SKOV3 in medium seem quite synergistic and effective. In particular, despite recovering the fact that the compounds are more effective in the cell-line OVCAR8, we observe higher synergy levels for combinations of SKOV3 cultured in medium. As expected, the presence of ascites reduces the effect of the drug combination in both cell lines, and is associated with higher levels of antagonism when compared to the experiments with medium only. Moreover, we observe that the combination of Nilotinib with Niclosamide is more effective in medium for both OVCAR8 and SKOV3, while the opposite is observed when the cell lines are treated with ascites (see Figure \ref{fig:OC_rVUS}(b)).

We now focus our analysis on the experiments characterized by cells cultured in medium alone, since these seem to be the most interesting from the point of view of study of the interaction surfaces. Analogous results for the experiments where ascites was dispensed are reported in Section 1.2 of the Supplementary Material. Figure \ref{fig:OC_contour_pij} shows the contour plots of the posterior mean of the surface $p_{ij}$ for the cell-lines OVCAR8 and SKOV3, for both combinations. We can appreciate the good fitting of the model in all the scenarios, with values of LPML in panels (a,b,c,d) equal to 159.626, 152.862, 104.170 and 99.351, respectively. In order to characterize the behaviour of the selected combinations, we can define:
\begin{equation}\label{eq:bi_EC50}
\text{bi-EC}_{50}(\delta) := \{(x,y) \in \mathbb{R}^2 | p(x,y) \in (0.5 \pm \delta)\}, \ \delta > 0,
\end{equation}
representing the set of $\log_{10}$-concentrations for which the surface $p(x,y)$ takes values in a small neighbourhood of 0.5. This quantity, interpretable as a bi-variate $\text{EC}_{50}$ set, can be estimated from the MCMC sample by inverting the posterior mean $\hat{p}_{ij}$, computed as Monte Carlo average. Thus, it can be very useful due to its interpretation, especially when communicating with biomedical researchers, who are very familiar with the concept of $\text{EC}_{50}$ in the univariate case. To the best of our knowledge, this quantity has not been proposed in the literature before, and its computation is an advantage of the joint Bayesian modelling approach. Figure \ref{fig:OC_contour_pij} reports the $\text{bi-EC}_{50}(\delta)$ sets in $\log_{10}$-scale as a red dots, for $\delta = 0.01$. The the red points identify regions of clinical relevance, where it is expected to find a good efficacy/toxicity trade-off.

\begin{figure}[!ht]
\hspace{-0.5cm}
\subfloat[OVCAR8]{\includegraphics[height=.4\textwidth]{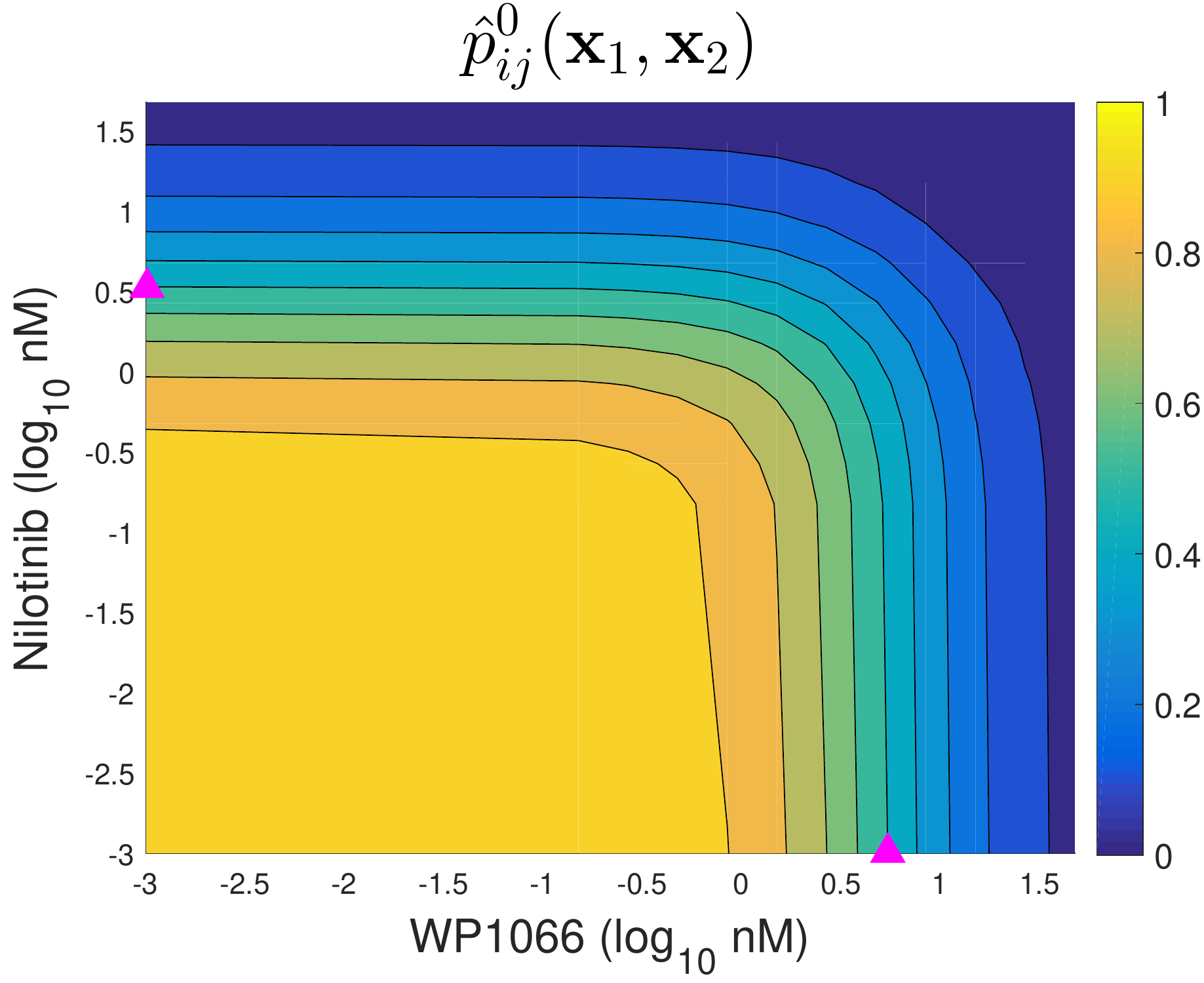}}
\subfloat[OVCAR8]{\includegraphics[height=.4\textwidth]{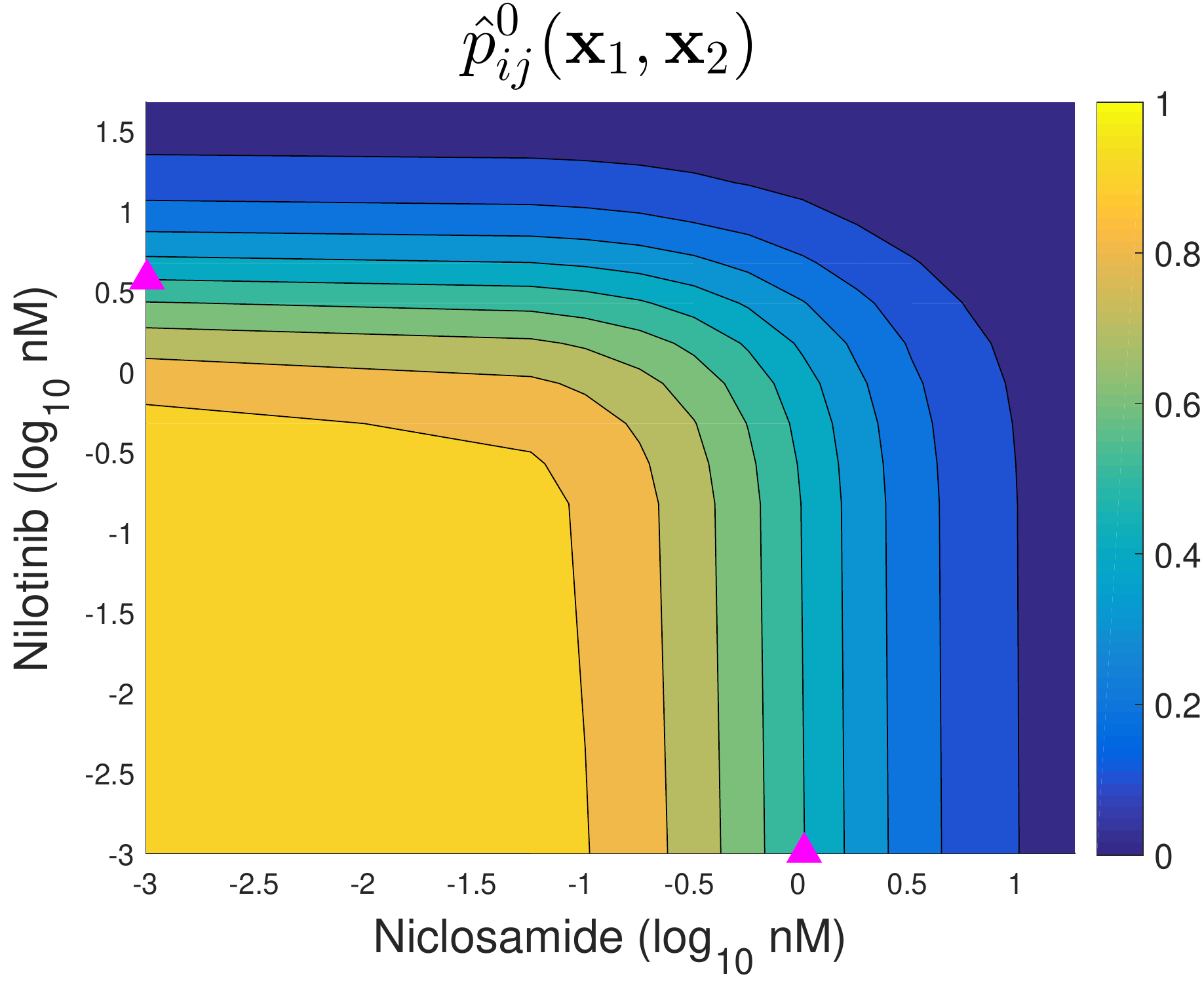}}\\

\hspace{-0.5cm}
\subfloat[SKOV3]{\includegraphics[height=.4\textwidth]{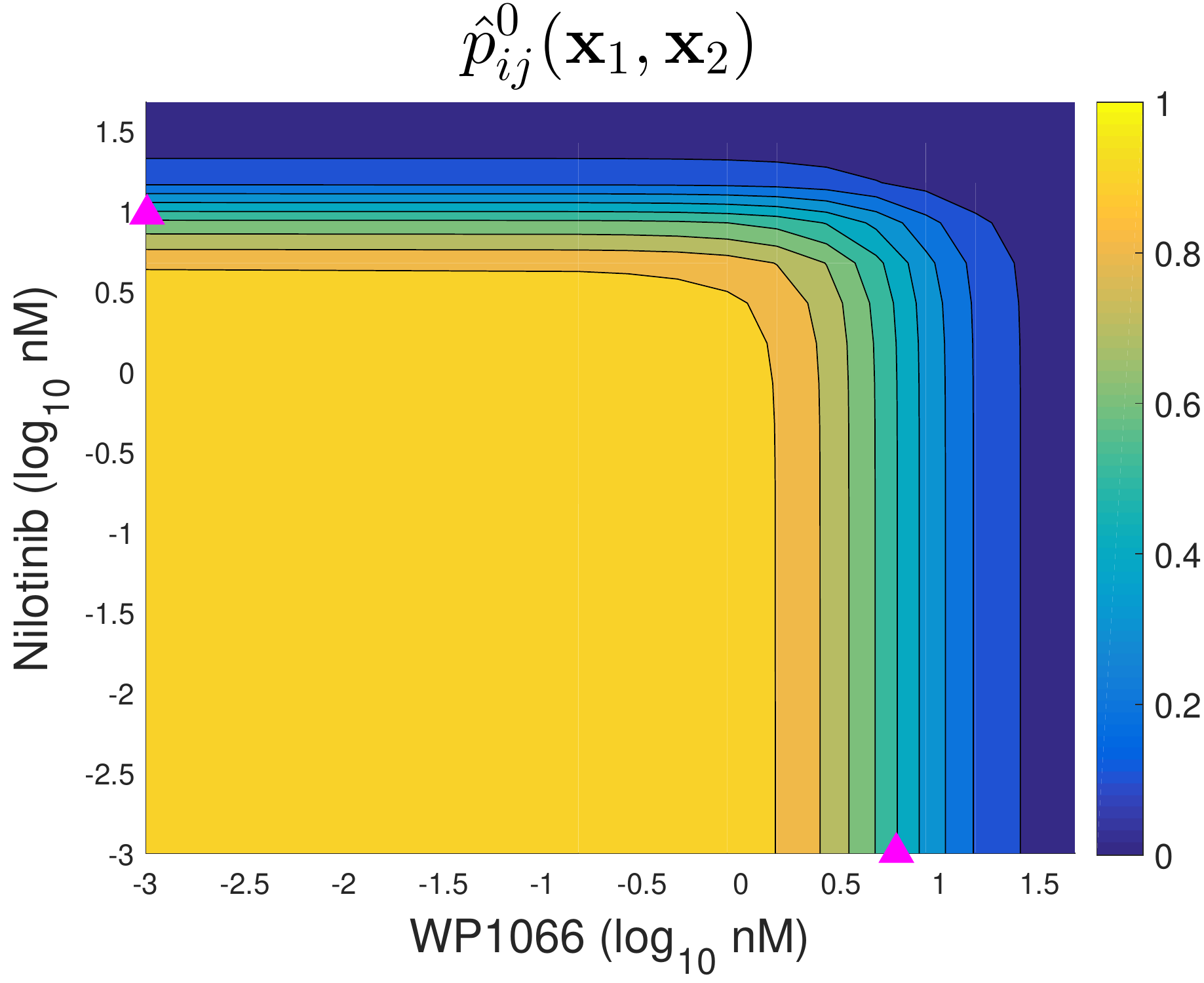}}
\subfloat[SKOV3]{\includegraphics[height=.4\textwidth]{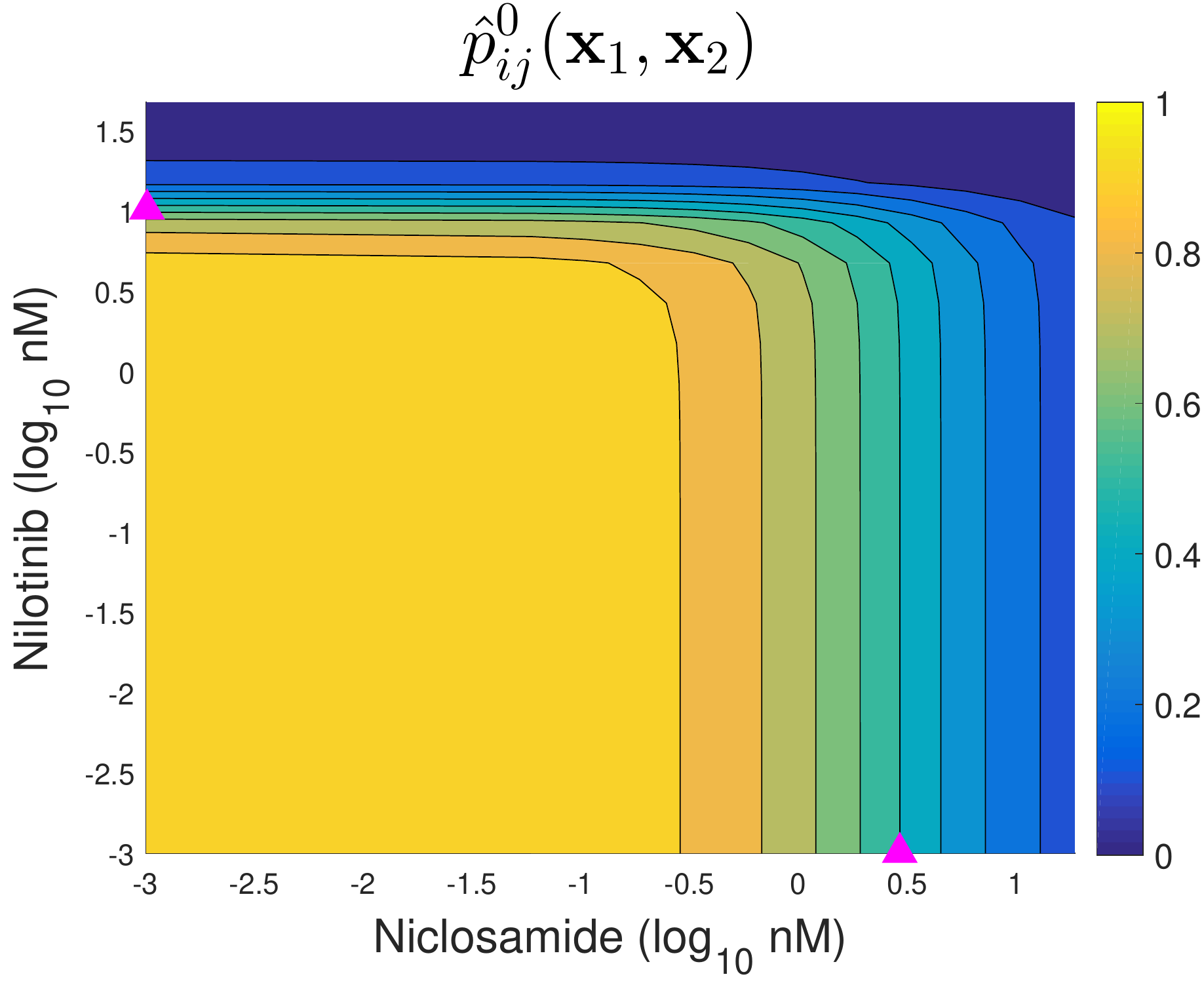}}
\caption{Application to Ovarian Cancer data -- Contour plots of the posterior mean of the zero-interaction surface $p^0_{ij}$, including the monotherapy $\text{EC}_{50}$ estimates as magenta triangles. Experiments with cell-lines treated in medium alone. Top row: OVCAR8 ; bottom row: SKOV3.}
  \label{fig:OC_contour_p0}
\end{figure}

\begin{figure}[!ht]
\hspace{-0.5cm}
\subfloat[OVCAR8]{\includegraphics[height=.4\textwidth]{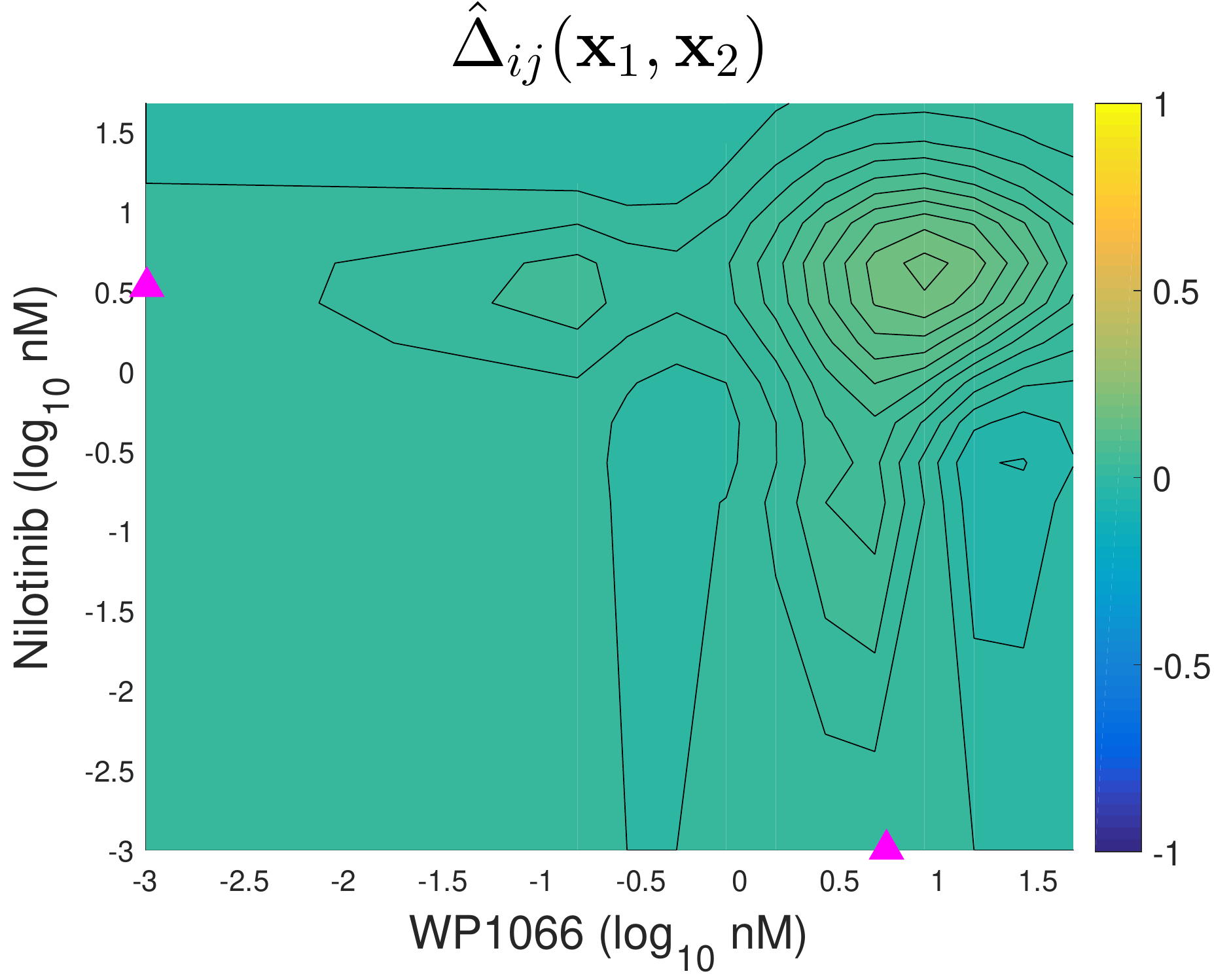}}
\subfloat[OVCAR8]{\includegraphics[height=.4\textwidth]{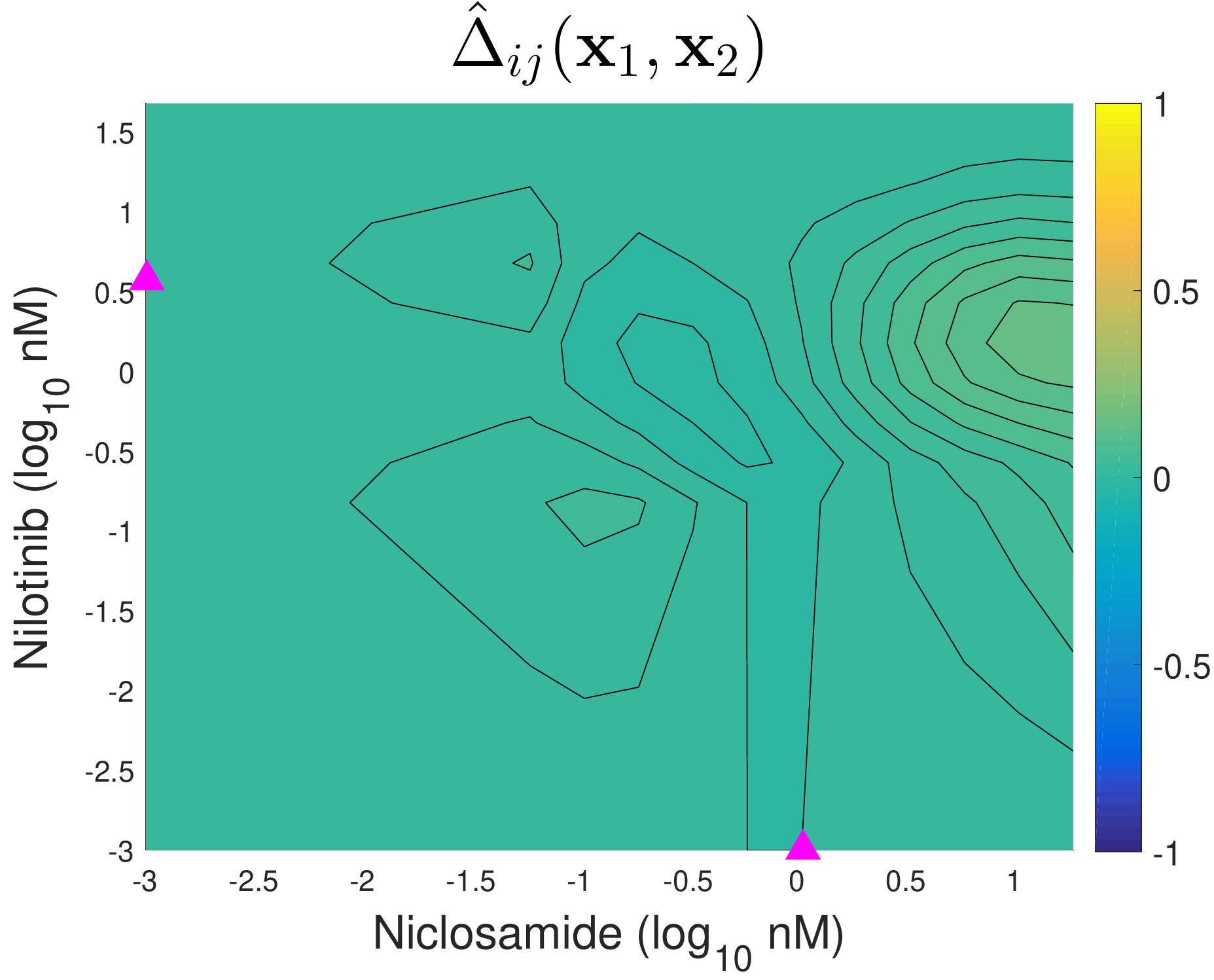}}\\

\hspace{-0.5cm}
\subfloat[SKOV3]{\includegraphics[height=.4\textwidth]{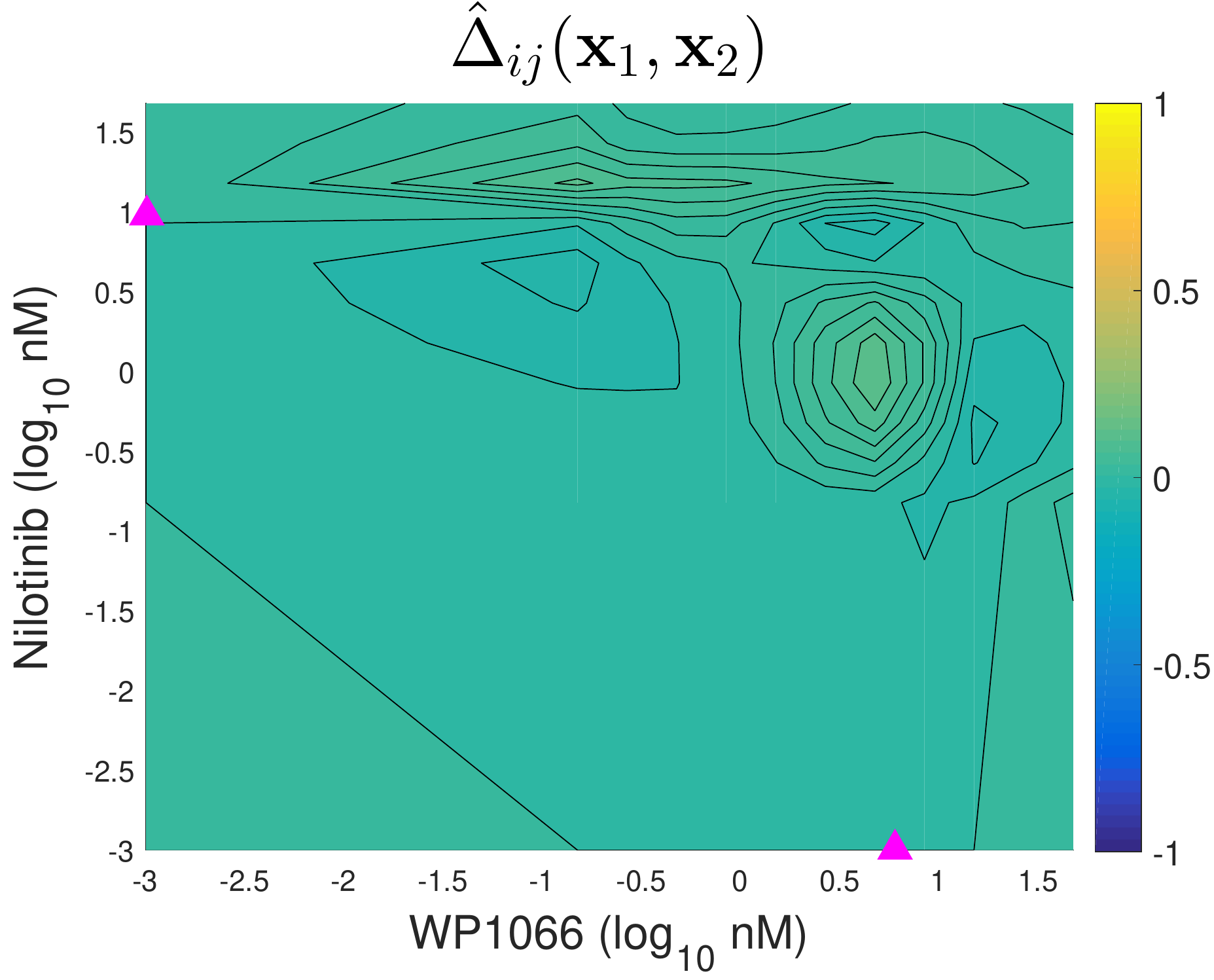}}
\subfloat[SKOV3]{\includegraphics[height=.4\textwidth]{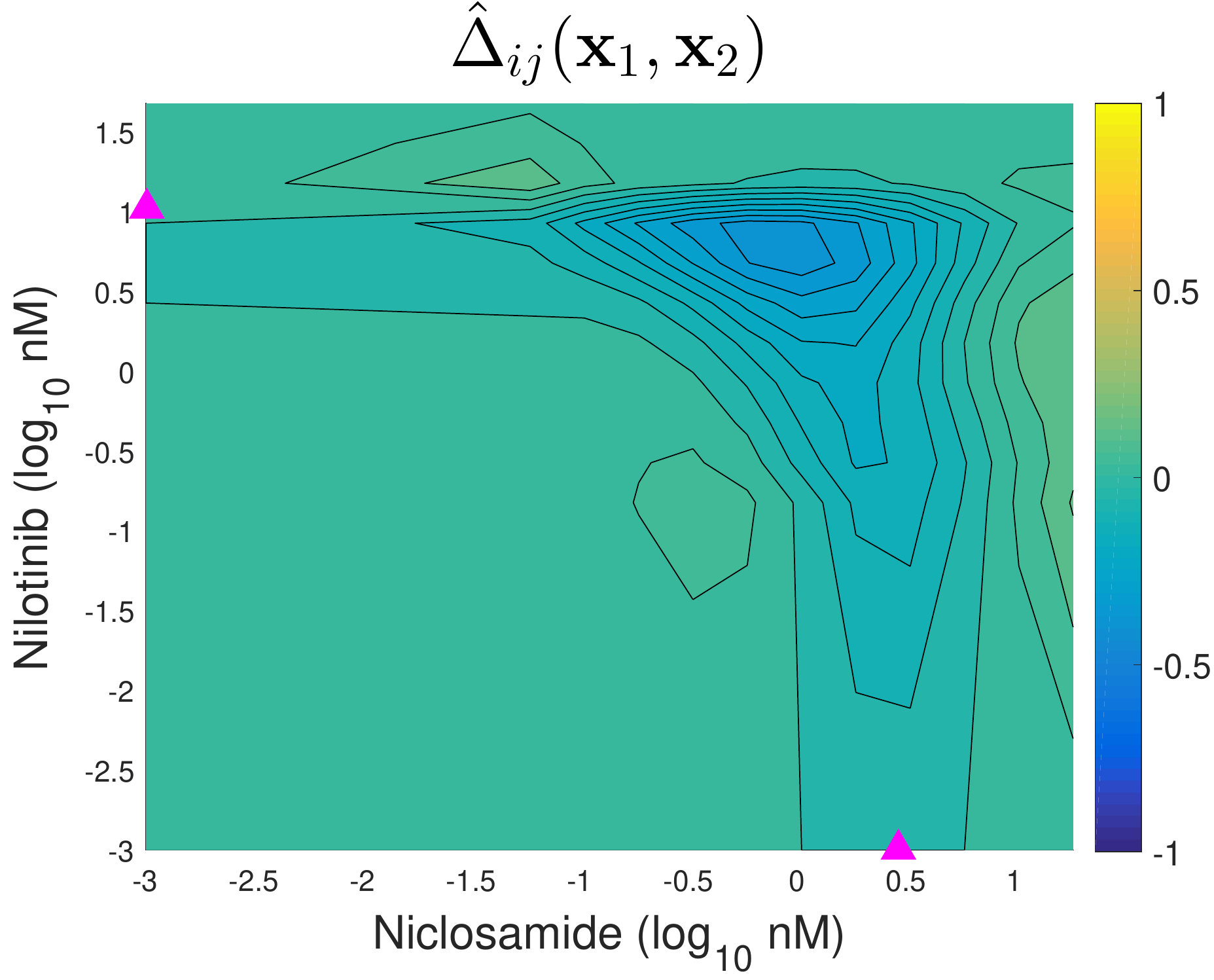}}
\caption{Application to Ovarian Cancer data -- Contour plots of the posterior mean of the interaction surface $\Delta_{ij}$, including the monotherapy $\text{EC}_{50}$ estimates as magenta triangles. Experiments with cell-lines treated in medium alone. Top row: OVCAR8 ; bottom row: SKOV3.}
  \label{fig:OC_contour_Delta}
\end{figure}

As far as the zero-interaction and interaction surfaces are concerned, we report similar contour plots in Figures \ref{fig:OC_contour_p0} and \ref{fig:OC_contour_Delta}, once again focusing our analysis on the experiments where only medium was dispensed. The posterior estimates of the zero-interaction surface shown in Figure \ref{fig:OC_contour_p0} present similar behaviour for each cell-line, indicating that the presence of Nilotinib (the common drug in all the experiments) might be the driver of the response when the two compounds do not interact, deducible from the steep decay on that side of the surface. However, differences can be observed when looking at the interaction surfaces of Figure \ref{fig:OC_contour_Delta}. In particular, the posterior estimates of the interaction surface for the cell-line SKOV3 show different behaviours for concentrations of Nilotinib around its estimated $\text{EC}_{50}$ (posterior mean, magenta triangle), indicating that some synergy can be found when combined with Niclosamide. Specifically, concentrations of Nilotinib slightly lower than its $\text{EC}_{50}$, induce synergistic interactions with Niclosamide around its $\text{EC}_{50}$ concentration.

\begin{figure}[!ht]
\hspace{-0.5cm}
\subfloat[OVCAR8]{\includegraphics[height=.4\textwidth]{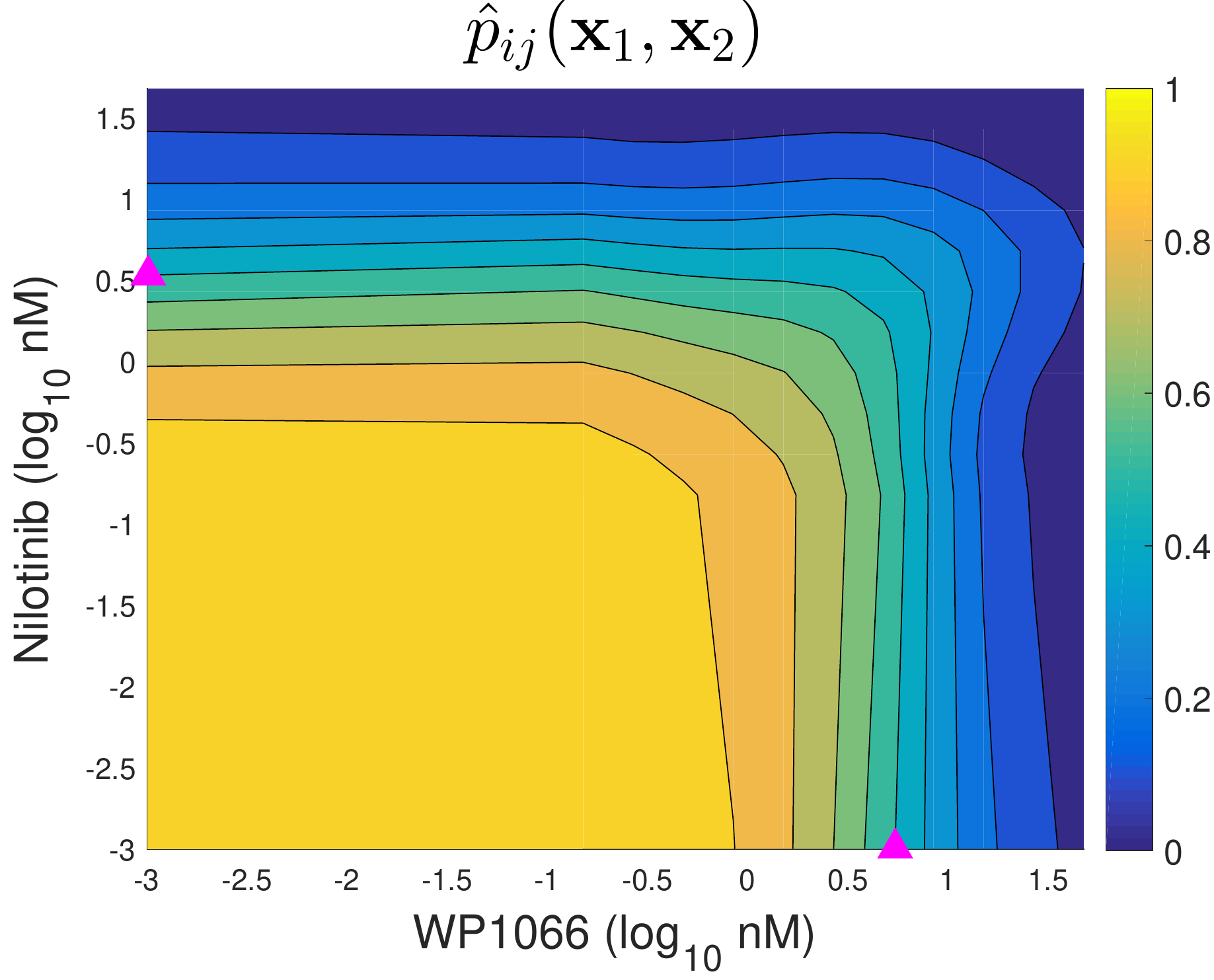}}
\subfloat[OVCAR8]{\includegraphics[height=.4\textwidth]{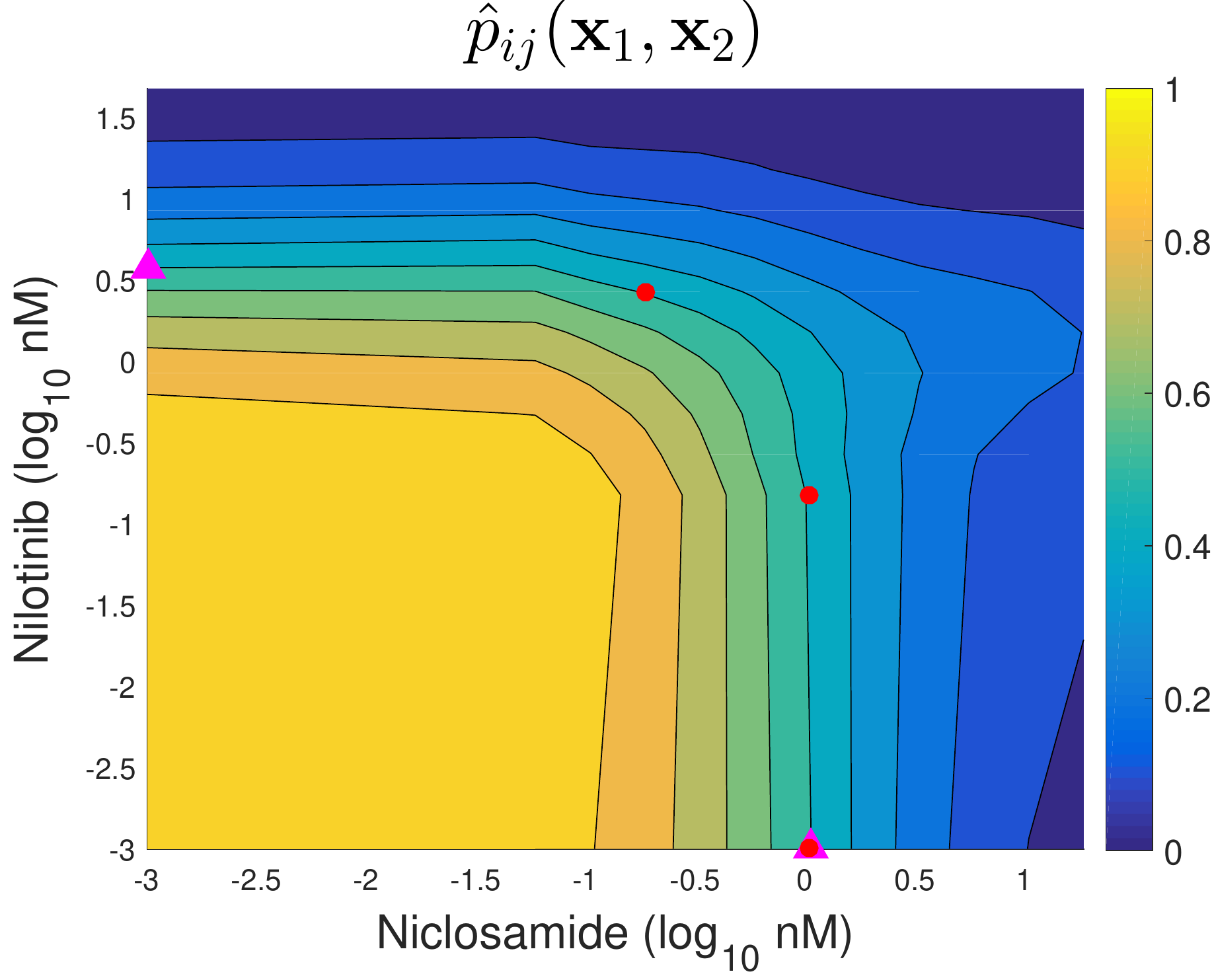}}\\

\hspace{-0.5cm}
\subfloat[SKOV3]{\includegraphics[height=.4\textwidth]{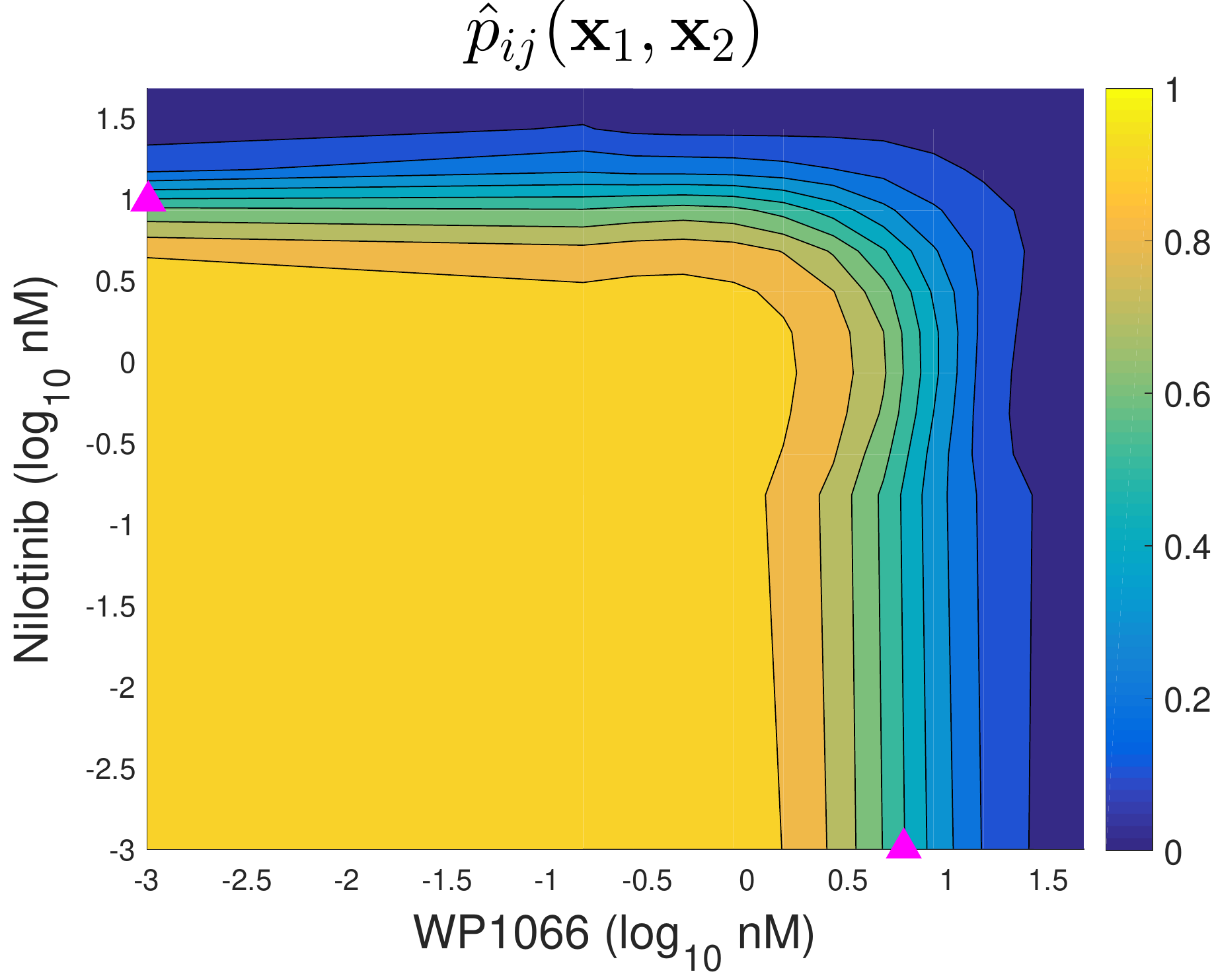}}
\subfloat[SKOV3]{\includegraphics[height=.4\textwidth]{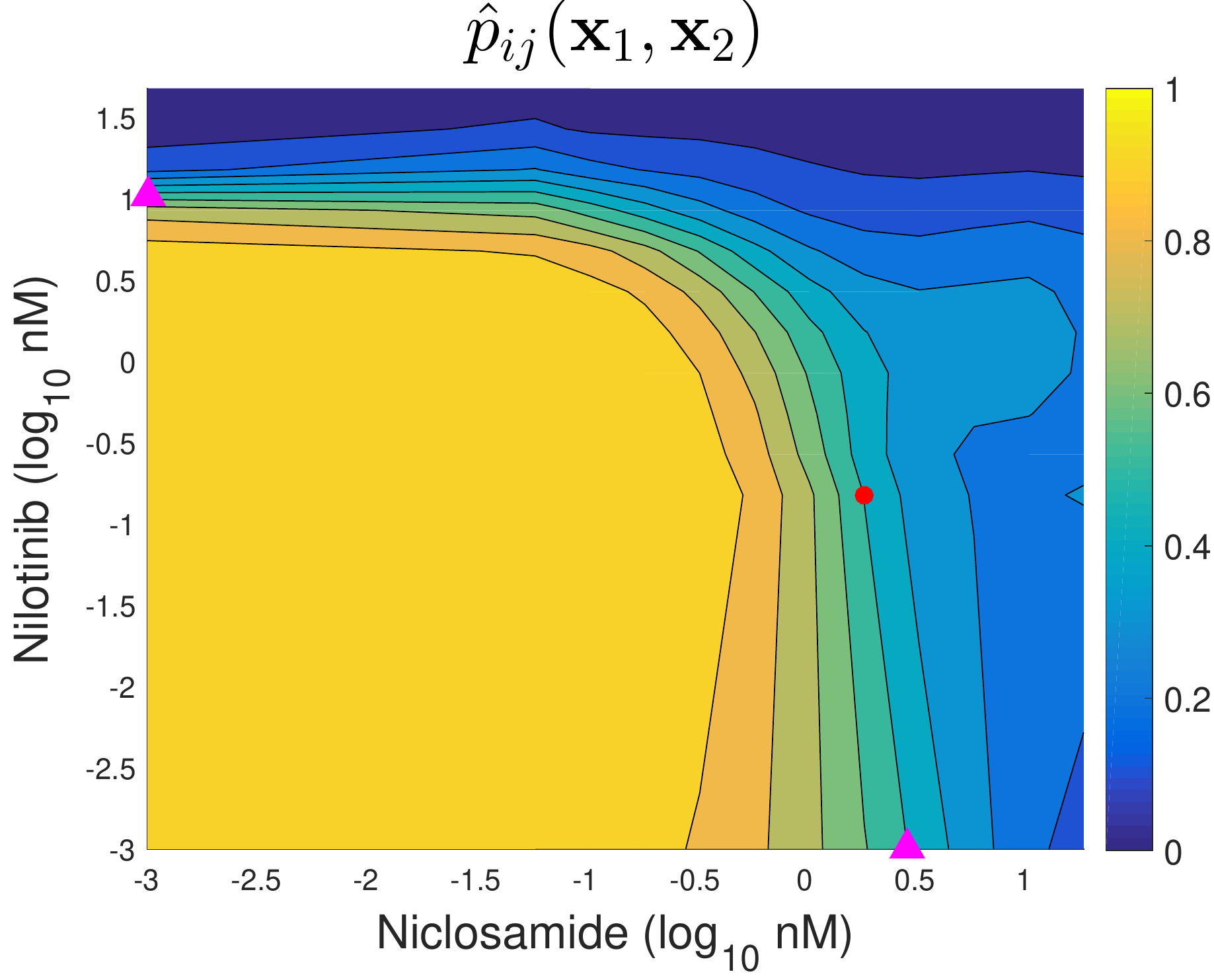}}
\caption{Application to Ovarian Cancer data -- Contour plots of the posterior mean of the surface $p_{ij}$, including the $\text{bi-EC}_{50}(0.01)$ estimates as red dots, and the monotherapy $\text{EC}_{50}$ estimates as magenta triangles. Experiments with cell-lines treated in medium alone. Top row: OVCAR8 ; bottom row: SKOV3. In order to obtain more self-explaining plots, the grid of concentrations used was extended between the observed values $\bm x_{1,1:n_1}$ and $\bm x_{2,1:n_2}$, i.e. excluding the zero-concentrations.}
  \label{fig:OC_contour_pij}
\end{figure}

\section{Conclusions}\label{sec:Concl}
Drug sensitivity screening is an important component in determining personalized therapies for cancer patients, and the practice of screening multiple compounds at a time is becoming increasingly relevant, since combination therapies are often preferred over treatment by individual drugs. If synergistic, drugs combinations can help reducing the individual dosage, thereby decreasing the risk of intolerable side effects. In addition, if the drugs have different mechanisms of action, they can decrease the likelihood that the tumor develops drug resistance. The availability of suitable technology, small molecule targeting compounds, and organic material, makes this task a feasible goal.

In this work, we provided a probabilistic interpretation of the quantities at play in viability assay experiments, by interpreting the viable state of each cell as the outcome of a Bernoulli experiment. We model the probability of success in order to distinguish between the zero-interaction and an interaction term. In the proposed Bayesian setting, the use of splines as generalised covariates allows the interaction term to present both synergistic and antagonistic features in the same combination study. This novel approach to the study of drug-drug interaction yields high flexibility, interpretability of the zero-interaction and interaction components, as well as the quantification of the uncertainty of the estimates.

We provided an extensive simulation study, highlighting the importance of the choice of the prior distributions for the model parameters. In the proposed study, the choice of vague prior distributions seemed to be able to mitigate the low number of replicates available. A comparison with standard approaches to calculate measures of drug interaction (Loewe, Bliss, ZIP and HSA scores) to the same datasets showed higher performance of the proposed method in terms of goodness-of-fit and MSE measures.

As an application to real life data, we fit the proposed model to an ovarian cancer dataset produced by in-vitro experiments in \cite{OC_inpreparation}, presenting various experimental conditions for two drug combinations. Namely, the ovarian cancer cell-lines OVCAR8 and SKOV3, cultured in medium alone or in medium and ascites (a fluid present in the abdominal cavity at later stages of the tumor progression), are exposed to the effect of the combination of Nilotinib and WP1066 or of Nilotinib and Niclosamide. These are compounds targeting small molecules along the same cell pathways, and already used for the treatment of other malignancies. In particular, we observe how the combination of drugs is more advantageous for fighting the malignancy in case of the cell line SKOV3, with some interaction found in the combination of Nilotinib with Niclosamide. On the other hand, the presence of ascites seems to counter-act the effect of the drugs at the given concentrations in all scenarios, already at the single drug level. This may be due to the very high levels of IL-6 in the ascites, which may override the effect of the inhibitors (in the cultures in medium there is some IL-6 secreted by the cell lines, but much less).

The proposed approach can be extended to accommodate different prior settings for the parameters. Of particular interest is the study of the prior distribution of the variances of the parameters, which can be specified to incorporate prior information (e.g., from clinicians and experts) or to be covariate-dependent (e.g., by including heteroscedasticity in the model). Further work concerns the possibility of extending the model, allowing the joint analysis of different combinations at a time, e.g. by formulating a joint hierarchical model, where appropriate hyper-prior distributions allow borrowing of information across the different drug combinations. Finally, note that while we specified our model for cell viability assays in this manuscript, it can be easily adapted to other cell counting assays such as cytotoxicity assays, measuring the counts of dead cells in the experiment.

\clearpage
\section*{Supplementary Materials}

The Supplementary Materials file: \\ \detokenize{Cremaschi_Frigessi_Tasken_Zucknick_Bayes_Synergy_Suppl.pdf} \\ referenced throughout the manuscript is made available with this paper.

\noindent
The zip folder: \\ \detokenize{Cremaschi_Frigessi_Tasken_Zucknick_Bayes_Synergy_Code.zip} \\ contains the Matlab codes used in the simulated examples of Section \ref{sec:Simul_Study}.
\vspace*{-8pt}

\section*{Acknowledgements}
Andrea Cremaschi performed this work while affiliated with the Oslo Centre for Biostatistics and Epidemiology (\textbf{OCBE}), University of Oslo, and with the Department of Cancer Immunology, Institute of Cancer Research, Oslo University Hospital.

\noindent
The authors acknowledge funding by the University of Oslo, Faculty for Mathematics and Natural Sciences and by the UiO Convergence Grant PerCaThe, the  Research Council of Norway (NFR Centre of Excellence BigInsight, No. 237718), The Norwegian Cancer Society and The Regional Health Authority for South-Eastern Norway.
\vspace*{-8pt}

\bibliographystyle{plainnat}
\bibliography{Biblio}

\begin{thebibliography}{33}
\providecommand{\natexlab}[1]{#1}
\providecommand{\url}[1]{\texttt{#1}}
\expandafter\ifx\csname urlstyle\endcsname\relax
  \providecommand{\doi}[1]{doi: #1}\else
  \providecommand{\doi}{doi: \begingroup \urlstyle{rm}\Url}\fi

\bibitem[Berenbaum(1989)]{Berenbaum_1989}
M.~C. Berenbaum.
\newblock \emph{What is synergy?}
\newblock \emph{Pharmacological reviews}, 41\penalty0 (2):\penalty0 93--141,
  1989.

\bibitem[Bliss(1939)]{Bliss_1939}
C.~I. Bliss.
\newblock \emph{The toxicity of poisons applied jointly}.
\newblock \emph{Annals of applied biology}, 26\penalty0 (3):\penalty0 585--615,
  1939.

\bibitem[Boik et~al.(2008)Boik, Newman, and Boik]{Boik_etal_2008}
J.~C. Boik, R.~A. Newman, and R.~J. Boik.
\newblock Quantifying synergism/antagonism using nonlinear mixed-effects
  modeling: A simulation study.
\newblock \emph{Statistics in medicine}, 27\penalty0 (7):\penalty0 1040--1061,
  2008.

\bibitem[Chou and Talalay(1984)]{Chou_Talalay_1984}
T.~C. Chou and P.~Talalay.
\newblock \emph{Quantitative analysis of dose-effect relationships: the
  combined effects of multiple drugs or enzyme inhibitors}.
\newblock \emph{Advances in enzyme regulation}, 22:\penalty0 27--55, 1984.

\bibitem[Coleman et~al.(2013)Coleman, Monk, Sood, and Herzog]{Coleman_2013}
R.~L. Coleman, B.~J. Monk, A.~K. Sood, and T.~J. Herzog.
\newblock Latest research and treatment of advanced-stage epithelial ovarian
  cancer.
\newblock \emph{Nature reviews Clinical oncology}, 10\penalty0 (4):\penalty0
  211, 2013.

\bibitem[de~Boor(2001)]{De_Boor_2001}
Carl de~Boor.
\newblock A practical guide to splines (applied mathematical sciences vol. 27),
  2001.

\bibitem[Eilers and Marx(2010)]{Eilers_2010}
P.~H.~C. Eilers and B.~D. Marx.
\newblock Splines, knots, and penalties.
\newblock \emph{Wiley Interdisciplinary Reviews: Computational Statistics},
  2\penalty0 (6):\penalty0 637--653, 2010.

\bibitem[Eroukhmanoff et~al.(2019)Eroukhmanoff, Cremaschi, Landskron,
  Flage-Larsen, Gade, Bj\o~rge, Urbanucci, and Task\'{e}n]{OC_inpreparation}
L.~Eroukhmanoff, A.~Cremaschi, J.~Landskron, L.~Flage-Larsen, A.~Gade,
  L.~Bj\o~rge, A.~Urbanucci, and K.~Task\'{e}n.
\newblock Ovarian cancer ascites promotes aberrant signalling activation
  plasticity and protection toward treatment options.
\newblock \emph{article in preparation}, 2019.

\bibitem[Fitzgerald et~al.(2006)Fitzgerald, Schoeberl, Nielsen, and
  Sorger]{Fitzgerald_2006}
J.~B. Fitzgerald, B.~Schoeberl, U.~B. Nielsen, and P.~K. Sorger.
\newblock Systems biology and combination therapy in the quest for clinical
  efficacy.
\newblock \emph{Nature chemical biology}, 2\penalty0 (9):\penalty0 458, 2006.

\bibitem[Fouquier and Guedj(2015)]{Fouquier_Guedj_2015}
J.~Fouquier and M.~Guedj.
\newblock \emph{Analysis of drug combinations: current methodological
  landscape}.
\newblock \emph{Pharmacology research and perspectives}, 3\penalty0 (3), 2015.

\bibitem[Geisser and Eddy(1979)]{Geisser_Eddy_1979}
S.~Geisser and W.~F. Eddy.
\newblock \emph{A predictive approach to model selection}.
\newblock \emph{Journal of the American Statistical Association}, 74\penalty0
  (365):\penalty0 153--160, 1979.

\bibitem[Greco et~al.(1995)Greco, Bravo, and Parsons]{Greco_etal_1995}
W.~R. Greco, G.~Bravo, and J.~C. Parsons.
\newblock The search for synergy: a critical review from a response surface
  perspective.
\newblock \emph{Pharmacological reviews}, 47\penalty0 (2):\penalty0 331--385,
  1995.

\bibitem[Griffin and Stephens(2013)]{Griffin_Stephens_2013}
J.~E. Griffin and D.~A. Stephens.
\newblock Advances in markov chain monte carlo.
\newblock pages 104--144. 2013.

\bibitem[Hautaniemi et~al.(2018)Hautaniemi, Koz{\l}owska,
  et~al.]{Hautaniemi_etal_2018}
S.~Hautaniemi, E.~Koz{\l}owska, et~al.
\newblock Mathematical modeling predicts response to chemotherapy and drug
  combinations in ovarian cancer.
\newblock \emph{Cancer Research}, pages canres--3746, 2018.

\bibitem[Hennessey et~al.(2010)Hennessey, Rosner, Bast~Jr, and
  Chen]{Hennessey_etal_2010}
V.~G. Hennessey, G.~L. Rosner, R.~C. Bast~Jr, and M.~Y. Chen.
\newblock A bayesian approach to dose--response assessment and synergy and its
  application to in vitro dose--response studies.
\newblock \emph{Biometrics}, 66\penalty0 (4):\penalty0 1275--1283, 2010.

\bibitem[Hill(1910)]{Hill_1910}
A.~V. Hill.
\newblock The possible effects of the aggregation of the molecules of
  haemoglobin on its dissociation curves.
\newblock \emph{J Physiol (Lond)}, 40:\penalty0 4--7, 1910.

\bibitem[Johnstone et~al.(2016)Johnstone, Bardenet, Gavaghan, and
  Mirams]{Johnstone_etal_2016}
R.~H. Johnstone, R.~Bardenet, D.~J. Gavaghan, and G.~R. Mirams.
\newblock Hierarchical bayesian inference for ion channel screening
  dose-response data.
\newblock \emph{Wellcome open research}, 1, 2016.

\bibitem[Kashif et~al.(2017)Kashif, Andersson, Larsson, Nygren, and
  Gustafsson]{Kashif_etal_2017}
M.~Kashif, S.~Andersson, C.a nd~Mansoori, R.~Larsson, P.~Nygren, and M.~G.
  Gustafsson.
\newblock Bliss and loewe interaction analyses of clinically relevant drug
  combinations in human colon cancer cell lines reveal complex patterns of
  synergy and antagonism.
\newblock \emph{Oncotarget}, 8\penalty0 (61), 2017.

\bibitem[Kim et~al.(2016)Kim, Kim, and Song]{Kim_2016}
S.~Kim, B.~Kim, and Y.~S. Song.
\newblock Ascites modulates cancer cell behavior, contributing to tumor
  heterogeneity in ovarian cancer.
\newblock \emph{Cancer science}, 107\penalty0 (9):\penalty0 1173--1178, 2016.

\bibitem[Lee and Kong(2009)]{Lee_Kong_2009}
J.~J. Lee and M.~Kong.
\newblock Confidence intervals of interaction index for assessing multiple drug
  interaction.
\newblock \emph{Statistics in biopharmaceutical research}, 1\penalty0
  (1):\penalty0 4--17, 2009.

\bibitem[Li et~al.(2007)Li, Yu, Chin, Lucksiri, Flockhart, and
  Hall]{Li_etal_2007}
L.~Li, M.~Yu, R.~Chin, A.~Lucksiri, D.~A. Flockhart, and S.~D. Hall.
\newblock Drug–drug interaction prediction: a bayesian meta‐analysis
  approach.
\newblock \emph{Statistics in medicine}, 26\penalty0 (20):\penalty0 3700--3721,
  2007.

\bibitem[Loewe(1953)]{Loewe_1953}
S.~Loewe.
\newblock \emph{The problem of synergism and antagonism of combined drugs}.
\newblock \emph{Arzneimittelforschung}, 3:\penalty0 285--290, 1953.

\bibitem[Loewe and Muischnek(1926)]{Loewe_Muischnek_1926}
S.~Loewe and H.~Muischnek.
\newblock \emph{\:{u}ber kombinationswirkungen}.
\newblock \emph{Naunyn-Schmiedebergs Archiv f\:{u}r experimentelle Pathologie
  und Pharmakologie}, 114:\penalty0 313--326, 1926.

\bibitem[O'Neil et~al.(2016)]{ONeil_etal_2016}
J.~O'Neil et~al.
\newblock \emph{An unbiased oncology compound screen to identify novel
  combination strategies}.
\newblock \emph{Molecular cancer therapeutics}, 15\penalty0 (6):\penalty0
  1155--1162, 2016.

\bibitem[Reid et~al.(2017)Reid, Permuth, and Sellers]{Reid_2017}
B.~M. Reid, J.~B. Permuth, and T.~A. Sellers.
\newblock Epidemiology of ovarian cancer: a review.
\newblock \emph{Cancer biology \& medicine}, 14\penalty0 (1):\penalty0 9, 2017.

\bibitem[Tallarida(1992)]{Tallarida_1992}
R.~J. Tallarida.
\newblock Statistical analysis of drug combinations for synergism.
\newblock \emph{Pain}, 49\penalty0 (1):\penalty0 93--97, 1992.

\bibitem[Tallarida et~al.(1989)Tallarida, Porreca, and
  Cowan]{Tallarida_Porreca_Cowan_1989}
R.~J. Tallarida, F.~Porreca, and A.~Cowan.
\newblock Statistical analysis of drug-drug and site-site interactions with
  isobolograms.
\newblock \emph{Life sciences}, 45\penalty0 (11):\penalty0 947--961, 1989.

\bibitem[Tang et~al.(2015)Tang, Wennerberg, and Aittokallio]{Tang_etal_2015}
J.~Tang, K.~Wennerberg, and T.~Aittokallio.
\newblock \emph{What is synergy? The Saariselk\"{a} agreement revisited}.
\newblock \emph{Frontiers in pharmacology}, 6, 2015.

\bibitem[Webb(1963)]{Webb_1963}
J.~L. Webb.
\newblock Effect of more than one inhibitor.
\newblock \emph{Enzyme and metabolic inhibitors}, 1:\penalty0 66--79, 1963.

\bibitem[Wheeler(2017)]{Wheeler_2017}
M.~W. Wheeler.
\newblock Bayesian additive adaptive basis tensor product models for modeling
  high dimensional surfaces: An application to high-throughput toxicity
  testing.
\newblock \emph{arXiv preprint arXiv:1702.04775}, 2017.

\bibitem[Whitehead et~al.(2013)Whitehead, Su, Thygesen, Sperrin, and
  Harbron]{Whitehead_etal_2013}
A.~Whitehead, T.~L. Su, H.~Thygesen, M.~Sperrin, and C.~Harbron.
\newblock \emph{Investigation of the robustness of two models for assessing
  synergy in pre‐clinical drug combination studies}.
\newblock \emph{Pharmaceutical statistics}, 12\penalty0 (5):\penalty0 300--308,
  2013.

\bibitem[Yadav et~al.(2015)Yadav, Wennerberg, Aittokallio, and
  Tang]{Yadav_etal_2015}
B.~Yadav, K.~Wennerberg, T.~Aittokallio, and J.~Tang.
\newblock \emph{Searching for Drug Synergy in Complex Dose-Response Landscapes
  Using an Interaction Potency Model}.
\newblock \emph{Computational and Structural Biotechnology Journal},
  13:\penalty0 504--513, 2015.

\bibitem[Yadav et~al.(2014)]{Yadav_etal_2014}
B.~Yadav et~al.
\newblock \emph{Quantitative scoring of differential drug sensitivity for
  individually optimized anticancer therapies}.
\newblock \emph{Scientific reports}, 4\penalty0 (5193), 2014.

\end{thebibliography}


\begin{thebibliography}{1}
\providecommand{\natexlab}[1]{#1}
\providecommand{\url}[1]{\texttt{#1}}
\expandafter\ifx\csname urlstyle\endcsname\relax
  \providecommand{\doi}[1]{doi: #1}\else
  \providecommand{\doi}{doi: \begingroup \urlstyle{rm}\Url}\fi

\bibitem[Griffin and Stephens(2013)]{Griffin_Stephens_2013}
J.~E. Griffin and D.~A. Stephens.
\newblock Advances in markov chain monte carlo.
\newblock pages 104--144. 2013.

\end{thebibliography}

\end{document}


\title{\vspace{-2cm}
 Supplementary Material for: \\ \bf A Bayesian approach for the to study synergistic interaction effects in \\ in-vitro drug combination experiments}
\date{}

\author[1,2,5]{Andrea Cremaschi}
\author[1,3]{Arnoldo Frigessi}
\author[2,4]{Kjetil Task\'{e}n}
\author[1]{Manuela Zucknick}

\affil[1]{\footnotesize Oslo Centre for Biostatistics and Epidemiology (\textbf{OCBE}), University of Oslo, Oslo, Norway}
\affil[2]{\footnotesize Department of Cancer Immunology, Institute of Cancer Research, Oslo University Hospital, Oslo, Norway}
\affil[3]{\footnotesize Oslo Centre for Biostatistics and Epidemiology (\textbf{OCBE}), Oslo University Hospital, Oslo, Norway}
\affil[4]{\footnotesize Institute of Clinical Medicine, University of Oslo, Oslo, Norway}
\affil[5]{\footnotesize Yale-NUS College, Singapore.}

\maketitle

\clearpage

\section{Applications}

\subsection{Simulation Study: additional results}\label{sec:Suppl_Results_Simul}

We report in this section some additional results concerning the wide simulation study presented in Section 3.2 of the main text. We report the tabular version of Figure 1 in Supplementary Table \ref{tab:MSE_INTER_Suppl}, the MSE values for the interaction surface estimated with standard methodologies in Supplementary Table \ref{tab:MSE_INTER_STD_Suppl}, the LPML values for all the simulation scenarios in Supplementary Table \ref{tab:LPML_Suppl}, and MSE for the mean surface for all the simulation scenarios in Supplementary Table \ref{tab:MSE_Suppl}.

From the point of view of the choice of the prior distribution for the variance parameters, we observe lowest LPML values and highest errors in the case of fixed variances (not reported here), and in the case of inverse-gamma distribution with parameters $(\alpha,\beta) = (3,2)$, consistently throughout the simulations. As expected, we can observe an increment in the performance when the number of replicates $n_{rep}$ increases. Interestingly, the choice of the interaction term used in the simulations did not intensively affect the results, supporting the decision of using a flexible object such as the tensor-product spline in the model. However, higher sensitivity is shown when tuning the prior distribution for the volatilities $\sigma^2_{\phi}$, $\phi \in \{m_1, m_2, \gamma_0, \gamma_1, \gamma_2\}$, including the one of the error terms, $\sigma^2_{\epsilon}$. More comments are reported in the main paper, in relation to Figure 1.

\begin{sidewaystable}
\centering
\caption[]{Mean square error ($10^{-3}$) for the estimation of the interaction term $\Delta$. \\ The notation $\text{IG}(\alpha)$ indicates an inv-gamma$(\alpha, \alpha)$ distribution. \\ The highlighted scenario corresponds to the simulation we shown in the main paper, see Figure 2}\label{tab:MSE_INTER_Suppl}
\begin{tabular}{cccccccccccc}
    
\hline
$\textbf{\text{MSE}}^{(1)}$ & $n_{rep}$ & $ \text{HC}(0.1)$ & $ \text{HC}(0.5)$ & $ \text{HC}(1)$ & $ \text{HC}(5)$ & $ \text{HC}(10)$ & $ \text{IG}(3,2)$ & $ \text{IG}(0.1)$ & $ \text{IG}(0.05)$ & $ \text{IG}(0.025)$ & $ \text{IG}(0.01)$ \\ \hline

\multirow{4}{*}{$N(p_{ij}, \sigma^2_{\epsilon})$} & 1 & 0.683 & 0.700 & 0.716 & 0.728 & 0.7060 & 33.216 & 1.157 & 0.910 & 0.793 & 0.726 \\ \cline{2-12}
 & 3 & 0.323 & 0.323 & 0.321 & 0.328 & 0.321 & 1.455 & 0.329 & 0.320 & 0.321 & 0.326 \\ \cline{2-12}
 & 5 & 0.187 & 0.203 & 0.195 & 0.199 & 0.201 & 0.473 & 0.202 & 0.203 & 0.199 & 0.196 \\ \cline{2-12}
 & 10 & 0.212 & 0.205 & 0.206 & 0.208 & 0.209 & 0.267 & 0.216 & 0.210 & 0.212 & 0.210 \\ \hline 

\multirow{4}{*}{$t_5(p_{ij}, \sigma^2_{\epsilon})$} & 1 & 1.101 & 1.229 & 1.281 & 1.347 & 1.381 & 32.725 & 1.680 & 1.457 & 1.366 & 1.293 \\ \cline{2-12}
 & 3 & 0.703 & 0.719 & 0.721 & 0.735 & 0.729 & 2.271 & 0.784 & 0.753 & 0.741 & 0.731 \\ \cline{2-12}
 & 5 & 0.414 & 0.424 & 0.425 & 0.415 & 0.419 & 0.764 & 0.431 & 0.427 & 0.422 & 0.413 \\ \cline{2-12}
 & 10 & 0.270 & 0.274 & 0.279 & 0.281 & 0.284 & 0.338 & 0.290 & 0.292 & 0.289 & 0.287 \\ \hline \hline

\hline
$\textbf{\text{MSE}}^{(2)}$ & $n_{rep}$ & $ \text{HC}(0.1)$ & $ \text{HC}(0.5)$ & $ \text{HC}(1)$ & $ \text{HC}(5)$ & $ \text{HC}(10)$ & $ \text{IG}(3,2)$ & $ \text{IG}(0.1)$ & $ \text{IG}(0.05)$ & $ \text{IG}(0.025)$ & $ \text{IG}(0.01)$ \\ \hline
 
\multirow{4}{*}{$N(p_{ij}, \sigma^2_{\epsilon})$} & 1 & 0.474 & 0.481 & 0.495 & 0.502 & 0.514 & 29.155 & 0.591 & 0.536 & 0.514 & 0.492 \\ \cline{2-12}
 & 3 & 0.254 & 0.252 & 0.251 & 0.248 & 0.256 & 0.805 & 0.259 & 0.251 & 0.253 & 0.249 \\ \cline{2-12}
 & 5 & 0.175 & 0.181 & 0.181 & 0.183 & 0.184 & 0.331 & 0.191 & 0.185 & 0.185 & 0.185 \\ \cline{2-12}
 & 10 & 0.202 & 0.205 & 0.201 & 0.203 & 0.205 & 0.249 & 0.207 & 0.204 & 0.204 & 0.204 \\ \hline 
 
\multirow{4}{*}{$t_5(p_{ij}, \sigma^2_{\epsilon})$} & 1 & 0.645 & 0.803 & 0.865 & 0.944 & 0.969 & 26.661 & 1.033 & 0.965 & 0.909 & 0.859 \\ \cline{2-12}
 & 3 & 0.485 & 0.486 & 0.485 & 0.478 & 0.475 & 1.084 & 0.485 & 0.486 & 0.481 & 0.480 \\ \cline{2-12}
 & 5 & 0.320 & 0.323 & 0.322 & 0.310 & 0.320 & 0.401 & 0.317 & 0.318 & 0.312 & 0.314 \\ \cline{2-12}
 & 10 & 0.233 & 0.245 & 0.244 & 0.248 & 0.251 & 0.279 & 0.251 & 0.254 & 0.251 & 0.251 \\ \hline \hline

\hline
$\textbf{\text{MSE}}^{(3)}$ & $n_{rep}$ & $ \text{HC}(0.1)$ & $ \text{HC}(0.5)$ & $ \text{HC}(1)$ & $ \text{HC}(5)$ & $ \text{HC}(10)$ & $ \text{IG}(3,2)$ & $ \text{IG}(0.1)$ & $ \text{IG}(0.05)$ & $ \text{IG}(0.025)$ & $ \text{IG}(0.01)$ \\ \hline

\multirow{4}{*}{$N(p_{ij}, \sigma^2_{\epsilon})$} & 1 & 0.734 & 0.748 & 0.758 & 0.775 & 0.800 & 39.697 & 1.049 & 0.893 & 0.829 & 0.782 \\ \cline{2-12}
 & 3 & 0.384 & 0.371 & {\color{red}0.385} & 0.389 & 0.391 & 0.976 & 0.404 & 0.398 & 0.379 & 0.399 \\ \cline{2-12}
 & 5 & 0.222 & 0.240 & 0.232 & 0.231 & 0.238 & 0.376 & 0.245 & 0.239 & 0.240 & 0.242 \\ \cline{2-12}
 & 10 & 0.246 & 0.231 & 0.235 & 0.245 & 0.234 & 0.283 & 0.237 & 0.240 & 0.240 & 0.244 \\ \hline 

\multirow{4}{*}{$t_5(p_{ij}, \sigma^2_{\epsilon})$} & 1 & 1.153 & 1.329 & 1.377 & 1.458 & 1.482 & 36.862 & 1.608 & 1.496 & 1.452 & 1.375 \\ \cline{2-12}
 & 3 & 0.669 & 0.670 & 0.672 & 0.673 & 0.665 & 1.516 & 0.679 & 0.682 & 0.675 & 0.686 \\ \cline{2-12}
 & 5 & 0.413 & 0.409 & 0.404 & 0.404 & 0.404 & 0.567 & 0.426 & 0.420 & 0.414 & 0.407 \\ \cline{2-12}
 & 10 & 0.283 & 0.301 & 0.305 & 0.302 & 0.304 & 0.354 & 0.321 & 0.316 & 0.306 & 0.309
 \\ \hline
\end{tabular}    
\end{sidewaystable}

\begin{table}
\begin{adjustwidth}{-1cm}{}
\centering
{\renewcommand{\arraystretch}{1.2}

    \begin{tabular}{ccccccc}
    \hline
    
$\textbf{\text{MSE}}^{(1)}_{\Delta^{(1)}} (10^{-3})$ & $n_{rep}$ & Bliss & HSA & Loewe & ZIP \\ \hline
\multirow{4}{*}{$N(p_{ij}, \sigma^2_{\epsilon})$} & 1 & 11.387 & 6.717 & 5.1115 & 13.912 \\ \cline{2-6}
 & 3 & 9.347 & 4.786 & 3.38 & 4.562 \\ \cline{2-6}
 & 5 & 6.83 & 4.19 & 2.582 & 5.143 \\ \cline{2-6}
 & 10 & 6.235 & 4.065 & 2.828 & 3.018 \\ \hline 

\multirow{4}{*}{$t_5(p_{ij}, \sigma^2_{\epsilon})$} & 1 & 16.071 & 9.868 & 6.968 & 41.59 \\ \cline{2-6}
 & 3 & 10.955 & 4.398 & 3.41 & 9.475 \\ \cline{2-6}
 & 5 & 11.785 & 5.046 & 3.51 & 4.137 \\ \cline{2-6}
 & 10 & 6.93 & 3.89 & 2.472 & 4.174 \\ \hline
    \end{tabular}

    \begin{tabular}{ccccccc}
    \hline

$\textbf{\text{MSE}}^{(2)}_{\Delta^{(2)}} (10^{-3})$ & $n_{rep}$ & Bliss & HSA & Loewe & ZIP \\ \hline
\multirow{4}{*}{$N(p_{ij}, \sigma^2_{\epsilon})$} & 1 & 11.416 & 6.572 & 4.9863  & 15.55 \\ \cline{2-6}
 & 3 & 9.403 & 4.647 & 3.324 & 8.528 \\ \cline{2-6}
 & 5 & 6.837 & 3.994 & 2.544 & 9.018 \\ \cline{2-6}
 & 10 & 6.253 & 3.887 & 2.823 & 7.39 \\ \hline 

\multirow{4}{*}{$t_5(p_{ij}, \sigma^2_{\epsilon})$} & 1 & 16.065 & 9.709 & 6.822 & 43.78 \\ \cline{2-6}
 & 3 & 10.97 & 4.21 & 3.226 & 11.653 \\ \cline{2-6}
 & 5 & 11.89 & 4.984 & 3.649 & 5.86 \\ \cline{2-6}
 & 10 & 6.974 & 3.76 & 2.515 & 5.143 \\ \hline
    \end{tabular}
 
    \begin{tabular}{ccccccc}
    \hline

$\textbf{\text{MSE}}^{(3)}_{\Delta^{(3)}} (10^{-3})$ & $n_{rep}$ & Bliss & HSA & Loewe & ZIP \\ \hline
\multirow{4}{*}{$N(p_{ij}, \sigma^2_{\epsilon})$} & 1 & 14.568 & 7.002 & 5.472 & 31.295 \\ \cline{2-6}
 & 3 & 12.928 & 5.253 & 4.27 & 25.02 \\ \cline{2-6}
 & 5 & 10.13 & 4.278 & 3.442 & 23.083  \\ \cline{2-6}
 & 10 & 9.151 & 4.046 & 3.558 & 26.565 \\ \hline 

\multirow{4}{*}{$t_5(p_{ij}, \sigma^2_{\epsilon})$} & 1 & 18.869 & 10.069 & 7.186 & 70.809 \\ \cline{2-6}
 & 3 & 15.68 & 5.196 & 4.317 & 33.763 \\ \cline{2-6}
 & 5 & 16.867 & 6.317 & 5.591 & 25.662 \\ \cline{2-6}
 & 10 & 10.322 & 4.061 & 3.403 & 24.332 \\ \hline
    \end{tabular}
    
            }
    \caption{Mean square errors errors with respect to the interaction terms for standard methods.}
  \label{tab:MSE_INTER_STD_Suppl}
 \end{adjustwidth}
\end{table}

\begin{sidewaystable}
\centering
\caption[]{LPML($10^3$) for the different simulation scenarios. \\ The notation $\text{IG}(\alpha)$ indicates an inv-gamma$(\alpha, \alpha)$ distribution.}\label{tab:LPML_Suppl}
\begin{tabular}{cccccccccccc}
    
\hline
$\textbf{\text{LPML}}^{(1)}$ & $n_{rep}$ & $ \text{HC}(0.1)$ & $ \text{HC}(0.5)$ & $ \text{HC}(1)$ & $ \text{HC}(5)$ & $ \text{HC}(10)$ & $ \text{IG}(3,2)$ & $ \text{IG}(0.1)$ & $ \text{IG}(0.05)$ & $ \text{IG}(0.025)$ & $ \text{IG}(0.01)$ \\ \hline
  
\multirow{4}{*}{$N(p_{ij}, \sigma^2_{\epsilon})$} & 1 & 0.128 & 0.128 & 0.129 & 0.128 & 0.128 & 0.005 & 0.110 & 0.120 & 0.125 & 0.128 \\ \cline{2-12}
 & 3 & 0.422 & 0.422 & 0.421 & 0.422 & 0.421 & 0.243 & 0.415 & 0.420 & 0.421 & 0.422 \\ \cline{2-12}
 & 5 & 0.684 & 0.683 & 0.682 & 0.683 & 0.682 & 0.502 & 0.680 & 0.682 & 0.683 & 0.683 \\ \cline{2-12}
 & 10 & 1.371 & 1.371 & 1.370 & 1.371 & 1.370 & 1.192 & 1.369 & 1.370 & 1.369 & 1.371 \\ \hline 

\multirow{4}{*}{$t_5(p_{ij}, \sigma^2_{\epsilon})$} & 1 & 0.114 & 0.114 & 0.115 & 0.115 & 0.114 & 0.004 & 0.102 & 0.110 & 0.113 & 0.113 \\ \cline{2-12}
 & 3 & 0.350 & 0.350 & 0.350 & 0.350 & 0.349 & 0.228 & 0.346 & 0.349 & 0.349 & 0.350 \\ \cline{2-12}
 & 5 & 0.569 & 0.568 & 0.568 & 0.569 & 0.569 & 0.453 & 0.566 & 0.567 & 0.568 & 0.568 \\ \cline{2-12}
 & 10 & 1.122 & 1.122 & 1.121 & 1.122 & 1.121 & 1.029 & 1.121 & 1.121 & 1.121 & 1.121 \\ \hline \hline

\hline
$\textbf{\text{LPML}}^{(2)}$ & $n_{rep}$ & $ \text{HC}(0.1)$ & $ \text{HC}(0.5)$ & $ \text{HC}(1)$ & $ \text{HC}(5)$ & $ \text{HC}(10)$ & $ \text{IG}(3,2)$ & $ \text{IG}(0.1)$ & $ \text{IG}(0.05)$ & $ \text{IG}(0.025)$ & $ \text{IG}(0.01)$ \\ \hline

\multirow{4}{*}{$N(p_{ij}, \sigma^2_{\epsilon})$} & 1 & 0.132 & 0.130 & 0.131 & 0.130 & 0.131 & 0.009 & 0.116 & 0.125 & 0.130 & 0.132 \\ \cline{2-12}
 & 3 & 0.423 & 0.423 & 0.422 & 0.422 & 0.422 & 0.253 & 0.416 & 0.420 & 0.422 & 0.423 \\ \cline{2-12}
 & 5 & 0.686 & 0.686 & 0.686 & 0.686 & 0.686 & 0.510 & 0.683 & 0.685 & 0.686 & 0.686 \\ \cline{2-12}
 & 10 & 1.374 & 1.373 & 1.373 & 1.373 & 1.372 & 1.196 & 1.371 & 1.372 & 1.372 & 1.373 \\ \hline 

\multirow{4}{*}{$t_5(p_{ij}, \sigma^2_{\epsilon})$} & 1 & 0.114 & 0.114 & 0.113 & 0.114 & 0.114 & 0.008 & 0.105 & 0.111 & 0.113 & 0.114 \\ \cline{2-12}
 & 3 & 0.353 & 0.353 & 0.352 & 0.352 & 0.352 & 0.235 & 0.350 & 0.352 & 0.353 & 0.353 \\ \cline{2-12}
 & 5 & 0.573 & 0.572 & 0.572 & 0.572 & 0.572 & 0.460 & 0.571 & 0.572 & 0.572 & 0.572 \\ \cline{2-12}
 & 10 & 1.126 & 1.126 & 1.126 & 1.126 & 1.126 & 1.034 & 1.125 & 1.125 & 1.126 & 1.125 \\ \hline \hline
    
\hline
$\textbf{\text{LPML}}^{(3)}$ & $n_{rep}$ & $ \text{HC}(0.1)$ & $ \text{HC}(0.5)$ & $ \text{HC}(1)$ & $ \text{HC}(5)$ & $ \text{HC}(10)$ & $ \text{IG}(3,2)$ & $ \text{IG}(0.1)$ & $ \text{IG}(0.05)$ & $ \text{IG}(0.025)$ & $ \text{IG}(0.01)$ \\ \hline

\multirow{4}{*}{$N(p_{ij}, \sigma^2_{\epsilon})$} & 1 & 0.126 & 0.125 & 0.124 & 0.124 & 0.125 & -0.002 & 0.107 & 0.117 & 0.123 & 0.125 \\ \cline{2-12}
 & 3 & 0.422 & 0.421 & 0.421 & 0.421 & 0.422 & 0.245 & 0.416 & 0.420 & 0.421 & 0.421 \\ \cline{2-12}
 & 5 & 0.684 & 0.684 & 0.683 & 0.682 & 0.684 & 0.502 & 0.680 & 0.683 & 0.683 & 0.683 \\ \cline{2-12}
 & 10 & 1.371 & 1.372 & 1.371 & 1.370 & 1.371 & 1.193 & 1.371 & 1.371 & 1.371 & 1.371 \\ \hline 

\multirow{4}{*}{$t_5(p_{ij}, \sigma^2_{\epsilon})$} & 1 & 0.114 & 0.114 & 0.113 & 0.110 & 0.113 & -0.002 & 0.100 & 0.108 & 0.111 & 0.112 \\ \cline{2-12}
 & 3 & 0.347 & 0.347 & 0.347 & 0.347 & 0.346 & 0.226 & 0.344 & 0.346 & 0.347 & 0.347 \\ \cline{2-12}
 & 5 & 0.567 & 0.567 & 0.566 & 0.566 & 0.566 & 0.452 & 0.565 & 0.566 & 0.566 & 0.566 \\ \cline{2-12}
 & 10 & 1.120 & 1.119 & 1.120 & 1.118 & 1.119 & 1.027 & 1.118 & 1.119 & 1.120 & 1.119
 \\ \hline 
\end{tabular}
\end{sidewaystable}

\begin{sidewaystable}
\centering
\caption[]{Mean square error ($10^{-3}$) for the estimation of the response surface $p_{ij}$. \\ The notation $\text{IG}(\alpha)$ indicates an inv-gamma$(\alpha, \alpha)$ distribution.}\label{tab:MSE_Suppl}
\begin{tabular}{cccccccccccc}
    
\hline
$\textbf{\text{MSE}}^{(1)}$ & $n_{rep}$ & $ \text{HC}(0.1)$ & $ \text{HC}(0.5)$ & $ \text{HC}(1)$ & $ \text{HC}(5)$ & $ \text{HC}(10)$ & $ \text{IG}(3,2)$ & $ \text{IG}(0.1)$ & $ \text{IG}(0.05)$ & $ \text{IG}(0.025)$ & $ \text{IG}(0.01)$ \\ \hline

\multirow{4}{*}{$N(p_{ij}, \sigma^2_{\epsilon})$} & 1 & 0.671 & 0.671 & 0.683 & 0.691 & 0.660 & 34.479 & 1.073 & 0.856 & 0.747 & 0.689 \\ \cline{2-12}
 & 3 & 0.320 & 0.322 & 0.318 & 0.329 & 0.319 & 1.201 & 0.324 & 0.319 & 0.321 & 0.323 \\ \cline{2-12}
 & 5 & 0.182 & 0.194 & 0.187 & 0.190 & 0.189 & 0.425 & 0.188 & 0.189 & 0.184 & 0.182 \\ \cline{2-12}
 & 10 & 0.152 & 0.149 & 0.149 & 0.153 & 0.151 & 0.204 & 0.158 & 0.151 & 0.156 & 0.150 \\ \hline 

\multirow{4}{*}{$t_5(p_{ij}, \sigma^2_{\epsilon})$} & 1 & 1.027 & 1.045 & 1.056 & 1.062 & 1.074 & 34.910 & 1.369 & 1.178 & 1.119 & 1.065 \\ \cline{2-12}
 & 3 & 0.650 & 0.667 & 0.669 & 0.687 & 0.679 & 1.956 & 0.734 & 0.705 & 0.691 & 0.679 \\ \cline{2-12}
 & 5 & 0.375 & 0.387 & 0.389 & 0.384 & 0.382 & 0.708 & 0.396 & 0.393 & 0.387 & 0.376 \\ \cline{2-12}
 & 10 & 0.191 & 0.189 & 0.191 & 0.191 & 0.189 & 0.231 & 0.191 & 0.192 & 0.193 & 0.191 \\ \hline \hline

\hline
$\textbf{\text{MSE}}^{(2)}$ & $n_{rep}$ & $ \text{HC}(0.1)$ & $ \text{HC}(0.5)$ & $ \text{HC}(1)$ & $ \text{HC}(5)$ & $ \text{HC}(10)$ & $ \text{IG}(3,2)$ & $ \text{IG}(0.1)$ & $ \text{IG}(0.05)$ & $ \text{IG}(0.025)$ & $ \text{IG}(0.01)$ \\ \hline
 
\multirow{4}{*}{$N(p_{ij}, \sigma^2_{\epsilon})$} & 1 & 0.461 & 0.439 & 0.441 & 0.439 & 0.436 & 32.388 & 0.438 & 0.415 & 0.422 & 0.424 \\ \cline{2-12}
 & 3 & 0.237 & 0.232 & 0.231 & 0.231 & 0.234 & 0.599 & 0.227 & 0.225 & 0.229 & 0.227 \\ \cline{2-12}
 & 5 & 0.162 & 0.165 & 0.163 & 0.166 & 0.163 & 0.297 & 0.170 & 0.162 & 0.165 & 0.163 \\ \cline{2-12}
 & 10 & 0.137 & 0.142 & 0.138 & 0.141 & 0.142 & 0.175 & 0.141 & 0.141 & 0.142 & 0.141 \\ \hline 
 
\multirow{4}{*}{$t_5(p_{ij}, \sigma^2_{\epsilon})$} & 1 & 0.566 & 0.570 & 0.573 & 0.578 & 0.578 & 31.112 & 0.587 & 0.563 & 0.562 & 0.561 \\ \cline{2-12}
 & 3 & 0.459 & 0.462 & 0.460 & 0.459 & 0.456 & 0.953 & 0.464 & 0.467 & 0.460 & 0.456 \\ \cline{2-12}
 & 5 & 0.298 & 0.302 & 0.301 & 0.292 & 0.300 & 0.383 & 0.297 & 0.299 & 0.293 & 0.291 \\ \cline{2-12}
 & 10 & 0.167 & 0.170 & 0.166 & 0.172 & 0.170 & 0.188 & 0.167 & 0.171 & 0.168 & 0.167 \\ \hline \hline

\hline
$\textbf{\text{MSE}}^{(3)}$ & $n_{rep}$ & $ \text{HC}(0.1)$ & $ \text{HC}(0.5)$ & $ \text{HC}(1)$ & $ \text{HC}(5)$ & $ \text{HC}(10)$ & $ \text{IG}(3,2)$ & $ \text{IG}(0.1)$ & $ \text{IG}(0.05)$ & $ \text{IG}(0.025)$ & $ \text{IG}(0.01)$ \\ \hline

\multirow{4}{*}{$N(p_{ij}, \sigma^2_{\epsilon})$} & 1 & 0.710 & 0.692 & 0.699 & 0.701 & 0.712 & 37.478 & 0.909 & 0.776 & 0.734 & 0.708 \\ \cline{2-12}
 & 3 & 0.377 & 0.366 & 0.377 & 0.384 & 0.379 & 0.821 & 0.395 & 0.387 & 0.369 & 0.390 \\ \cline{2-12}
 & 5 & 0.223 & 0.234 & 0.226 & 0.225 & 0.228 & 0.357 & 0.235 & 0.225 & 0.230 & 0.233 \\ \cline{2-12}
 & 10 & 0.189 & 0.173 & 0.177 & 0.185 & 0.176 & 0.215 & 0.179 & 0.180 & 0.182 & 0.184 \\ \hline 

\multirow{4}{*}{$t_5(p_{ij}, \sigma^2_{\epsilon})$} & 1 & 1.071 & 1.092 & 1.088 & 1.105 & 1.100 & 37.522 & 1.212 & 1.128 & 1.117 & 1.085 \\ \cline{2-12}
 & 3 & 0.633 & 0.634 & 0.638 & 0.641 & 0.634 & 1.386 & 0.648 & 0.655 & 0.641 & 0.649 \\ \cline{2-12}
 & 5 & 0.377 & 0.373 & 0.368 & 0.370 & 0.370 & 0.536 & 0.393 & 0.386 & 0.380 & 0.371 \\ \cline{2-12}
 & 10 & 0.204 & 0.209 & 0.216 & 0.211 & 0.207 & 0.247 & 0.221 & 0.215 & 0.207 & 0.210
 \\ \hline
\end{tabular}
\end{sidewaystable}

\clearpage
\subsection{Ovarian cancer cell test data: experiements with ascites}\label{sec:OC_Ascites_Suppl}
We present in this section the results of the analysis of the ovarian cancer dataset for those experiments where both medium and ascites were dispensed. As shown in Section 3.5 of the main text, we report the contour plots of the posterior estimates of the mean response surface $p_ij$ in Supplementary Figure \ref{fig:OC_Ascites_contour_pij}, the zero-interaction surface $p^0_{ij}$ in Supplementary Figure \ref{fig:OC_Ascites_contour_p0}, and the interaction surface $\Delta_{ij}$ in Supplementary Figure \ref{fig:OC_Ascites_contour_Delta}. Each plot is composed of four panels, one for each combination/cell-line tested. The results show generally flat response surfaces, dominated by the contribution of the zero-interactions $p^0_{ij}$. However, we can observe a decrease in the estimated response surfaces for the combination experiments involving cell-line OVCAR8 (top row in Supplementary Figure \ref{fig:OC_Ascites_contour_pij}). In particular, the model is able to estimate the $\text{EC}_{50}$ of the drug WP1066 within the observed range of concentrations. These results are confirmed when looking at the interaction surfaces $\Delta_{ij}$, presenting negative areas for high concentrations of WP1066 or Nilotinib.

\begin{figure}[!ht]
\hspace{-0.5cm}
\subfloat[OVCAR8]{\includegraphics[height=.4\textwidth]{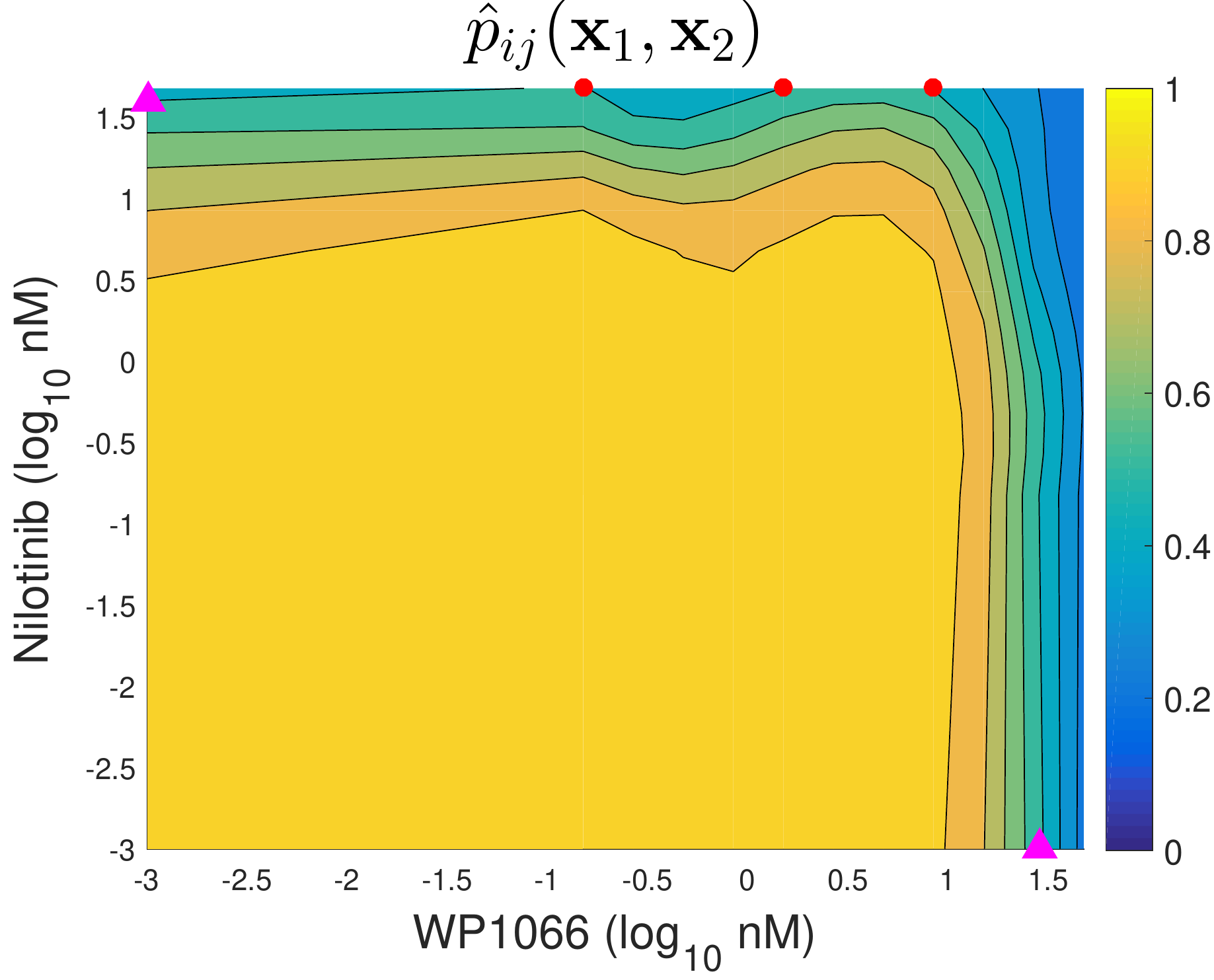}}
\subfloat[OVCAR8]{\includegraphics[height=.4\textwidth]{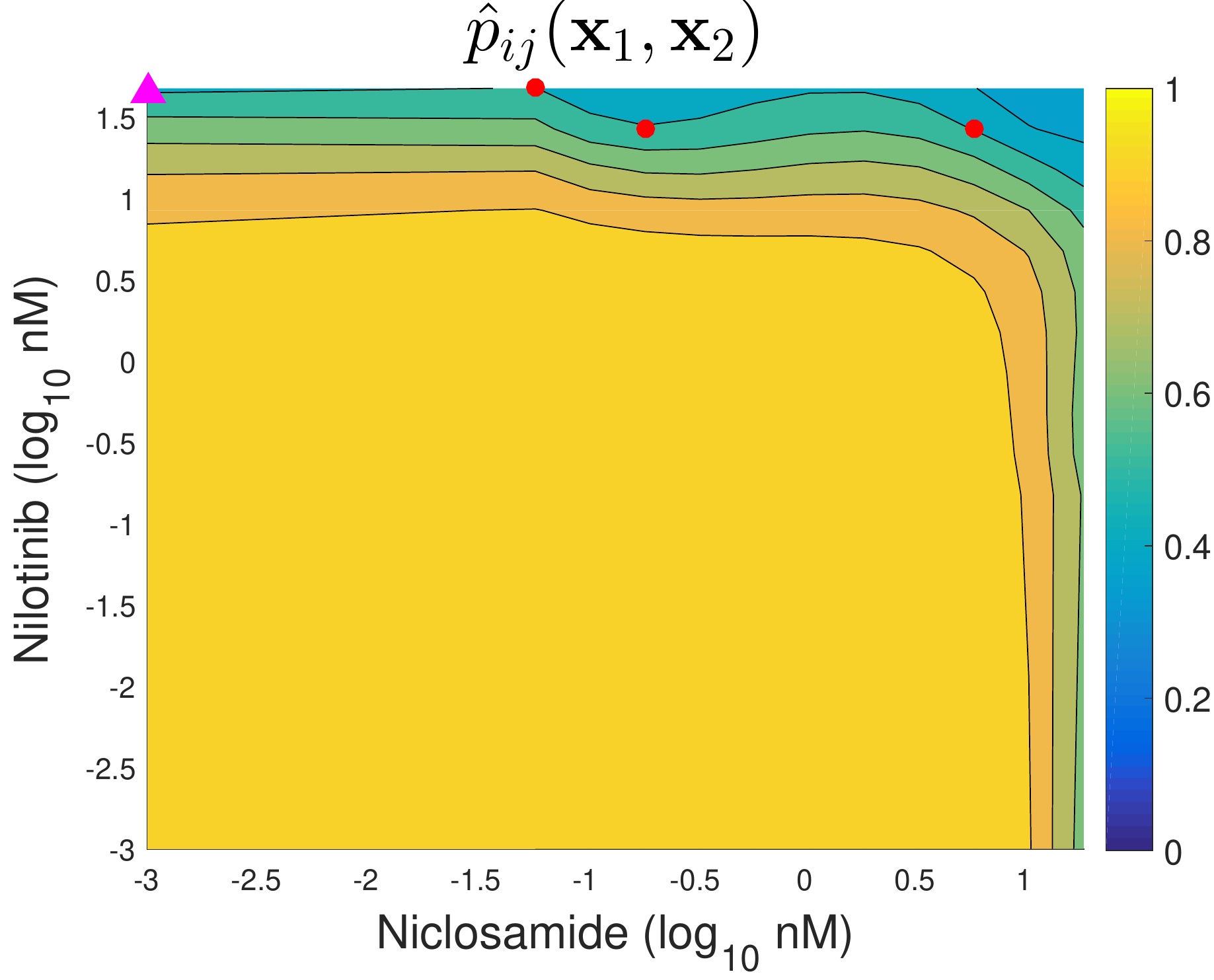}}\\

\hspace{-0.5cm}
\subfloat[SKOV3]{\includegraphics[height=.4\textwidth]{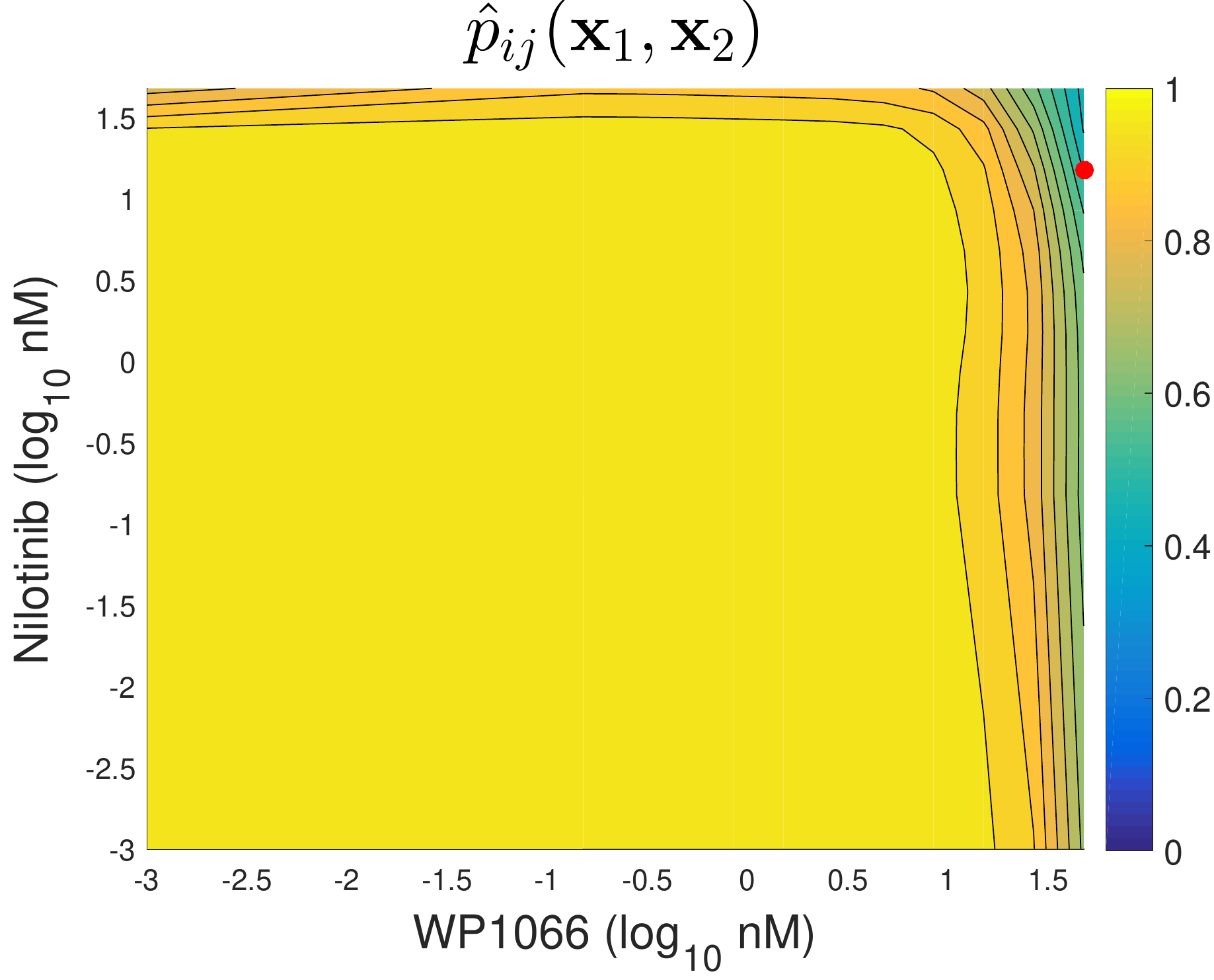}}
\subfloat[SKOV3]{\includegraphics[height=.4\textwidth]{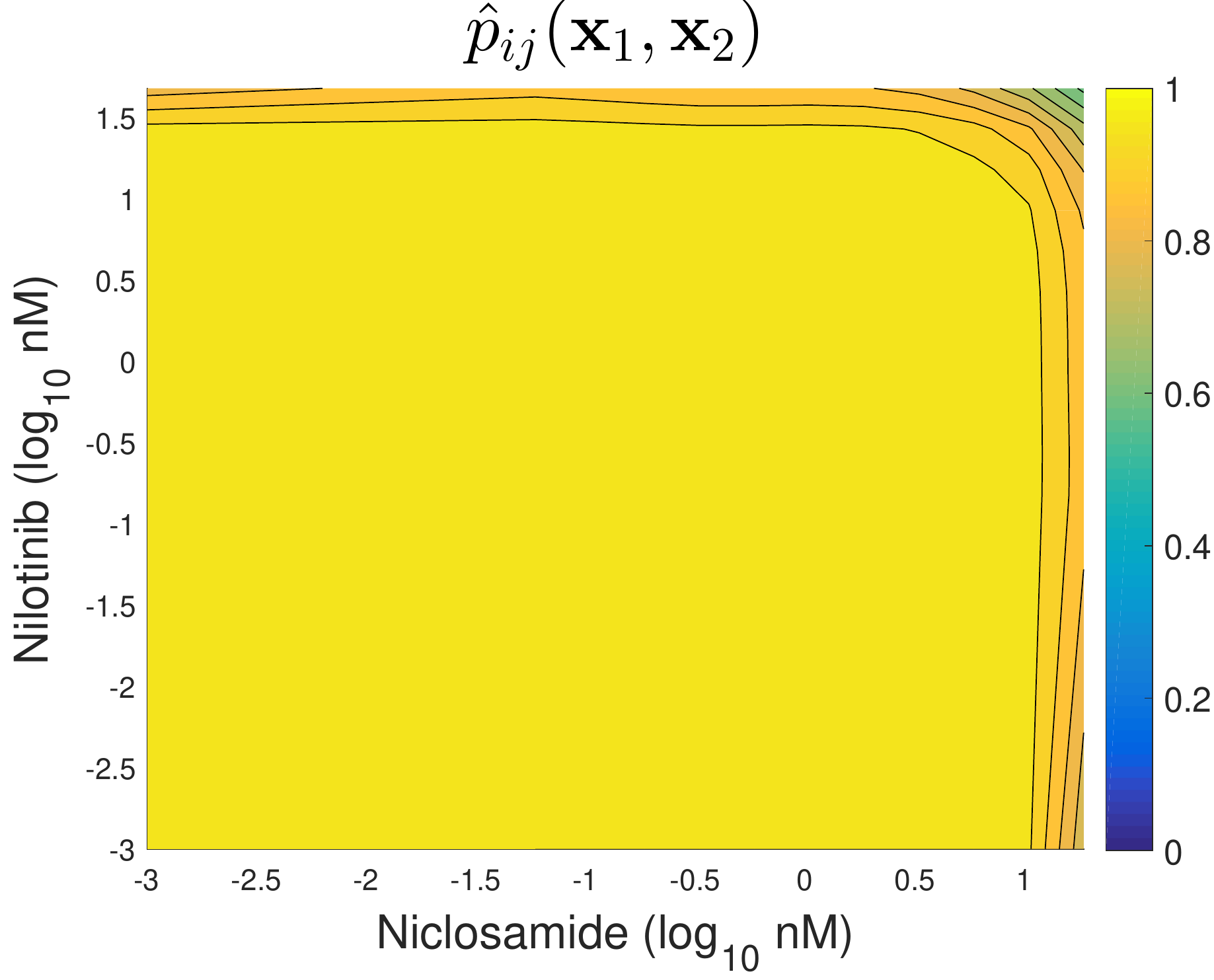}}
\caption{Application to Ovarian Cancer data -- Contour plots of the posterior mean of the surface $p_{ij}$, including the $\text{bi-EC}_{50}(0.01)$ estimates as red dots, and the monotherapy $\text{EC}_{50}$ estimates as magenta triangles. The triangles are not visible when the estimated $\text{EC}_{50}$'s are outside the considered range of concentrations. Experiments with cell-lines treated in medium and ascites. Top row: OVCAR8 ; bottom row: SKOV3.}
  \label{fig:OC_Ascites_contour_pij}
\end{figure}

\begin{figure}[!ht]
\hspace{-0.5cm}
\subfloat[OVCAR8]{\includegraphics[height=.4\textwidth]{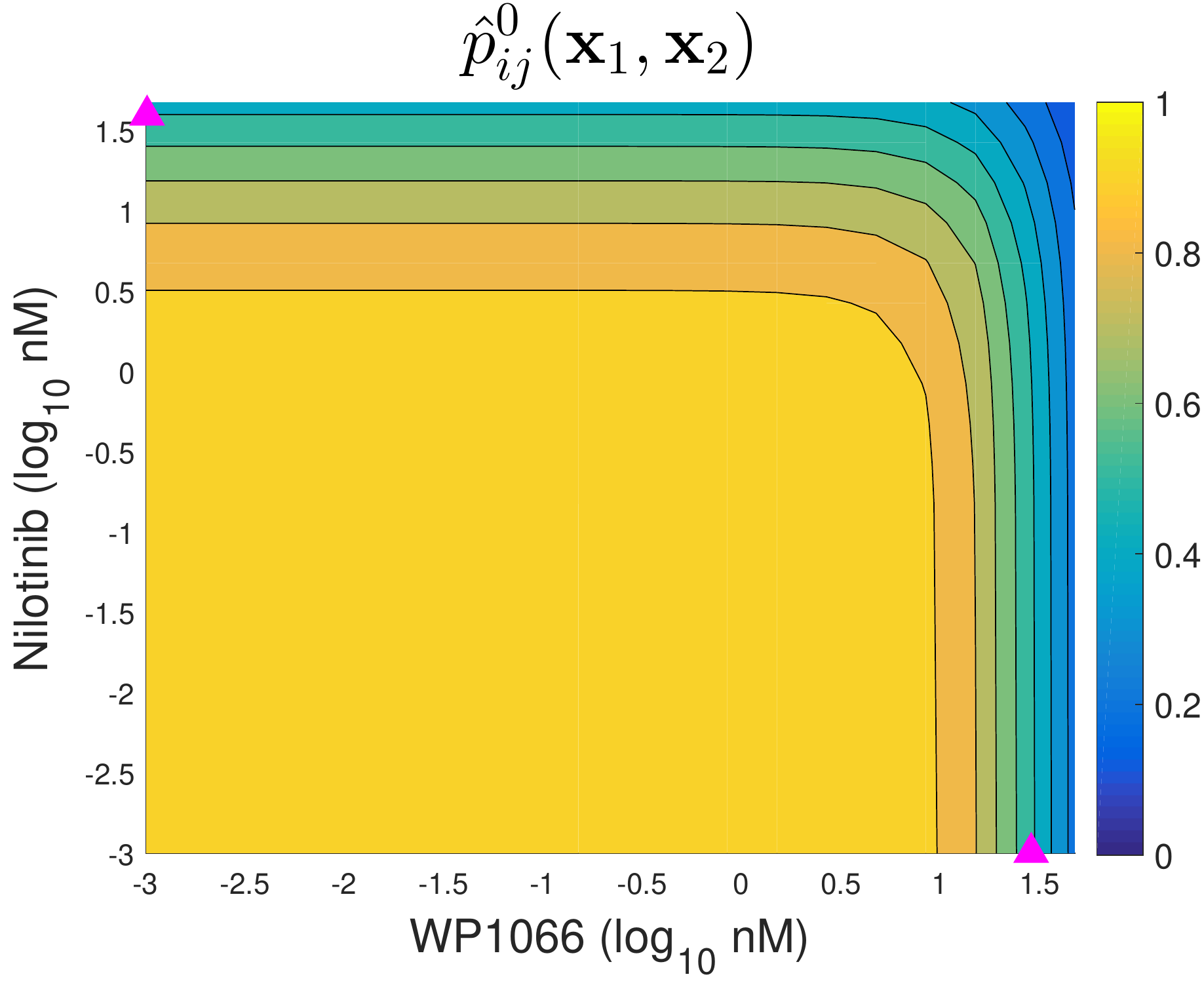}}
\subfloat[OVCAR8]{\includegraphics[height=.4\textwidth]{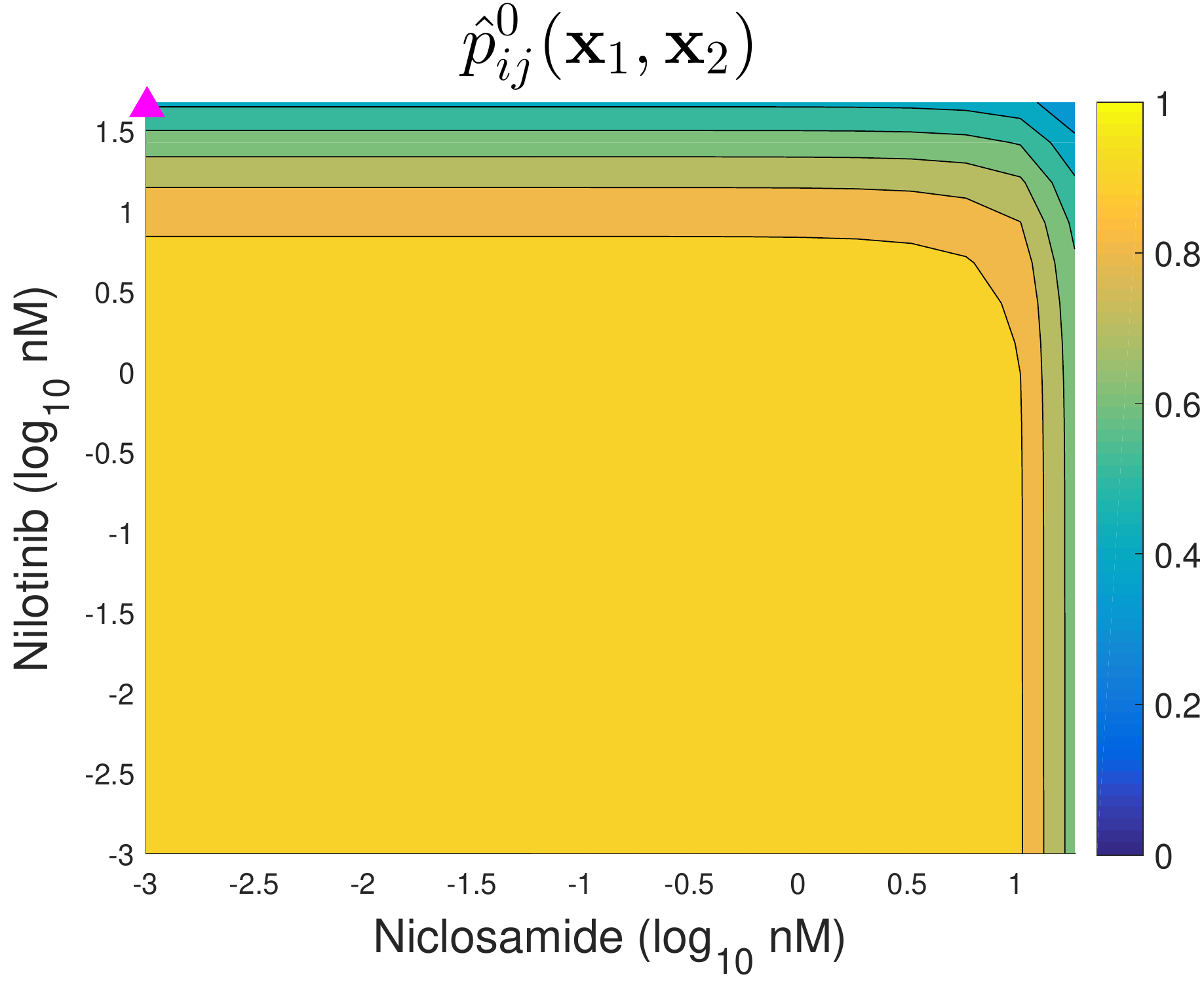}}\\

\hspace{-0.5cm}
\subfloat[SKOV3]{\includegraphics[height=.4\textwidth]{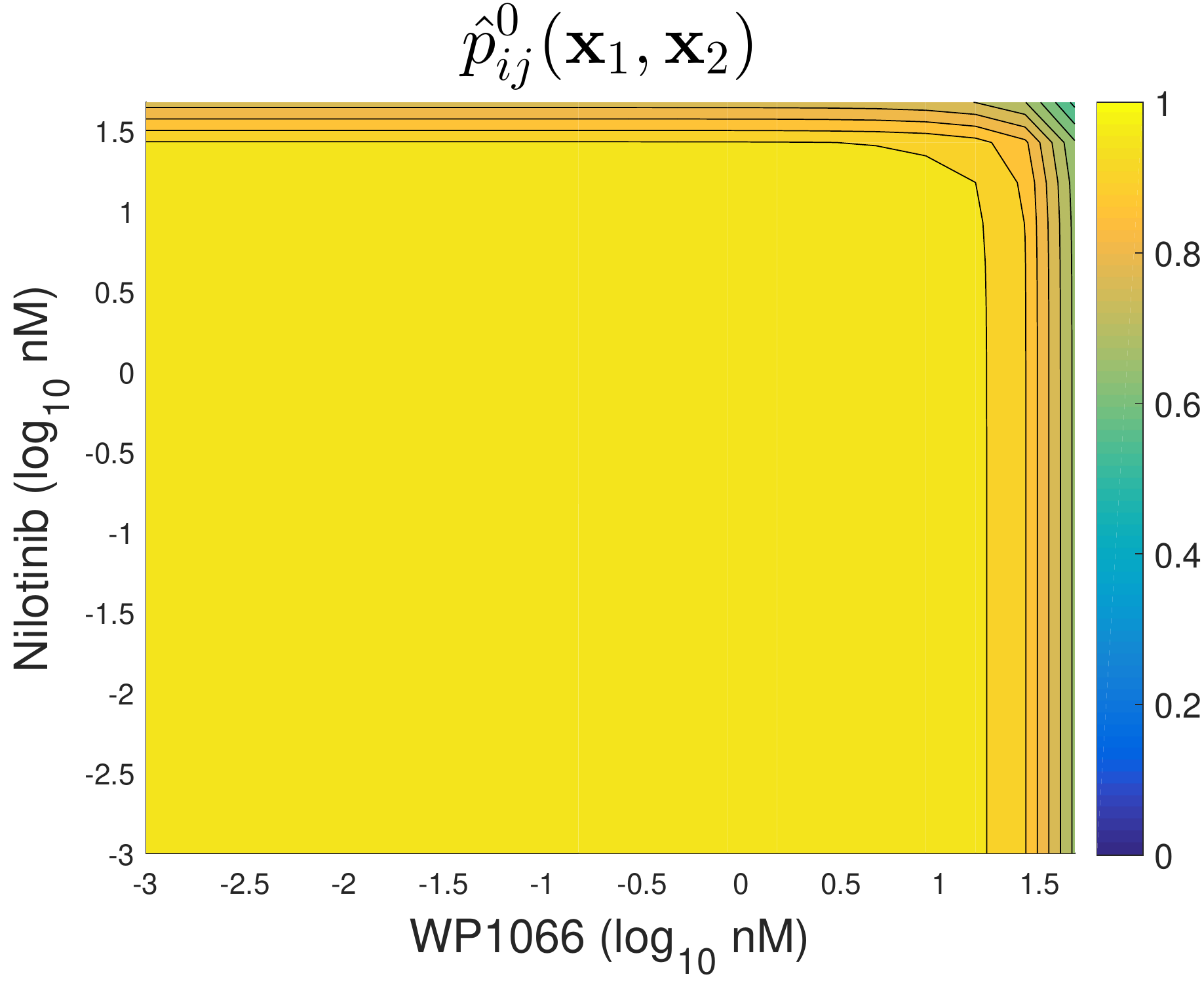}}
\subfloat[SKOV3]{\includegraphics[height=.4\textwidth]{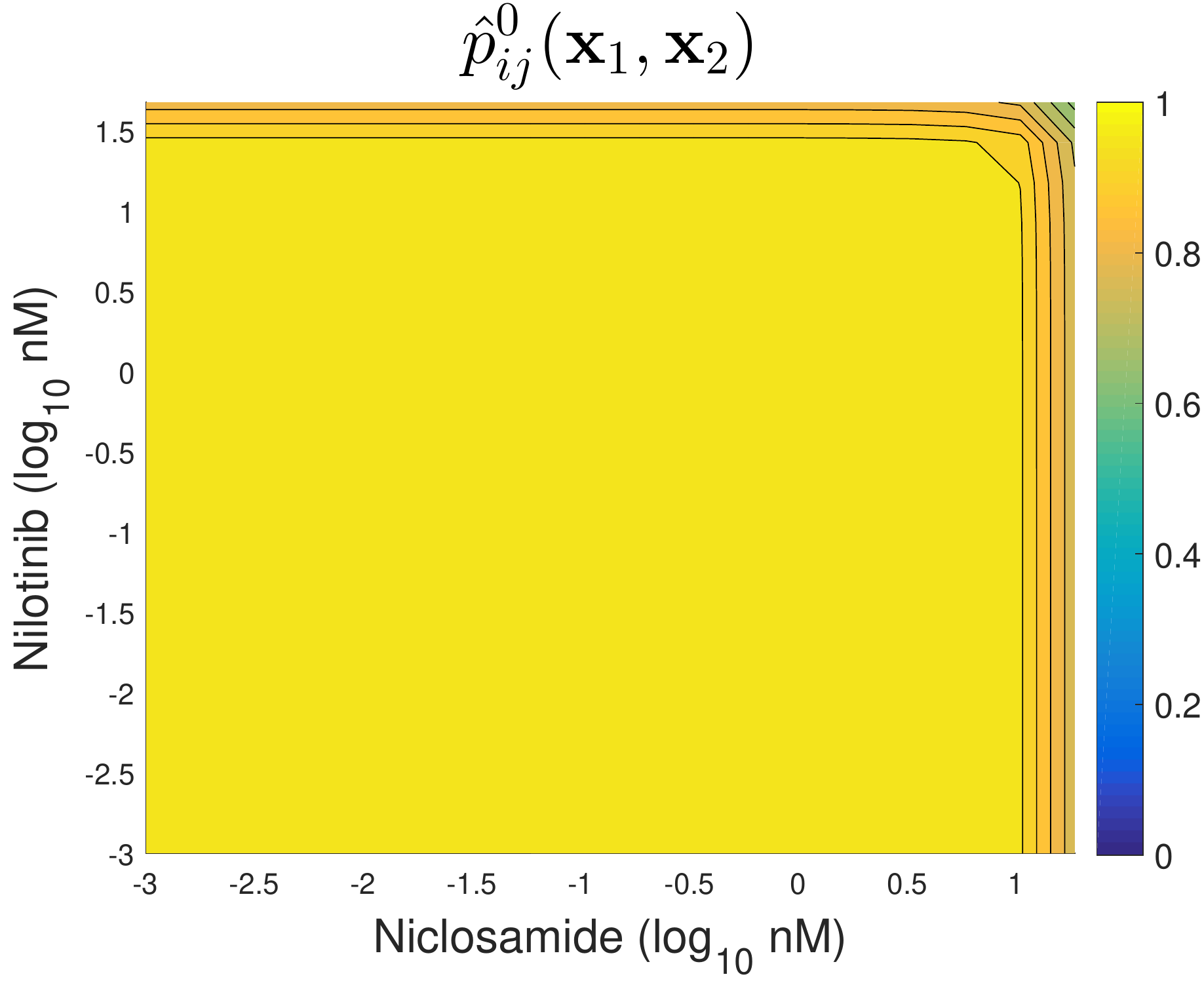}}
\caption{Application to Ovarian Cancer data -- Contour plots of the posterior mean of the zero-interaction surface $p^0_{ij}$, including the monotherapy $\text{EC}_{50}$ estimates as magenta triangles. The triangles are not visible when the estimated $\text{EC}_{50}$'s are outside the considered range of concentrations. Experiments with cell-lines treated in medium and ascites. Top row: OVCAR8 ; bottom row: SKOV3.}
  \label{fig:OC_Ascites_contour_p0}
\end{figure}

\begin{figure}[!ht]
\hspace{-0.5cm}
\subfloat[OVCAR8]{\includegraphics[height=.4\textwidth]{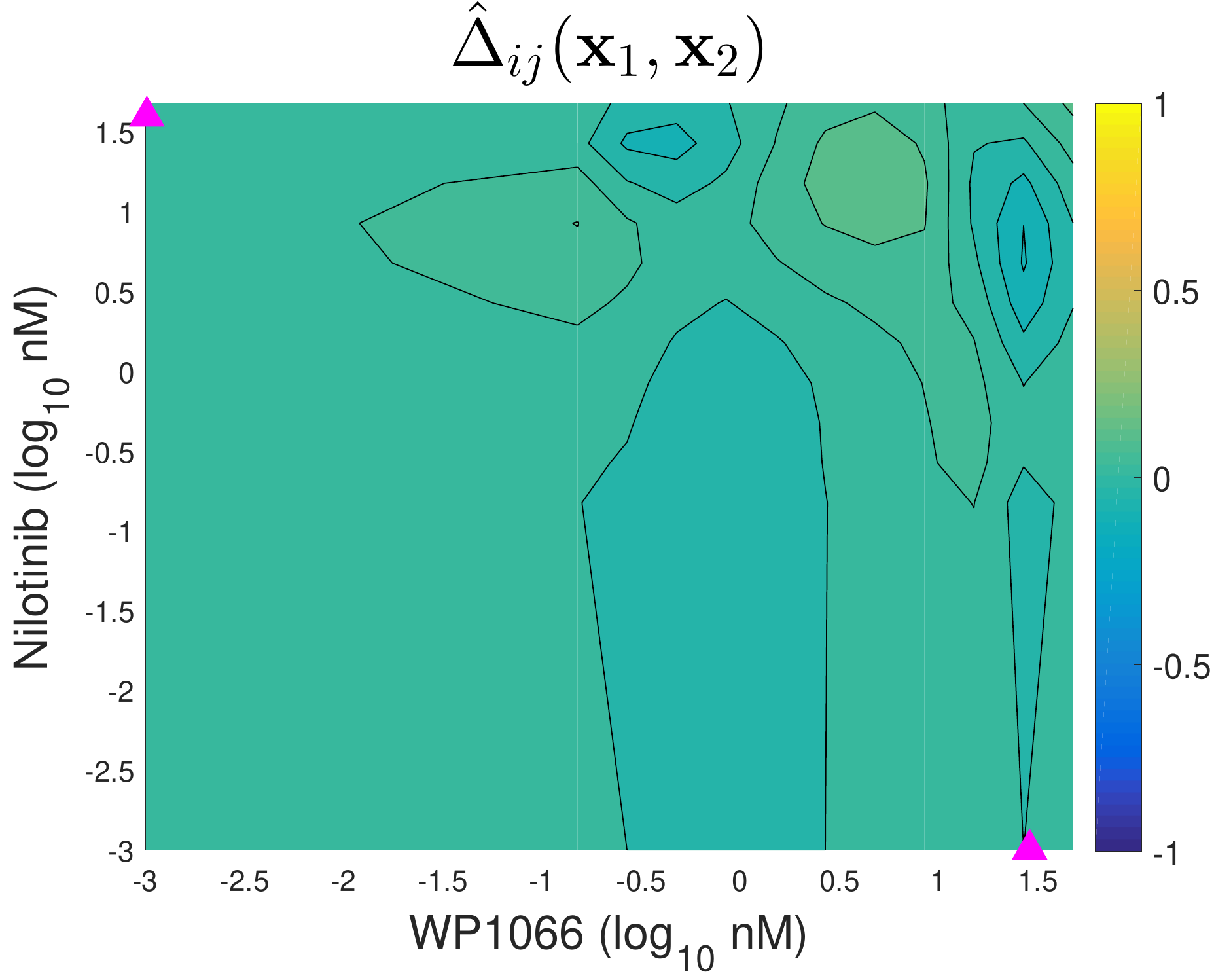}}
\subfloat[OVCAR8]{\includegraphics[height=.4\textwidth]{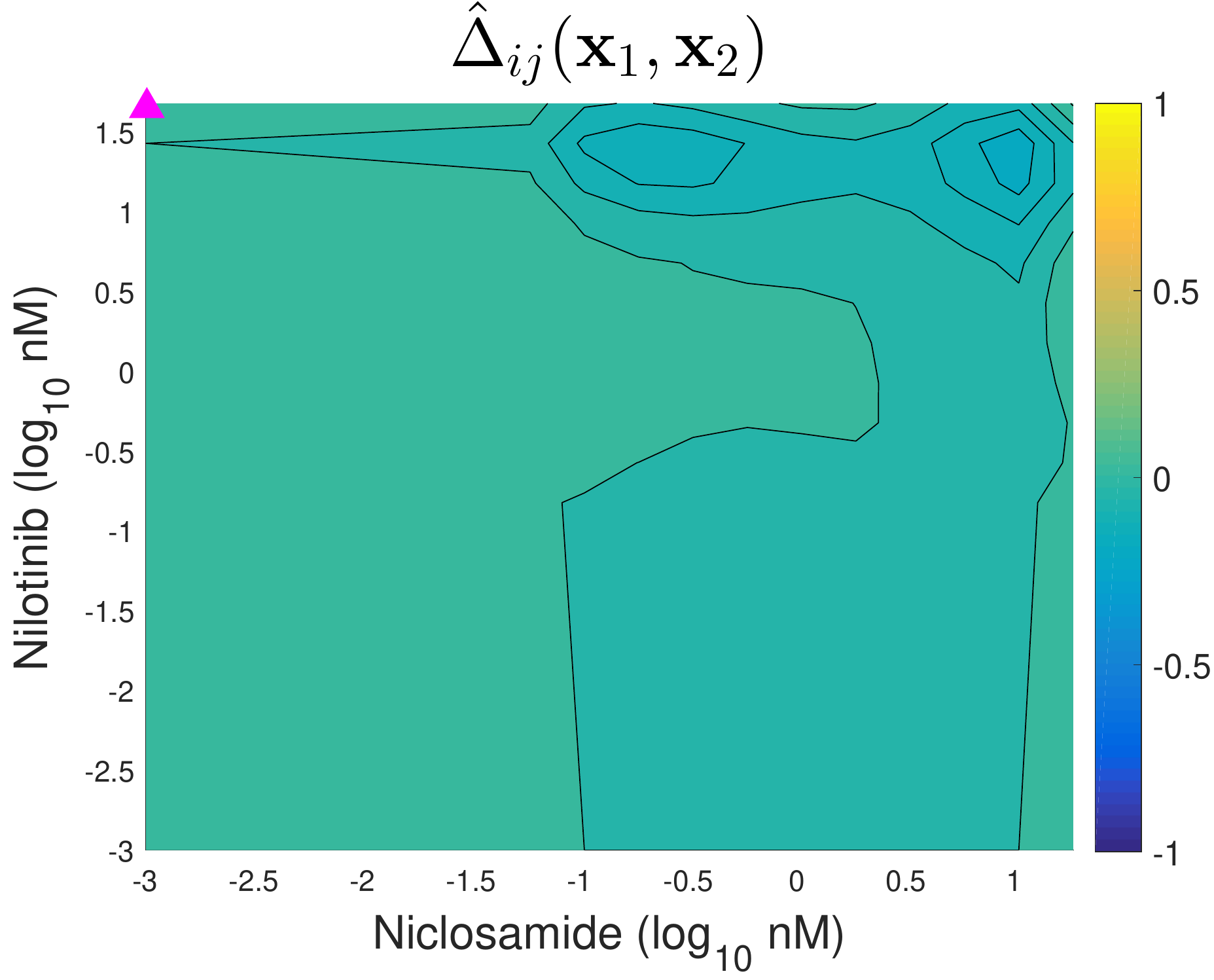}}\\

\hspace{-0.5cm}
\subfloat[SKOV3]{\includegraphics[height=.4\textwidth]{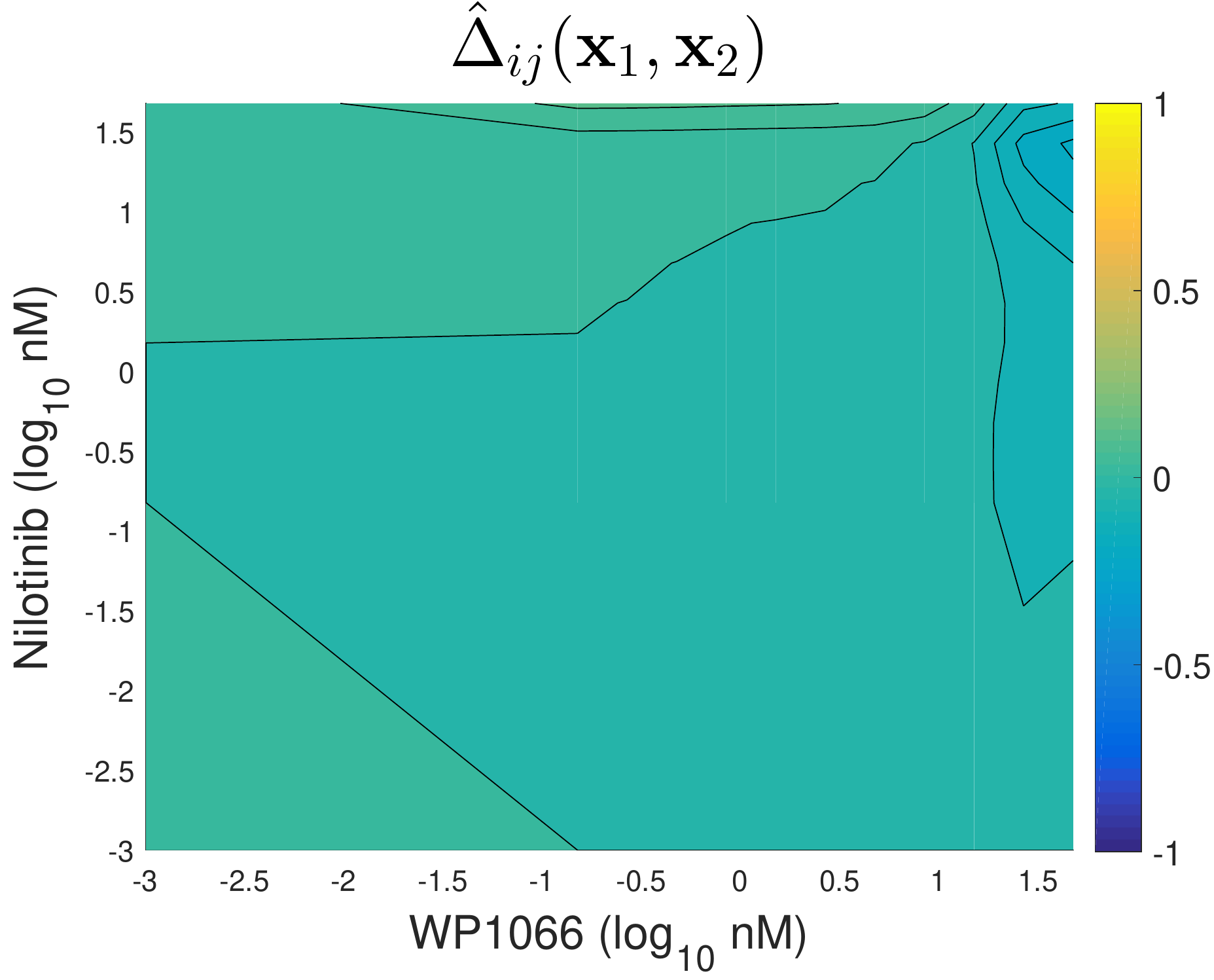}}
\subfloat[SKOV3]{\includegraphics[height=.4\textwidth]{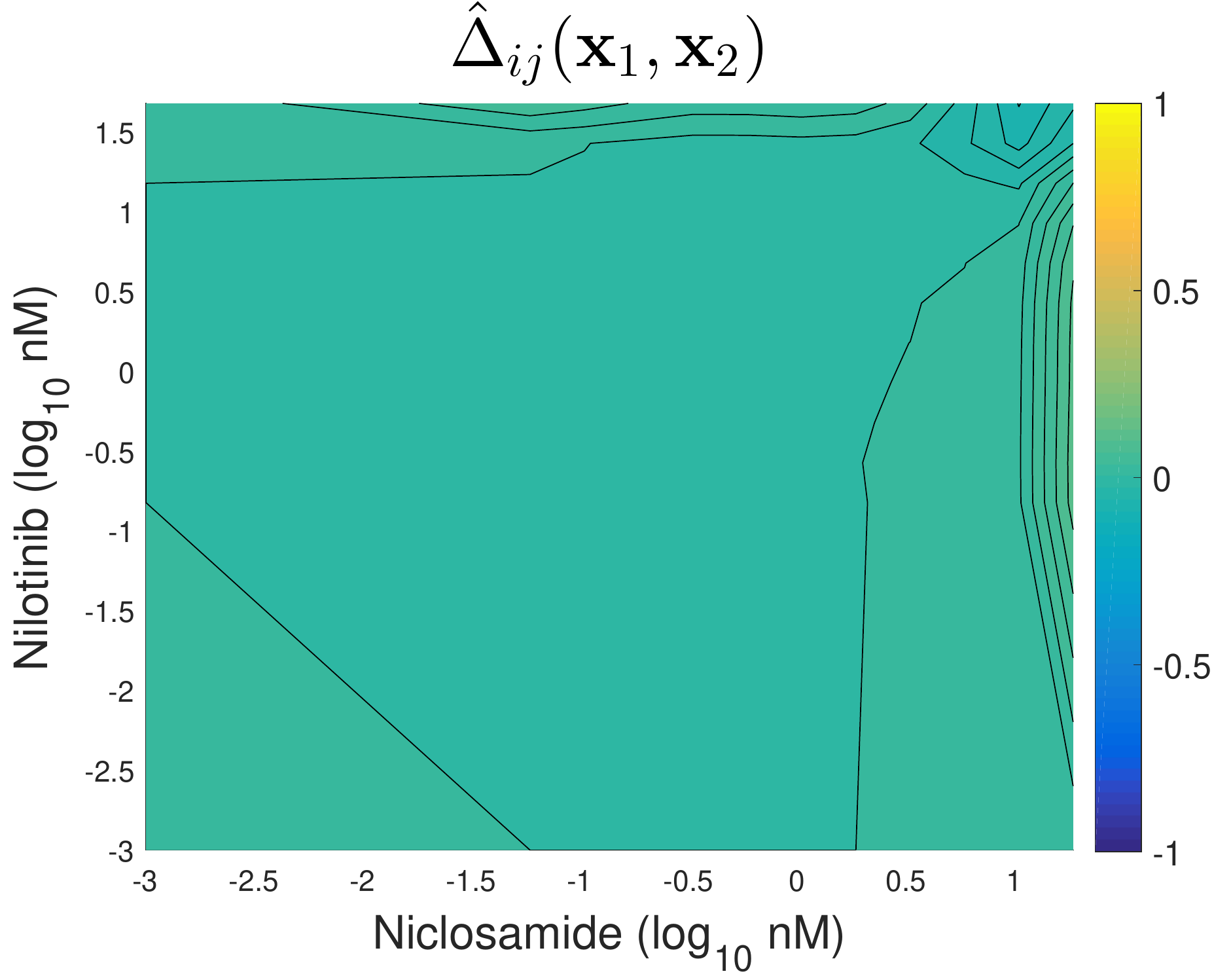}}
\caption{Application to Ovarian Cancer data -- Contour plots of the posterior mean of the interaction surface $\Delta_{ij}$, including the monotherapy $\text{EC}_{50}$ estimates as magenta triangles. The triangles are not visible when the estimated $\text{EC}_{50}$'s are outside the considered range of concentrations. Experiments with cell-lines treated in medium and ascites. Top row: OVCAR8 ; bottom row: SKOV3.}
  \label{fig:OC_Ascites_contour_Delta}
\end{figure}

\clearpage
\section{MCMC Algorithm}\label{sec:MCMC_Suppl}

As introduced in the main text, we resort to Metropolis-Hastings type updates in throughout the MCMC posterior sampling, since most of the parameters involved in this study are not conjugate to the model. In particular, to ensure faster convergence and good mixing properties of the posterior MCMC chains, we use adaptive algorithms \cite{Griffin_Stephens_2013}. The main idea at the basis of such algorithms is to re-calculates the parameters of the proposal distributions using the MCMC samples previously produced by the sampling process. Let $\bm \theta \in \Theta$, with $\Theta \subset \mathbb{R}^d$, be a random vector of dimension $d$, and suppose we want to sample from its posterior distribution with probability density function $p(\bm \theta | \bm y) \propto f(\bm y | \bm \theta)\pi(\bm \theta)$. As usual, we refer to $f(\bm y | \bm \theta)$ and $\pi(\bm \theta)$ as the likelihood and prior distribution, respectively. First, we consider a transformation $t: \Theta \rightarrow \mathbb{R}^d$, such that $t(\bm \theta)$ has full support on $\mathbb{R}^{d}$. If $\Theta = \mathbb{R}^d$, then we resort to the identity transformation $t(\bm \theta) = \bm \theta$. At the $g$-th iteration of the algorithm, we propose a new value $\bm \theta^*$ such that:
\begin{equation}\label{eq:AMHproposal}
t(\bm \theta^*)= t(\bm \theta^{(g)}) + \bm \epsilon, \quad \bm \epsilon\sim N(t(\bm \theta^{(g)}), \bm \xi^{(g)})
\end{equation}
and accept $\bm \theta^{(g+1)} = \bm \theta^*$ with probability:
\begin{equation}\label{eq:AMHratio}
a(\bm \theta^*,\bm \theta^{(g)}) = \min \left( 1, \frac{p(\bm \theta^* | \bm y)}{p(\bm \theta^{(g)} | \bm y)} \frac{|\mathcal{J}_t(\bm \theta^{(g)})|}{|\mathcal{J}_t(\bm \theta^*)|} \right),
\end{equation}
where $|\mathcal{J}_t(\bm \theta)|$ denotes the determinant of the Jacobian of the transformation $t$ evaluated at $\bm \theta$. After an initial number $g_0$ of standard (i.e., non-adaptive) Metropolis-Hastings steps ($g_0 = 1.000$ in the simulations), the algorithm updates the covariance matrix of the random walk proposal, for which $\bm \xi^{(g+1)}\equiv \bm \xi^{(0)}$, in this way:
\begin{align*} \bm \xi^{(g+1)} = \frac{s^{(g+1)}}{g-1} \left[ \sum\limits_{g'=1}^{g} t_{g'} {t_{g'}} ^\top -  \frac{1}{g}\left( \sum\limits_{g'=1}^{g} t_{g'} \right) \left( \sum\limits_{g'=1}^{g} t_{g'} \right)^\top \right] + s^{(g+1)} \bm \epsilon \mathbb{I}_{d}
\end{align*}
where $t_g \equiv t(\bm \theta^{(g)})$, $x^\top$ denotes the transposed of the vector $x$, and $\mathbb{I}_{d}$ denotes the $d$-dimensional identity matrix. The magnitude of the adaptation is determined by the quantity $s^{(g+1)}$, which is tuned via the acceptance probability $a(\bm \theta^*,\bm \theta^{(g)})$ at each iteration. Further details can be found in \citep{Griffin_Stephens_2013}. We use this approach within a Gibbs sampling scheme to update the parameters of the proposed model. In the following section, we provide the detailed expressions of the full-conditionals distributions used in MCMC algorithm.

\subsection{Additional details}

Let $f(\bm y | \bm \theta)$ denote the likelihood distribution for the proposed model, which in the proposed model corresponds to:
\begin{equation}
f(\bm y | \bm p, \sigma^2_{\epsilon}) = \prod_{i = 0}^{n_1}\prod_{j = 0}^{n_2}\prod_{r = 0}^{n_{rep}}N(y^r_{ij} | p_{ij}, \sigma^2_{\epsilon}),
\end{equation}
where we address the reader to formulas (4) in the main manuscript for the expression of the mean parameter $p_{ij}$. However, details on the parts that are modified in each proposal step are highlighted in the following part. In what follows, $N$ and $N_d$ represents a univariate and a $d$-variate Gaussian distribution, respectively.

\begin{itemize}

\item Update $m_1$ (analogously for $m_2$): propose $m_1^* \sim N(m_1^{(g)}, \xi^{(g)}_{m_1})$ and accept with probability \eqref{eq:AMHratio}, where $|\mathcal{J}_t(x)| = 1$ for $x = m_1^{(g)}, m_1^*$, and:
\begin{align*}
p^{0*}_{ij} &= f(x_{1i}|m^*_1, \lambda_1)f(x_{2j}|m_2, \lambda), \\
p^*_{ij} &=  p^{0*}_{ij} + \Delta_{ij},
\end{align*}
for $i = 0, \dots, n_1$ and $j = 0, \dots, n_2$. Including the likelihood and the prior distribution, we obtain the acceptance ratio:
$$
a(m_1^*,m_1^{(g)}) = \frac{f(\bm y | \bm p^*, \sigma^2_{\epsilon})}{f(\bm y | \bm p, \sigma^2_{\epsilon})} e^{-\frac{1}{2\sigma^2_{m_1}} \left( (m_1^*)^2 - (m_1^{(g)})^2 \right)}
$$

\item Update $\lambda_1$ (analogously for $\lambda_2$): propose $\log(\lambda_1^*) \sim N(\log(\lambda_1^{(g)}), \xi^{(g)}_{\lambda_1})$ and accept with probability \eqref{eq:AMHratio}, where $|\mathcal{J}_t(x)| = x^{-1}$ for $x = \lambda_1^{(g)}, \lambda_1^*$, and:
\begin{align*}
p^{0*}_{ij} &= f(x_{1i}|m_1, \lambda^*_1)f(x_{2j}|m_2, \lambda), \\
p^*_{ij} &=  p^{0*}_{ij} + \Delta_{ij},
\end{align*}
for $i = 0, \dots, n_1$ and $j = 0, \dots, n_2$. Including the likelihood and the prior distribution, we obtain the acceptance ratio:
$$
a(\lambda_1^*,\lambda_1^{(g)}) = \frac{f(\bm y | \bm p^*, \sigma^2_{\epsilon})}{f(\bm y | \bm p, \sigma^2_{\epsilon})} \left( \frac{\lambda_1^*}{\lambda_1^{(g)}} \right)^{\alpha_{\lambda_1}} e^{-\beta_{\lambda_1}(\lambda_1^* - \lambda_1^{(g)})}.
$$

\item Update $\bm b = (b_1, b_2)$ (jointly): propose $\log(\bm b^*) \sim N_2(\log(\bm b), \xi^{(g)}_{\bm b})$ and accept with probability \eqref{eq:AMHratio}, where $|\mathcal{J}_t(\bm x)| = \prod x^{-1}$ for $\bm x = \bm b, \bm b^*$, and:
\begin{align*}
g^*(B_{ij}) &= - p^0_{ij} (1 + e^{b^*_1 B_{ij}})^{-1} + (1 - p^0_{ij}) (1 + e^{-b^*_2 B_{ij}})^{-1}, \nonumber\\
\Delta^*_{ij} &= g^*(B_{ij})\mathbb{I}_{\{i,j > 0\}}(B_{ij}), \nonumber\\
p^*_{ij} &=  p^0_{ij} + \Delta^*_{ij},
\end{align*}
for $i = 0, \dots, n_1$ and $j = 0, \dots, n_2$. Including the likelihood and the prior distribution, we obtain the acceptance ratio:
$$
a(\bm b^*,\bm b^{(g)}) = \frac{f(\bm y | \bm p^*, \sigma^2_{\epsilon})}{f(\bm y | \bm p, \sigma^2_{\epsilon})} \left( \frac{b_1^*}{b_1^{(g)}} \right)^{\alpha_{b_1}} \left( \frac{b_2^*}{b_2^{(g)}} \right)^{\alpha_{b_2}} e^{-\beta_{b_1}(b_1^* - b_1^{(g)}) -\beta_{b_2}(b_2^* - b_2^{(g)})}.
$$

\item Update $\gamma_0$ (analogously for $\gamma_1$ and $\gamma_2$): \\ propose $\log(\gamma_0^*) \sim N(\log(\gamma_0^{(g)}), \xi^{(g)}_{\gamma_0})$ and accept with probability \eqref{eq:AMHratio}, where $|\mathcal{J}_t(x)| = x^{-1}$ for $x = \gamma_0^{(g)}, \gamma_0^*$, and:
\begin{align*}
B^*_{ij} &= \gamma^*_0 + \gamma_1 x_{1i} + \gamma_2 x_{2j} + B(x_{1i} , x_{2j}), \\
g(B^*_{ij}) &= - p^0_{ij} (1 + e^{b_1 B^*_{ij}})^{-1} + (1 - p^0_{ij}) (1 + e^{-b_2 B^*_{ij}})^{-1}, \nonumber\\
\Delta^*_{ij} &= g(B^*_{ij})\mathbb{I}_{\{i,j > 0\}}(B^*_{ij}), \nonumber\\
p^*_{ij} &=  p^0_{ij} + \Delta^*_{ij},
\end{align*}
for $i = 0, \dots, n_1$ and $j = 0, \dots, n_2$. Including the likelihood and the prior distribution, we obtain the acceptance ratio:
$$
a(\gamma_0^*,\gamma_0^{(g)}) = \frac{f(\bm y | \bm p^*, \sigma^2_{\epsilon})}{f(\bm y | \bm p, \sigma^2_{\epsilon})}e^{-\frac{1}{2\sigma^2_{\gamma_0}} \left( (\gamma_0^*)^2 - (\gamma_0^{(g)})^2 \right)}.
$$

\item Update $\bm C$ (jointly): propose $\bm C^* \sim N_{K_1K_2}(\text{vec}(\bm C^{(g)}), \xi^{(g)}_{\bm C})$ and accept with probability \eqref{eq:AMHratio}, where $|\mathcal{J}_t(\bm x)| = \prod x^{-1}$ for $\bm x = \bm C^{(g)}, \bm C^*$, and:
\begin{align*}
B^*(x_{1i} , x_{2j}) &= \sum\limits_{l,m} \bm C^*_{lm} B_l(x_{1i})B_m(x_{2j}), \nonumber\\
B^*_{ij} &= \gamma_0 + \gamma_1 x_{1i} + \gamma_2 x_{2j} + B^*(x_{1i} , x_{2j}), \\
g(B^*_{ij}) &= - p^0_{ij} (1 + e^{b_1 B^*_{ij}})^{-1} + (1 - p^0_{ij}) (1 + e^{-b_2 B^*_{ij}})^{-1}, \nonumber\\
\Delta^*_{ij} &= g(B^*_{ij})\mathbb{I}_{\{i,j > 0\}}(B^*_{ij}), \nonumber\\
p^*_{ij} &=  p^0_{ij} + \Delta^*_{ij},
\end{align*}
for $i = 0, \dots, n_1$ and $j = 0, \dots, n_2$. Including the likelihood and the prior distribution, we obtain the acceptance ratio:
$$
a(\bm C^*,\bm C^{(g)}) = \frac{f(\bm y | \bm p^*, \sigma^2_{\epsilon})}{f(\bm y | \bm p, \sigma^2_{\epsilon})}e^{-\frac{1}{2} \left( \text{tr}(\Psi_1^{-1} (\bm C^*)^\top \Psi_2^{-1} \bm C^*) - \text{tr}(\Psi_1^{-1} (\bm C^{(g)})^\top \Psi_2^{-1} \bm C^{(g)})) \right)}.
$$

\item Update $\sigma^2_{\phi}$, $\phi \in \{ m_1, m_2, \gamma_0, \gamma_1, \gamma_2 \}$. This step depends on the choice of the prior distribution, that in this work is chosen to be either a half-Cauchy$(h)$, for $h >0$, or an inverse-Gamma IG$(\alpha, \beta)$. 

\begin{itemize} 
\item[HC$(h)$] Propose $\log(\sigma_{\phi}^*) \sim N(\log(\sigma_{\phi}^{(g)}), \xi^{(g)}_{\sigma_{\phi}})$ and accept with probability \eqref{eq:AMHratio}, where $|\mathcal{J}_t(x)| = x^{-1}$ for $x = \sigma_{\phi}^{(g)}, \sigma_{\phi}^*$, and we obtain the acceptance ratio:
$$
a(\sigma_{\phi}^*,\sigma_{\phi}^{(g)}) = \ e^{-\frac{\phi^2}{2} \left( \frac{1}{(\sigma_{\phi}^*)^2} - \frac{1}{(\sigma_{\phi}^{(g)})^2} \right)} \frac{(\sigma_{\phi}^{(g)})^2 + h^2}{(\sigma_{\phi}^*)^2 + h^2}.
$$

\item[IG$(\alpha, \beta)$] The conjugate sampler has full conditional:
\begin{align*}
p(\sigma_{\phi} | \bm y, \bm \theta) &\sim \text{IG}(\alpha_{\phi}, \beta_{\phi}), \\
\alpha_{\phi} &= \alpha + \frac{1}{2}, \\
\beta_{\phi} &= \beta + \frac{1}{2} \phi^2.
\end{align*}

\end{itemize}

\item Update $\sigma^2_{\epsilon}$. This step depends on the choice of the prior distribution, that in this work is chosen to be either a half-Cauchy$(h)$, for $h >0$, or an inverse-Gamma IG$(\alpha, \beta)$. 

\begin{itemize} 
\item[HC$(h)$] Propose $\log(\sigma_{\epsilon}^*) \sim N(\log(\sigma_{\epsilon}^{(g)}), \xi^{(g)}_{\sigma^2_{\epsilon}})$ and accept with probability \eqref{eq:AMHratio}, where $|\mathcal{J}_t(x)| = x^{-1}$ for $x = \sigma_{\epsilon}^{(g)}, \sigma_{\epsilon}^*$, and we obtain the acceptance ratio:
\begin{align*}
&a(\sigma_{\epsilon}^*,\sigma_{\epsilon}^{(g)}) = \\
&\left( \frac{(\sigma_{\epsilon}^*)^2}{(\sigma_{\epsilon}^{(g)})^2} \right)^{(1 - n_{rep} n_1 n_2)/2} \ e^{-\sum\limits_{i,j,r}\frac{(y^r_{ij} - p_{ij})^2}{2} \left( \frac{1}{(\sigma_{\epsilon}^*)^2} - \frac{1}{(\sigma_{\epsilon}^{(g)})^2} \right)} \frac{(\sigma_{\epsilon}^{(g)})^2 + h^2}{(\sigma_{\epsilon}^*)^2 + h^2}.
\end{align*}

\item[IG$(\alpha, \beta)$] The conjugate sampler has full conditional:
\begin{align*}
p(\sigma_{\epsilon} | \bm y, \bm \theta) &\sim \text{IG}(\alpha_{\epsilon}, \beta_{\epsilon}), \\
\alpha_{\epsilon} &= \alpha + \frac{n_{rep} n_1 n_2}{2}, \\
\beta_{\epsilon} &= \beta + \frac{1}{2} \sum\limits_{i,j,r}(y^r_{ij} - p_{ij})^2.
\end{align*}
\end{itemize}

\end{itemize}

\clearpage
\bibliographystyle{plainnat}
\bibliography{Biblio}